\pdfoutput=1

\documentclass[reprint,aps,prb,twocolumn,floatfix,superscriptaddress]{revtex4-2}
\usepackage{graphicx}
\usepackage{amssymb}
\usepackage{bm}
\usepackage{amsmath}
\usepackage{dsfont}
\usepackage{lineno}
\usepackage[english]{babel}
\usepackage[utf8]{inputenc}
\usepackage{xcolor}
\RequirePackage[T1]{fontenc}
\RequirePackage{lmodern}
\newcommand{\red}[1]{\textcolor{black}{#1}}

\usepackage{appendix}
\usepackage{float}
\usepackage{algpseudocode}
\usepackage{algorithm}
\usepackage{mathtools}
\usepackage{filecontents}
\usepackage{multibib}
\usepackage{comment}
\usepackage{gensymb}
\usepackage{siunitx}

\begin{document}

\def\E{\mathbb E}
\def\H{\mathcal H}
\def\N{\mathcal N}
\def\P{\mathbb P}
\def\Q{\mathbb Q}
\def\R{\mathbb R}
\def\bfp{\mathbf p}
\def\bfq{\mathbf q}
\def\bfx{\mathbf x}
\def\V{\mathbb V}
\def\argmax{\mathrm{argmax}}
\def\argmin{\mathrm{argmin}}
\def\sgn{\mathrm{sgn}}
\newcommand\given[1][]{\:#1\vert\:}

\title{Practical Strategies for Enhancing the Valley Splitting in Si/SiGe Quantum Wells}

\author{Merritt P. Losert}
\affiliation{Department of Physics, University of Wisconsin-Madison, Madison, WI 53706 USA}
\author{M. A. Eriksson}
\affiliation{Department of Physics, University of Wisconsin-Madison, Madison, WI 53706 USA}
\author{Robert Joynt}
\affiliation{Department of Physics, University of Wisconsin-Madison, Madison, WI 53706 USA}
\author{Rajib Rahman}
\affiliation{School of Physics, University of New South Wales, Sydney, Australia}
\author{Giordano Scappucci}
\affiliation{QuTech and Kavli Institute of Nanoscience, Delft University of Technology, Delft, The Netherlands}
\author{S. N. Coppersmith}
\affiliation{School of Physics, University of New South Wales, Sydney, Australia}
\author{Mark Friesen}
\affiliation{Department of Physics, University of Wisconsin-Madison, Madison, WI 53706 USA}

\date{\today}

\begin{abstract}
Silicon/silicon-germanium heterostructures have many important advantages for hosting spin qubits. 
However, controlling the valley splitting (the energy splitting between the two low-lying conduction-band valleys) remains a critical challenge for ensuring qubit reliability. 
Broad distributions of valley splittings are commonplace, even among quantum dots formed on the same chip. 
\red{
In this work, we theoretically explore the interplay between quantum-well imperfections that suppress the valley splitting and cause variability, such as broadened interfaces and atomic steps at the interface, while self-consistently accounting for germanium concentration fluctuations.
We consider both conventional and unconventional approaches for controlling the valley splitting, and present concrete strategies for implementing them.
Our results provide a clear path for achieving qubit uniformity in a scalable silicon quantum computer.}
\end{abstract}

\maketitle

\section{Introduction} \label{introduction}

Qubits formed from quantum dots in Si/SiGe heterostructures are promising candidates for large-scale quantum computing~\cite{Loss:1998p120, Zwanenburg:2013p961, Zhang:2018p32}. Naturally abundant spin-zero nuclear isotopes and the highly developed infrastructure of the semiconductor industry lend a particular advantage to silicon-based qubits. Recently, one and two-qubit gates in spin qubits in Si/SiGe quantum dots have yielded fidelities above 99\% \cite{Xue:2022p343, Noiri:2022p338, Mills:2022p5130}, attesting to the viability of this materials platform. 

However, qubits in Si suffer from the degeneracy of low-energy features in the conduction band, known as valleys. 
Although the sixfold valley degeneracy of bulk silicon is lifted by tensile strain in the quantum well, a nearly degenerate excited valley state remains accessible to electrons in quantum dots~\cite{Zwanenburg:2013p961}. The energy splitting between these low-lying valley states, $E_v$, can range widely from 20 to 300~$\mu$eV \cite{Borselli:2011p123118, Shi:2011p233108, Zajac:2015p223507, Scarlino:2017p165429, Mi:2017p176803, Ferdous:2018p26, Mi:2018p161404, Neyens:2018p243107, Borjans:2019p044063, Hollmann:2020p034068, Oh:2021p125122}, even for devices fabricated on the same chip~\cite{Chen:2021p044033,Wuetz:2022p7730}. 
When the valley splitting is too low, the excited state provides a leakage path outside the logical space of the spins, posing a significant threat for qubit operations~\cite{Zwanenburg:2013p961}. To date, it has not been possible to engineer devices with reliably high valley splittings.

In conventional SiGe/Si/SiGe heterostructures, the valley splitting is determined by the quantum well confinement potential. Accordingly, the variability of $E_v$ is attributed to the variability of the interfaces. Such behavior has been well studied theoretically, using tight-binding~\cite{Boykin:2004p115, Kharche:2007p092109, Klimeck:2007p1079, Klimeck:2007p2090, Abadillo-Uriel:2018p165438, Klos:2018p155320, Ercan:2022p247701, Ercan:2021p235302,Dodson:2022p146802} and effective-mass methods~\cite{Friesen:2007p115318, Chutia:2008p193311, Saraiva:2011p155320, Gamble:2013p035310, Gamble:2016p253101, Tariq:2019p125309, Hosseinkhani:2020p043180}. Studies have focused on heterostructure parameters such as the width of the interface~\cite{Chen:2021p044033} or quantum well~\cite{Boykin:2004p115, Friesen:2007p115318}. Additional variability is caused by imperfections and disorder. Tilted interfaces and single-atom steps, in particular, have been studied extensively~\cite{Friesen:2006p202106, Goswami:2007p41, Kharche:2007p092109, Culcer:2010p205315, Friesen:2010p115324, Gamble:2013p035310, Gamble:2016p253101, Boross:2016p035438, Tariq:2019p125309, Hosseinkhani:2020p043180, Hosseinkhani:2021p085309, Dodson:2022p146802}. 
Experimental work has validated some of these predictions. For example in Si metal-oxide-semiconductor (Si-MOS) stacks, where the semiconductor-dielectric interface is characteristically sharp, interface roughness has been shown to correlate with the valley splitting~\cite{Gamble:2016p253101}.

An additional type of disorder is present in heterostructures containing SiGe alloy. 
In this case, the crystal lattice sites are filled randomly with Si or Ge atoms, as determined by the average concentration profile.
For Si/SiGe heterostructures, it has recently been shown that such uncorrelated, random alloy disorder can have a dominant effect on the intervalley coupling $\Delta$~\cite{Wuetz:2022p7730}. 
Specifically, $\Delta$ can be decomposed into two components: (1) an average, `deterministic’ component $\Delta_0$, which is largely uniform across a sample, and (2) a random component $\delta \Delta$, which varies significantly by location. Here, $\Delta_0$ is determined by the smooth quantum well confinement potential, while $\delta\Delta$ arises from local Ge fluctuations caused by alloy disorder. Since $E_v = 2|\Delta|$, large variations in $\delta\Delta$ lead to large variations in $E_v$, as verified experimentally in quantum dots \cite{McJunkin:2022p7777, Wuetz:2022p7730}.

\red{
In this paper, we show that sharp features in the heterostructure profile, like a sharp interface, can enhance the valley splitting, while random alloy disorder strongly suppresses this effect.
The crossover between these two types of behavior occurs in a regime where heterostructure features are abrupt and difficult to achieve in the laboratory.
Deterministically enhanced valley splittings are therefore difficult to achieve by sharp interfaces alone.}

\red{
To better understand this crossover, we consider several `conventional' heterostructures, where we characterize competing effects like sharp interfaces vs.\ interface steps.
We find that steps can be detrimental to valley splitting; however they have essentially no effect for interface widths of three or more atomic monolayers.
In the randomly dominated regime, we show that when the electron is exposed to more Ge, it experiences a larger average valley splitting  and a larger variability.}

\red{
We also characterize unconventional geometries like the Wiggle Well, which yields the greatest improvements to the valley splitting, but is challenging to grow in the laboratory.
We compare this to an alternative geometry, with uniform Ge in the quantum well, where the mean and standard deviation of the valley splitting are both enhanced.
We argue that such structures provide a more reliable approach for improving qubits, if they can be electrostatically tuned to locations with desirable valley splittings.
We finally argue that both of these approaches are superior to sharp interfaces, and show they are optimal in certain operating regimes.}

The paper is organized as follows. 
In Sec.~\ref{sec:intuitive}, we study the dominant sources of $E_v$ variation and how they interact, and we explain these behaviors in the context of a universal theory of valley splitting. 
Expanding on ideas first presented in Refs.~\cite{Wuetz:2022p7730} and \cite{McJunkin:2022p7777}, we show that valley splitting depends fundamentally on the strength of the quantum well confinement potential at the special reciprocal-space wavevector $2 k_0$ (we refer to this as `$2k_0$ theory'), which is the distance between the two $z$ valleys in the first Brillouin zone. We go on to show that deterministic and random-alloy effects can both contribute to this $2k_0$ wavevector. Interface width, atomic steps, alloy disorder, and other features can therefore be studied and compared within a single analytic framework, providing intuition as well as quantitative predictions.

\begin{figure}[t] 
	\includegraphics[width=8cm]{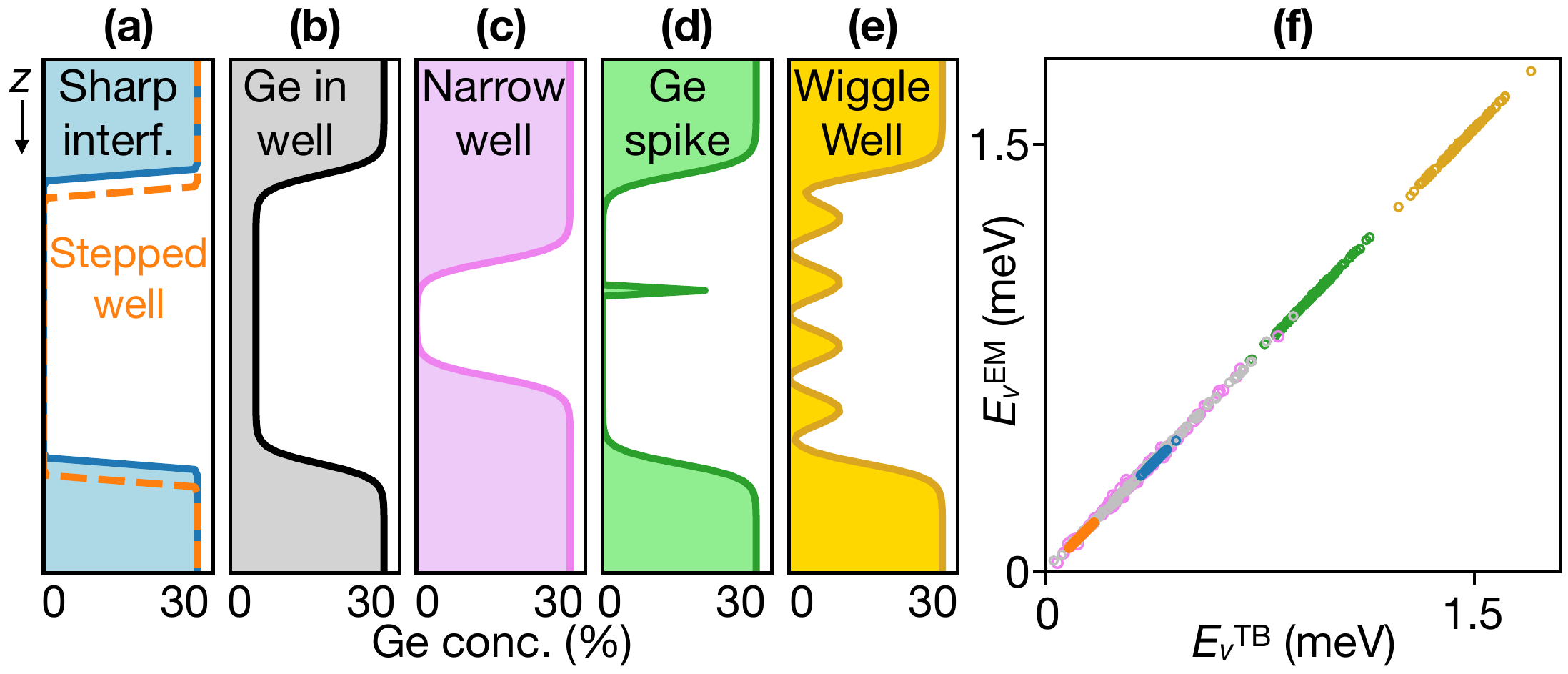}
	\centering
	\caption{ Confirmation of the universal $2k_0$ theory of valley splitting. 
(a)-(e) Schematic illustrations of several of the Si/SiGe heterostructures considered in this work: 
(a) a quantum well with a sharp interface and no interface steps (blue), or with one interface step (orange);
(b) a quantum well with a uniform concentration of Ge;
(c) a narrow quantum well;
(d) a quantum well with a single-monolayer spike of Ge;
(e) a Wiggle Well.
(f) Valley-splitting correlation plot for the structures shown in (a)-(e), with the same color coding used in those panels. 
On one axis, we plot tight-binding (TB) results for the valley splitting, $E_v^\text{TB}$.
On the other axis, we plot the universal $2k_0$ prediction, based on effective-mass (EM) theory, $E_v^\text{EM}$, as defined in Eqs.~(\ref{discreteDelta}) and (\ref{discreteDelta2D}), where $E_v=2|\Delta|$.
(See main text for simulation details.)
For each different heterostructure, we generate 100 instances of alloy disorder for a given, average Ge concentration profile, as described in the main text, using the same disorder profiles for both the TB and EM simulations.
All results fall onto a universal curve, demonstrating the validity of the $2k_0$ theory.
}
	\label{fig:heterostructureCartoon}
\end{figure}

In Sec.~\ref{sec:methods}, we outline the theoretical methods employed here, including tight-binding models and effective-mass theory.
We investigate the effects of alloy disorder on the valley splitting in dots formed in SiGe/Si/SiGe quantum wells. 
This disorder leads to large $E_v$ variations, which we show depend on the amount of Ge the wavefunction is exposed to. 
Such variations also increase the mean value of the valley splitting, and they significantly reduce the fraction of dots with low valley splittings. 
Here, we make a crucial distinction between quantum wells in which all quantum dots (except a vanishingly small subset) have valley splittings that are large enough for qubit applications, and dots with a wide distribution of valley splittings, extending all the way to zero energy. 
We refer to the former category as `deterministically enhanced’ and the latter as `randomly dominated.' 
By simulating many random instances of alloy disorder, we show that there is a sharp transition between these two types of behavior. 
We also obtain an analytical expression for the crossover, using the statistical properties of random alloys.
We then show that nearly all recent experiments are of the randomly dominated type, with important implications for scaling up to large numbers of qubits.
We further show that when physical limitations, such as growth constraints, do not permit the formation of heterostructures with very sharp features (on the order of 1-2 atomic layers), the resulting devices fall into the randomly dominated category. 
For such structures, it is generally more effective to increase the average valley splitting by increasing the wavefunction exposure to Ge.

In Sec.~\ref{sec:conventional}, we use our theoretical toolbox to examine conventional Si/SiGe heterostructures.
We study the interactions between alloy disorder and interface steps in conventional Si/SiGe heterostructures as a function of the interface width [Fig.~\ref{fig:heterostructureCartoon}(a)]. 
For devices with sharp interfaces, steps are found to strongly suppress the valley splitting, as is well known. However, for devices with wider interfaces, the steps are found to have little or no effect on the valley splitting. 
In this regime, the valley splitting depends mainly on the local alloy disorder, and we show that this disorder leads to large $E_v$ variations as a function of dot position. 

In Sec.~\ref{sec:specialized}, we consider unconventional heterostructures proposed to boost the valley splitting [Figs.~\ref{fig:heterostructureCartoon}(b)-\ref{fig:heterostructureCartoon}(e)], by adding Ge to the interior or the boundary of the quantum well. These include Ge-rich barrier layers~\cite{Neyens:2018p243107}, and other more-complicated superlattice barrier structures~\cite{Zhang:2013p2396, Wang:2022p165308}, single-atom spikes of Ge inside the quantum well~\cite{McJunkin:2021p085406}, narrow quantum wells~\cite{Boykin:2004p115, Friesen:2007p115318, Chen:2021p044033}, and oscillating Ge concentrations with specially chosen oscillation wavelengths (e.g., the `Wiggle Well’~\cite{McJunkin:2022p7777, Feng:2022p085304}). 
We analyze these designs and characterize their deterministic and random-alloy contributions to valley splitting, which allows us to compare ideal performance to actual operation.

To close this section, we apply optimization procedures to determine Ge concentration profiles that maximize the valley splitting, using two different optimization strategies. 
First, we maximize the deterministic valley splitting $E_{v0}=2|\Delta_0|$, without including alloy disorder. This approach yields heterostructures with concentration oscillations very similar to the short-period Wiggle Well, confirming the optimality of that structure. 
In the second approach, we maximize the standard deviation $\sigma_\Delta$, which can be shown to maximize the average valley splitting in the randomly dominated regime.
This approach yields smooth Ge concentration profiles centered in the middle of the quantum well.

In Sec.~\ref{sec:summary}, we summarize our main results, and finally in Sec.~\ref{sec:conclusions}, we describe the best forward-looking strategies for enhancing the valley splitting, which can be used to guide future experiments in Si/SiGe heterostructures.
Here we argue that the Wiggle Well is the preferred approach, in the deterministically enhanced regime.
In the randomly dominated regime, we argue that the best approach is to introduce uniform Ge into the quantum well and then electrostatically tune the dot position, to find a location where the valley splitting is suitable.

\begin{figure*}[t] 
	\includegraphics[width=15cm]{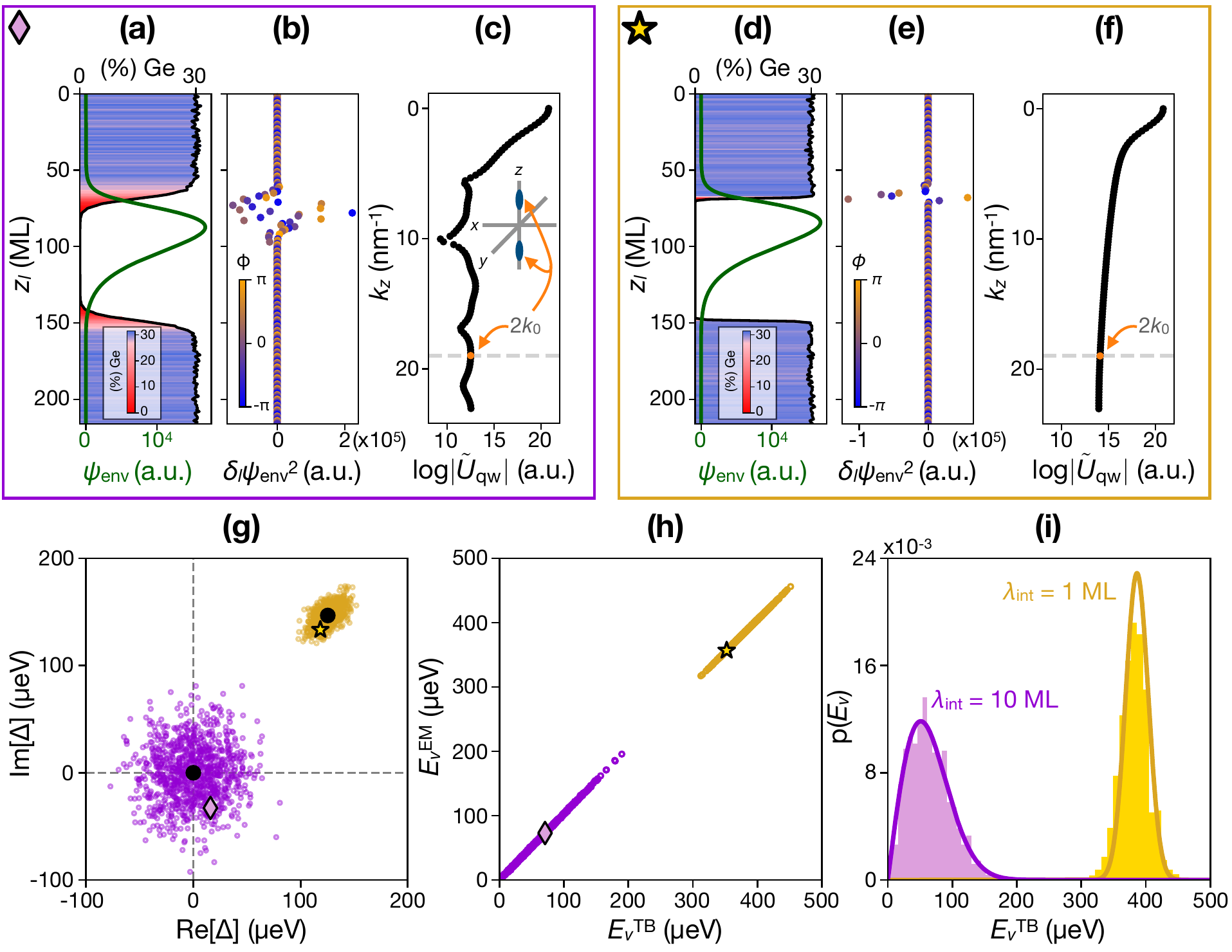}
	\centering
	\caption{
 Effective-mass analysis of the effects of alloy disorder for quantum wells with (a)-(c) wide interfaces, or (d)-(f) super-sharp interfaces.
 (a), (d) Typical Ge concentration profiles along the growth direction ($\hat z$).
 The local Ge concentration is also shown as a color scale.
 Here, the average Si concentration profiles are given by sigmoid functions [Eq.~(\ref{eq:sigmoid})], with quantum well and substrate concentrations given by $X_w = 1$ and $X_s = 0.7$, respectively, and interface widths given by $\lambda_\text{int} = 10$~ML for (a) or $\lambda_\text{int} = 1$~ML for (d), where ML stands for atomic monolayers, corresponding to the length $a_0/4$.
Small fluctuations in Ge concentration occur from location to location due to alloy disorder and the finite size of the quantum dot.
 We also show here the envelope functions $\psi_\text{env}(z)$ (solid green curves), used to calculate the intervalley matrix element $\Delta$ [Eq.~(\ref{discreteDelta})], obtained by solving a 1D Schr\"{o}dinger equation for the quantum well confinement potential $U_\text{qw}(z)$ [Eq.~(\ref{Uqw})].
 (b), (e) Individual terms in the the $\delta\Delta$ sum [Eq.~(\ref{deltaDelta})], plotted as a function of the vertical position $z_l$.
 The corresponding complex phases, $\phi=-2k_0z_l$, are indicated as a color scale. 
 (c), (f) Discrete Fourier transform $\tilde U_\text{qw}(k_z)$ of the weighted confinement potential $\tilde U_\text{qw}(z_l)=U_\text{qw}(z_l)|\psi_\text{env}(z_l)|^2$ appearing in Eq.~(\ref{discreteDelta}), for the quantum wells shown in (a) and (d).
 The $2k_0$ wavevector component of $\tilde U_\text{qw}$ (gray dashed line) couples valleys within the first Brillouin zone, causing valley splitting.
 (g) Such $\Delta$ calculations are repeated for 1,000 different disorder realizations, for the same two interfaces (wide vs.\ sharp), obtaining the purple and gold distributions, which are plotted here on the complex plane.
The highlighted points (purple diamond, gold star) correspond to the specific results shown in (a)-(f).
The central black points correspond to the deterministic centers of the distributions, $\Delta_0$. 
For wide interfaces (purple), we observe that the $\Delta$ distributions are broader and centered much closer to the origin.
In these cases, the standard deviation of the distribution, $\sigma_\Delta$, is much larger than the mean value, $\Delta_0$, while the opposite is true for sharp interfaces (gold).
 (h) Effective-mass valley splittings ($E_v^\text{EM}=2|\Delta|$) are obtained from (g) and plotted against 1D tight-binding calculations ($E_v^\text{TB}$), using the same disorder realizations.
The collapse of the data onto a line of slope 1 indicates nearly perfect correlations between the two methods.
 (i) Histogram plot of the tight-binding results shown in (h).
 Solid lines show the computed Rice distributions [Eq.~(\ref{riceDistrib})].}
	\label{fig:introFourier}
\end{figure*}

\section{A universal picture of valley splitting} \label{sec:intuitive}

A primary result of this paper is the unified understanding of how all key features in a heterostructure, from deterministic to random, work together to determine the valley splitting in Si/SiGe devices. In this section, we present an intuitive outline of the physics, with the details left to later sections.

In Si/SiGe quantum wells, the degenerate $\pm z$ valleys are separated in the first Brillouin zone by the wavevector $2 k_0$, as indicated schematically in the inset of Fig.~\ref{fig:introFourier}(c). The degeneracy of the valley states is lifted in a process known as `valley splitting,' which occurs when the quantum well confinement potential, determined by the Ge concentration profile, has Fourier components at this special wavevector. More precisely, effective-mass theory states that the valley splitting energy $E_v$ is proportional to the Fourier transform of the quantum well potential, weighted by the dot probability density $|\psi(\mathbf{r})|^2$, evaluated at the wavevector $2k_0$. We call this the `$2k_0$ theory.'

This simple description of valley splitting has wide-ranging explanatory power, which is both qualitative and quantitative. For example, it is the basis for the short-period Wiggle Well~\cite{McJunkin:2022p7777, Feng:2022p085304}, which exploits heterostructures where the $2k_0$ wavevector is intentionally engineered into the quantum well.
It further explains how other engineered sharp features in the confinement potential, such as sharp interfaces or Ge spikes, can enhance $E_v$. In the latter case, sharp features in real space produce broad Fourier spectra in $k$ space, including components at the wavevector $2k_0$. The $2k_0$ theory also explains the random effects of alloy disorder. Here, since the heterostructure is composed of individual atoms, there will be fluctuations of the Ge concentration from layer to layer inside a finite-size dot. These random fluctuations alter the confinement potential slightly at each layer, creating a small random Fourier component at wavevector $2k_0$. Despite being small, we show in this work that such fluctuations accurately predict the wide range of valley splittings observed in recent experiments. 

Finally, the $2k_0$ theory also explains the reduction of valley splitting due to interface steps. When steps are present, the $z$ confinement potential naturally varies in different portions of the dot. Averaging over the plane of the quantum well, the step effectively causes the dot to experience a wider interface. The Fourier component of this broadened confinement potential at wavevector $2k_0$ is correspondingly reduced.

To demonstrate the universal nature of the $2k_0$ theory, in Fig.~\ref{fig:heterostructureCartoon}(f) we show simulation results for several types of engineered heterostructures, including conventional heterostructures with sharp interfaces, heterostructures with Ge spikes, narrow quantum wells, and heterostructures with additional, uniform Ge concentration added to the well.
We also include the effects of disorder in the form of interface steps and random alloys. (Details of the specific geometries and simulations are presented in later Sections.) For each geometry, we perform 100 simulations with different realizations of random alloy, using disorder models that are consistent with atomic-scale characterization based on atom probe tomography~\cite{Dyck:2017p1700622, Wuetz:2022p7730}. For each simulation, we compute the valley splitting using the $2k_0$ theory (the effective-mass approximation, $E_v^\text{EM}$). Then, using the same disorder profiles, we also compute the valley splitting using a two-band tight-binding model, $E_v^\text{TB}$, described below. The results show nearly perfect correlations, with all data points falling onto a universal curve. This demonstrates that the same physics governs valley splitting in deterministic vs.\ random structures, and it validates the $2k_0$ theory.

\section{Modeling the quantum dot} \label{sec:methods}

In this section, we describe the various theoretical approaches used in this work.
Our main tools are effective-mass theory~\cite{Friesen:2007p115318,Friesen:2010p115324,Gamble:2015p235318} and tight-binding theory, including the two-band model of Boykin et al.~\cite{Boykin:2004p165325} and the NEMO-3D 20-band sp$^3$d$^5$s$^*$ model~\cite{Klimeck:2007p1079, Klimeck:2007p2090}. 
Effective-mass theory provides an intuitive understanding of valley splitting behavior in many problems of interest. Although the approach can be applied to more complicated problems~\cite{Gamble:2016p253101}, analytical applications are most effective for systems than can be reduced, approximately, to one dimension (1D).
As we shall see, this includes many problems involving alloy disorder. In this work, we use effective-mass theory to clarify and characterize the distinct types of behavior associated with deterministic vs.\ randomly dominated valley splitting. 
For geometries that are intrinsically higher-dimensional, such as those involving atomic steps, tight-binding approaches are more effective. NEMO-3D is a sophisticated tool that provides quantitatively accurate results, over a wide energy range, and makes it possible to include atomistic details and strain. However, NEMO-3D is computationally expensive compared to effective-mass and two-band tight-binding theories. We show here that most valley-splitting physics is well described by simplified models, and that many problems of interest can be reduced to lower-dimensional systems that are more amenable to simple approaches. We now outline the details of these different methods.

\subsection{NEMO-3D}

The most rigorous model we use to simulate Si/SiGe heterostructures is the 20-band spin-resolved sp$^3$d$^5$s$^*$ nearest-neighbour tight-binding model, known as NEMO-3D \cite{Klimeck:2007p1079, Klimeck:2007p2090}. Although these simulations are computationally expensive, they are truly atomistic and, therefore, the most physically accurate. 

To specify a model geometry, we first define the heterostructure concentration profile, Si$_{\bar X_l}$Ge$_{1-\bar X_l}$, in the (nominal) growth direction [001], where $\bar X_l\in [0,1]$ represents the Si concentration averaged over the entire atomic layer, with layer index $l$.
We also use the notation $X_l$ to refer to the Si concentration averaged over just the lateral extent of the dot in layer $l$, as explained in Appendix~\ref{appendix:2Dalloy}.
Due to the finite size of the dot, $X_l$ is therefore a gaussian random variable fluctuating about its mean value, $\bar X_l$.
(Note that we use the notations $Y_l=1-X_l$ and $\bar Y_l=1-\bar X_l$ interchangeably, as convenient.)
In NEMO-3D, each atom in the lattice must be assigned as Si or Ge. 
In systems without interface steps, we therefore assign Si atoms in layer $l$ with probability $\bar X_l$. In systems with interface steps, this probability also depends on the lateral position of the step ($x_\text{step}$) as
\begin{equation} \label{SiConcStep}
    \bar X_l (x, y) = \bar X_l \Theta(x \leq x_\mathrm{step}) + \bar X_{l+1} \Theta(x > x_\mathrm{step}),
\end{equation}
where $\Theta$ is the Heaviside step function, and we take the concentration profiles to be identical on either side of the step, except for the single-atom shift, $l\rightarrow l+1$. 
Here, and throughout this work, we use lower-case $(x,y)$ notation to refer to spatial positions and upper-case $(X,Y)$ notation to refer to (Si,Ge) concentrations.
Note that atoms in our NEMO-3D simulations are actually located on diamond lattice sites, although we often specify their positions in continuum notation $\mathbf{r}=(x,y,z)$, for brevity.
Also note that we assume the step position in Eq.~(\ref{SiConcStep}) is independent of $y$, as consistent with a linear step oriented along [010].
This choice is made for convenience, although more nontrivial geometries may also be considered.
In this way, we generate realistic 3D lattice geometries consistent with the desired, average heterostructure concentration profile.
Repeating this procedure over and over yields disorder realizations that correctly reproduce the statistics of a random alloy.

Several other contributions to on-site energy terms are included in the simulation model.
Local bond lengths are incorporated using a strain optimization procedure in a valence-force-field Keating model.
A simple, separable lateral confinement potential is used to describe the dot, taking the parabolic form
\begin{equation}
U_\text{conf}(x,y) = {1 \over 2} m_t \omega_\text{orb}^2 \left[ (x-x_c)^2 + (y-y_c)^2 \right],
\label{eq:Uiso}
\end{equation}
where $m_t\approx 0.19m_e$ is the transverse effective mass, $m_e$ is the bare mass of the electron, $\hbar\omega_\text{orb}$ is the orbital excitation energy, and $(x_c, y_c)$ is the center of the dot.
This simple model is chosen for convenience, but is unlikely to affect the qualitative features of our results.
We also include a uniform vertical electric field, with the corresponding potential energy $e E_z z$. 

Simulations proceed by computing the energy eigenstates and eigenvalues for a given simulation geometry.
The valley splitting is obtained as the energy difference between the two lowest valley states. The simulation procedure is then repeated many times, with different realizations of alloy disorder, to build up a distribution of results.

\begin{figure*}[t] 
	\includegraphics[width=6in]{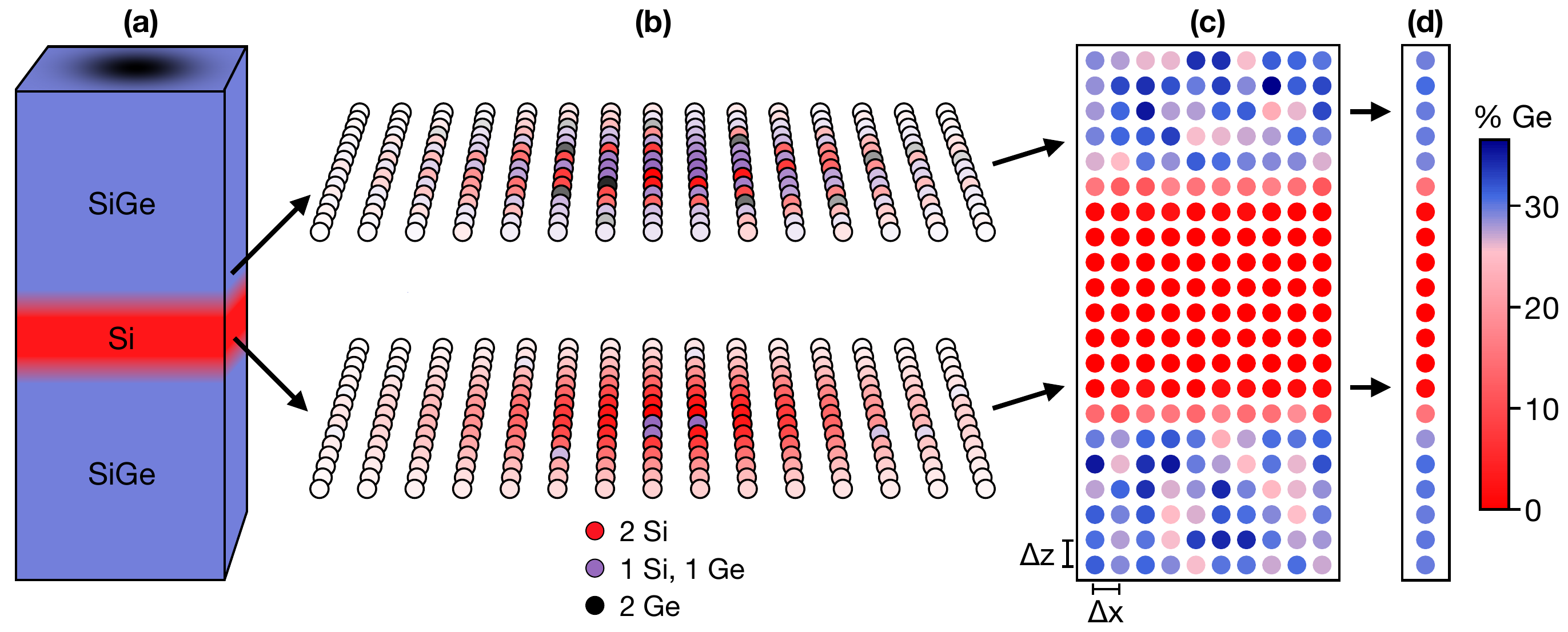}
	\centering
	\caption{ 
Cellular models of alloy disorder in one, two, and three dimensions.
High-Ge concentration regions ($\sim 30$\%) are indicated by blue, while low-Ge concentration regions are indicated by red.
(a) Schematic illustration of a Si/SiGe quantum well with diffused top and bottom interfaces.
The quantum dot is located in the Si quantum well below the dark-shaded region on the top of the illustration.
(b) The heterostructure is divided into rectangular 3D cells containing one or more atoms.
The illustration shows two layers of cells in the high-Ge (upper) and low-Ge (lower) regions, with color intensities proportional to the wavefunction probability. 
(c) Effective 2D cell geometry obtained by taking a weighted average of the Ge concentration of the 3D cell geometry along the $y$ axis, 
as explained in Appendix~\ref{appendix:2Dalloy}.
The 2D cell dimensions $\Delta x$ and $\Delta z$ are indicated, and the corresponding concentration fluctuations are evident.
For the minimal tight-binding models used in this work, we choose $\Delta z = a_0/4$ and $\Delta x = a_0/2$, where $a_0$ is the width of the conventional cubic unit cell.
(d) We further obtain an effective 1D cell geometry by taking weighted averages along $\hat x$. 
 }
	\label{fig:intro1}
\end{figure*}

\subsection{Modeling the quantum well potential} \label{sec:qwPotential}

While NEMO-3D allows us to perform accurate, atomistic simulations of quantum dots, it is computationally expensive, making it challenging to accumulate large statistical distributions for characterizing random-alloy disorder. 
Numerical methods also do not provide the same theoretical guidance as analytical theories. 
To overcome these problems, we also make use of minimal tight-binding and effective-mass models. 
In Appendix~\ref{appendix:tb}, we show that all three approaches yield consistent results.
However to facilitate comparisons, we need to define a mapping between the atomistic model used in NEMO-3D, and the other model schemes.
This can be done for one, two, or three-dimensional models, although we will focus on 1D and 2D geometries here, taking advantage of their computational efficiency.

The mapping between simulation models uses coarse graining.
We first define the notion of `cells,' which may include one or more atoms.
Cells may be defined in one, two, or three dimensions.
Cells may also have different sizes; however to correctly describe valley splitting in the tight-binding theory, we require a vertical cell height of one atom~\cite{Boykin:2004p165325}, $\Delta z=a_0/4$, where $a_0=0.543$~nm is the width of the conventional cubic unit cell.
Within a given cell, we employ the virtual-crystal approximation, for which the cell as a whole takes the average properties of the atoms contained within.
For a cell located at $\mathbf{r}_\text{cell}$, the potential energy describing the heterostructure is given by
\begin{equation} \label{Uqw}
U_\text{qw}(\mathbf{r}_\text{cell}) = \Delta E_c \frac{X_\text{cell} - X_w}{X_s - X_w} ,
\end{equation}
where $X_\text{cell}=X(\mathbf{r}_\text{cell})$ is the average Si concentration within the cell, $X_s$ is the average Si concentration deep in the SiGe substrate or barrier region, and $X_w$ is the average Si concentration in the center of the quantum well.
The conduction-band offset $\Delta E_c$ describes the potential energy difference between the barrier region and the quantum well region. 
For simplicity, we assume the barrier is fully strain-relaxed, and that the strain state of the quantum well is reflected in $\Delta E_c$~\cite{Schaffler:1997p1515}.
Following \cite{Wuetz:2022p7730}, we model the band offset as
\begin{multline}
    \Delta E_c = \left(X_w - X_s \right) \left[ {\frac{X_w}{1 - X_s}} \Delta E_{\Delta_2}^\mathrm{Si}(X_s) \right. \\ \left. - {\frac{1 - X_w}{X_s}} \Delta E_{\Delta_2}^\mathrm{Ge}(X_s) \right] ,
\end{multline}
where the functions $\Delta E_{\Delta_2}^\mathrm{Si (Ge) }(X)$ correspond to the $\Delta_2$ conduction band offsets for strained Si (Ge) grown on an unstrained $\mathrm{Si}_X\mathrm{Ge}_{1-X}$ substrate, making use of the following linear approximations, which are approximately valid over the concentration range of interest~\cite{Schaffler:1997p1515}:
\begin{equation}
\begin{split}
    \Delta E_{\Delta_2}^\mathrm{Si} (X) &\approx -0.502 (1-X) \; \mathrm{(eV)} , \\
    \Delta E_{\Delta_2}^\mathrm{Ge} (X) &\approx 0.743 - 0.625 (1-X) \; \mathrm{(eV)} .
\end{split}
\end{equation}

For certain quantum well geometries, where the vertical confinement along $z$ is much stronger than the lateral confinement, it is a good approximation to treat the total confinement potential as separable~\cite{Ando:1982}, such that $U_\text{total}(\mathbf{r})\approx U_\text{qw}(z)+U_\text{conf}(x,y)$, with the wavefunction given by $\psi(\mathbf{r}) \approx \psi_{xy}(x,y) \psi_{z}(z)$.
When 3D alloy disorder is present, this separable approach requires performing a 2D average over the lateral extent of the dot, in each plane $l$.
The vertical confinement potential is then given by $U_\text{qw}(X_l)$, where the physics of the valleys is contained in the vertical wavefunction $\psi_{z}(z_l)$.
For 1D calculations, the lateral confinement potential and ground-state wavefunction $|\psi_{xy}(x,y)|^2$ still play a role in computing the average Si concentrations $X_l$, as described in Appendix~\ref{appendix:2Dalloy}.

In systems with interface steps, the lateral and vertical wavefunctions are no longer separable. 
However, if the step is straight and oriented along $\hat y$, we can write $\psi(\mathbf{r}) \approx \psi_{xz}(x,z) \psi_{y}(y)$. 
Assuming a confinement potential of form $U_\text{conf}(y) = {1 \over 2} m_t \omega_\text{orb}^2 (y-y_c)^2$, the wavefunction $\psi_y(y)$ is a gaussian, which we use to perform the averaging procedure along $\hat y$.
Defining the lateral cell index along $\hat x$ as $j$, we then have $X_\text{cell}=X_{j,l}$.
Here we adopt the lateral cell dimension $\Delta x = a_0/2$.
Although this particular choice is not required for $\Delta x$, we have found that it gives results consistent with other computational schemes, as described in Appendix~\ref{appendix:tb}.
We also note that the averaging procedure described here converts the Si diamond crystal lattice to an effective, rectangular cell structure.

The full averaging procedure is illustrated in Fig.~\ref{fig:intro1}.
A typical Si/SiGe heterostructure is shown in Fig.~\ref{fig:intro1}(a).
Averaging is first performed within individual cells, as shown in Fig.~\ref{fig:intro1}(b).
Here, blue shading indicates dominantly Ge cells, red shading indicates dominantly Si cells, intermediate shading indicates cells with mixed Si-Ge content, and the color intensity indicates the wavefunction probability density, which is used in later steps to obtain weighted averages of the Si-Ge concentrations.
Cells with higher Ge concentrations have higher potential energies, as per Eq.~(\ref{Uqw}).
The wavefunction probability distribution is used to reduce the 3D cell geometry successively to 2D [Fig.~\ref{fig:intro1}(c)] or 1D [Fig.~\ref{fig:intro1}(d)] geometries, following the procedure described in Appendix~\ref{appendix:2Dalloy}.
Figure~\ref{fig:intro1} clarifies how the random nature of the original SiGe lattice is transferred to the different cell geometries -- through the fractional Ge content.
Although these local Ge concentration fluctuations are small, they can ultimately have a strong effect on the valley splitting.

Finally we note that, while it is possible to generate a new atomic lattice for every disorder realization (in fact, this is necessary in NEMO-3D), such 3D procedures are inefficient and unnecessary, since reduced-dimensional cell geometries may also be generated using the statistical properties of alloy disorder~\cite{Wuetz:2022p7730}.
To do this, we assume the Si concentration in each cell follows a binomial probability distribution with a known mean and variance, as supported by an atomistic analysis of actual Si/SiGe heterostructures using atom probe tomography \cite{Wuetz:2022p7730}. 
For a given cell, with a given dimensionality, the mean of the distribution is given by the expected Si concentration in the cell $\bar X_\text{cell}$ (for example, based on experimental characterization), and the variance is derived from $\bar X_\text{cell}$ and the probability density of the quantum dot in the $x$-$y$ plane.
Unless otherwise specified, we use this statistical approach to generate the 1D and 2D cell geometries for the minimal tight-binding and effective-mass calculations described below.
Full details of the method are described in Appendix~\ref{appendix:2Dalloy}. 

\subsection{Minimal tight-binding model}

In this work, we consider a two-band tight-binding model that accounts for physics at the very bottom of the conduction band, including the location of the valley minima ($\pm k_0\hat z$) and the band curvature (i.e., the longitudinal effective mass).
For a 1D model geometry oriented along [001], these parameters are given by $k_0=0.82\,(2\pi/a_0)$ and $m_l=0.916 m_e$. 
This minimal band structure can be mapped onto a minimal 1D tight-binding model, containing only nearest and next-nearest-neighbor hopping terms~\cite{Boykin:2004p165325}, given by $t_1 = 0.68$~eV and $t_2 = 0.61$~eV, respectively.
For a 2D model geometry in the $x$-$z$ plane, the valley minima are located at $\mathbf{k}_{0\pm}=(0,\pm k_0)$.
In this case, we use the same two hopping parameters along $\hat z$, and we include a nearest-neighbor hopping term along $\hat x$, $t_3 = -2.72$~eV, which gives the correct transverse effective mass for a cell of width $\Delta x=a_0/2$. 
We note that this minimal model assumes a rectangular lattice geometry~\cite{Saraiva:2010p245314}, with cells of size $(\Delta x,\Delta z)$.

In addition to the off-diagonal hopping terms, our minimal model Hamiltonian also includes on-site potential terms $U_\text{total}(X_{j,l})$, where $(j,l)$ are the 1D-2D cell indices. 
For 2D geometries, we are particularly interested in comparing the effects of interface steps, defined in Eq.~(\ref{SiConcStep}) and included in the Hamiltonian via Eq.~(\ref{Uqw}), to alloy disorder, defined in the coarse-grained cell potentials $U_\text{qw}(X_{j,l})$.
As for NEMO-3D, the simulations are typically repeated for many realizations of alloy disorder to obtain statistical distributions of results, as described in Appendix~\ref{appendix:2Dalloy}.
When alloy disorder is not included in the simulations, we simply replace the locally fluctuating Si concentration $X_{j,l}$ by its average value $\bar X_l$. 

\subsection{Effective-mass theory} \label{sec:effectiveMass}

The effective-mass theory is similar to the minimal tight-binding theory in that it incorporates the physics of the bottom of the conduction band.
The most important difference between the two approaches is that valley couplings do not arise naturally in the effective-mass theory, and must be included perturbatively.
The perturbation theory is straightforward however~\cite{Friesen:2007p115318}, and we summarize it here for completeness.

We consider as basis states the quantum dot wavefunctions formed of Bloch states localized near the $\pm k_0\hat z$ valleys in reciprocal space.
For our purposes, it is a good approximation to write the real-space expressions for these wavefunctions as
\begin{equation}
\psi_\pm(\mathbf{r})\approx e^{\pm ik_0z}\psi_\text{env}(\mathbf{r}) , \label{eq:psipm}
\end{equation}
where $\psi_\text{env}$ is the effective-mass envelope of the ground-state wavefunction in the total confinement potential $U_\text{total}(\mathbf{r})$.
This approximation assumes weak valley-orbit coupling, so both valley states have the same envelope function.
Such perturbative treatment is appropriate for many problems of interest.
In the limit of large valley-orbit coupling, Eq.~(\ref{eq:psipm}) should be modified to account for the differing envelope functions in the ground and excited valley states.
However the simplicity of Eq.~(\ref{eq:psipm}) provides considerable intuition, as discussed below.
We note that the exponential phase term oscillates rapidly in Eq.~(\ref{eq:psipm}), over a length scale of $2\pi/k_0$, while the confinement potential and envelope function vary slowly over this same length scale.

The intervalley-coupling matrix element is given by
\begin{equation}
\Delta = \langle \psi_-| U_\text{qw}|\psi_+\rangle
= \int\! dr^3 e^{-2 i k_0 z} U_\text{qw}(\mathbf{r}) |\psi_\text{env}(\mathbf{r})|^2 ,
\label{eq:EMDelta}
\end{equation}
and the valley splitting is given by $E_v=2|\Delta|$.
Because of the separation of length scales in Eq.~(\ref{eq:EMDelta}), we see that $\Delta$ and $E_v$ should approximately vanish in the limit of slowly varying $U_\text{qw}(\mathbf{r})$.
A nonvanishing valley splitting therefore requires some type of sharp feature to be present in the confinement potential.
The conventional `sharp feature' in many valley-spitting proposals is an abrupt quantum well interface.

Although effective-mass equations are conventionally expressed in a continuum description, they may also be discretized; here, such an approach helps to make contact with the tight-binding theories discussed in previous sections.
We may consider 1D or 2D expressions for the intervalley matrix element:
\begin{gather} \label{discreteDelta}
\Delta_\text{1D} = \frac{a_0}{4} \sum_l e^{-2 i k_0 z_l} U_\text{qw}(z_l) |\psi_\text{env}(z_l)|^2,\\
\label{discreteDelta2D}
\Delta_\text{2D} = \frac{a_0}{4} \sum_l e^{-2 i k_0 z_l} \frac{a_0}{2}\sum_j U_\text{qw}^\text{2D}(x_j, z_l) |\psi_\text{env}(x_j, z_l)|^2,
\end{gather}
where $(j,l)$ are cell indices corresponding to $X_{j,l}$ in the tight binding theories, and we assume proper normalization, given by $\sum_l (a_0/4) |\psi_\text{env}(z_l)|^2 =\sum_{j,l}(a_0^2/8) |\psi_\text{env}(x_j, z_l)|^2=1$.

A key, take-away message from Eq.~(\ref{eq:EMDelta}) is that the valley splitting can be understood, quite literally, as the $2k_0$ Fourier component (along $k_z$) of the quantity 
$U_\text{qw}(\mathbf{r}) |\psi_\text{env}(\mathbf{r})|^2$. More simply, it is the $2k_0$ Fourier component of $U_\text{qw}(\mathbf{r})$, weighted by the electron probability at the location where the oscillations occur.
This is a powerful statement that transcends effective-mass theory.
As we demonstrate in this work, such a universal description of valley splitting is quantitatively accurate for all quantum well geometries studied here, including interface steps, broadened interfaces, alloy disorder, Wiggle Wells, and other phenomena.

\subsection{Alloy-disorder analysis} \label{sec:alloyDisorderTheory}

In this section, we use effective-mass methods to characterize the deterministic vs.\ random components of the valley splitting using an approach similar to Ref.~\cite{Wuetz:2022p7730}, for effectively 1D geometries. 
Figures~\ref{fig:introFourier}(a) and \ref{fig:introFourier}(d) show typical concentration fluctuations in quantum wells with wide or narrow interfaces, respectively.
Below, we show that, even when Si concentrations vary only slightly from their average values, as in these examples, such small fluctuations can have an outsize effect on the valley splitting. 

The intervalley coupling matrix element in Eq.~(\ref{discreteDelta}) can be decomposed into its deterministic ($\Delta_0$) and fluctuating ($\delta \Delta$) components: $\Delta_\text{1D}=\Delta_0+\delta \Delta$.
This assignment is unambiguous when the Si concentration in layer $l$, defined as $X_l = \bar X_l + \delta_l$, can be decomposed into its average ($\bar X_l$) and fluctuating ($\delta_l$) contributions. 
The deterministic part of the valley splitting, $E_{v0} = 2|\Delta_0|$, is determined by the average heterostructure profile, including geometrical features like interface steps. 
It is computed by setting $\delta_l = 0$ in all layers. 
Substituting Eq.~(\ref{Uqw}) into Eq.~(\ref{discreteDelta}), we obtain
\begin{equation}
\Delta_0 = \frac{a_0}{4} \frac{\Delta E_c}{X_w - X_s} \sum_l e^{-2 i k_0 z_l} (\bar X_l - X_s) |\psi_\text{env}(z_l)|^2.
\label{eq:D0}
\end{equation}

The random component of the intervalley matrix element, $\delta \Delta$, arises from the alloy disorder, and is given by
\begin{equation} \label{deltaDelta}
    \delta \Delta = \frac{a_0}{4} \frac{\Delta E_c}{ X_w - X_s} \sum_l e^{-2 i k_0 z_l} \delta_l |\psi_\text{env} (z_l)|^2 .
\end{equation}
Here, the fluctuations are contained within $\delta_l=X_l-\bar X_l$, which describes the concentration variations in layer $l$ weighted by the dot probability.
Since SiGe is a completely random alloy, $\delta_l$ has a binomial probability distribution, given by
\begin{equation}
\delta_l \sim \frac{1}{N_\text{eff}} \text{Binom}(N_\text{eff}, \bar X_l)
\label{eq:deltal}
\end{equation}
where $N_\text{eff} = 4\pi a_\text{dot}^2 / a_0^2$ is the approximate number of atoms in a dot, in a given layer, and we have assumed a circular quantum dot as defined in Eq.~(\ref{eq:Uiso}), with orbital excitation energy $\hbar \omega_\text{orb}$ and characteristic size $a_\text{dot} = \sqrt{\hbar / m_t \omega_\text{orb}}$. 
See Appendix~\ref{appendix:2Dalloy} for further details on the derivation of Eq.~(\ref{eq:deltal})

The individual amplitudes $\delta_l |\psi_\text{env}(z_l)|^2$ contributing to the sum in Eq.~(\ref{deltaDelta}) are plotted in Figs.~\ref{fig:introFourier}(b) and \ref{fig:introFourier}(e), for the same disorder realizations shown in Figs.~\ref{fig:introFourier}(a) and \ref{fig:introFourier}(d).
Here, the complex phase, $-2 k_0 z_l$, is also indicated by the color scale. 
In both figures, we see that amplitudes are maximized when the wavefunction strongly overlaps with Ge. 
For the wide-interface geometry in Fig.~\ref{fig:introFourier}(b), we find that many layers contribute significantly to the sum, while for the narrow interface shown in Fig.~\ref{fig:introFourier}(e), only a few layers contribute. 
The total intervalley coupling $\Delta$ is therefore complex, with a large random component determined by the details of the Ge distribution. 
Figure~\ref{fig:introFourier}(g) shows the resulting distribution of $\Delta$ values in the complex plane, for many different realizations of the alloy disorder, corresponding to the wide (purple) or narrow (gold) interfaces. 
The black dots indicate the deterministic components $\Delta_0$, which are generally located at the center of the distributions. 
For wide interfaces, we see that $|\Delta_0|$ can be much smaller than the standard deviation of $|\Delta|$. 
For narrow interfaces, on the other hand, $|\Delta_0|$ can be large enough for all $\Delta$ results to be well separated from the origin.
However, we note that the interface is extremely sharp in this example, with an average width of just 1 atomic monolayer (ML).

This effective-mass description of alloy disorder agrees very well with tight-binding simulations. 
Figure~\ref{fig:introFourier}(h) shows a correlation plot of effective-mass results $E_v^\text{EM}$, obtained from Eq.~(\ref{discreteDelta}), vs.\ tight-binding results $E_v^\text{TB}$, obtained for the same disorder realizations. 
The correlations between these independent methods is nearly perfect for both the purple and gold data sets in Fig.~\ref{fig:introFourier}(h), emphasizing the accuracy of this analytic interpretation.

\subsection{Statistical properties of $E_v$} \label{sec:statisticalProperties}

\begin{figure}[] 
	\includegraphics[width=6cm]{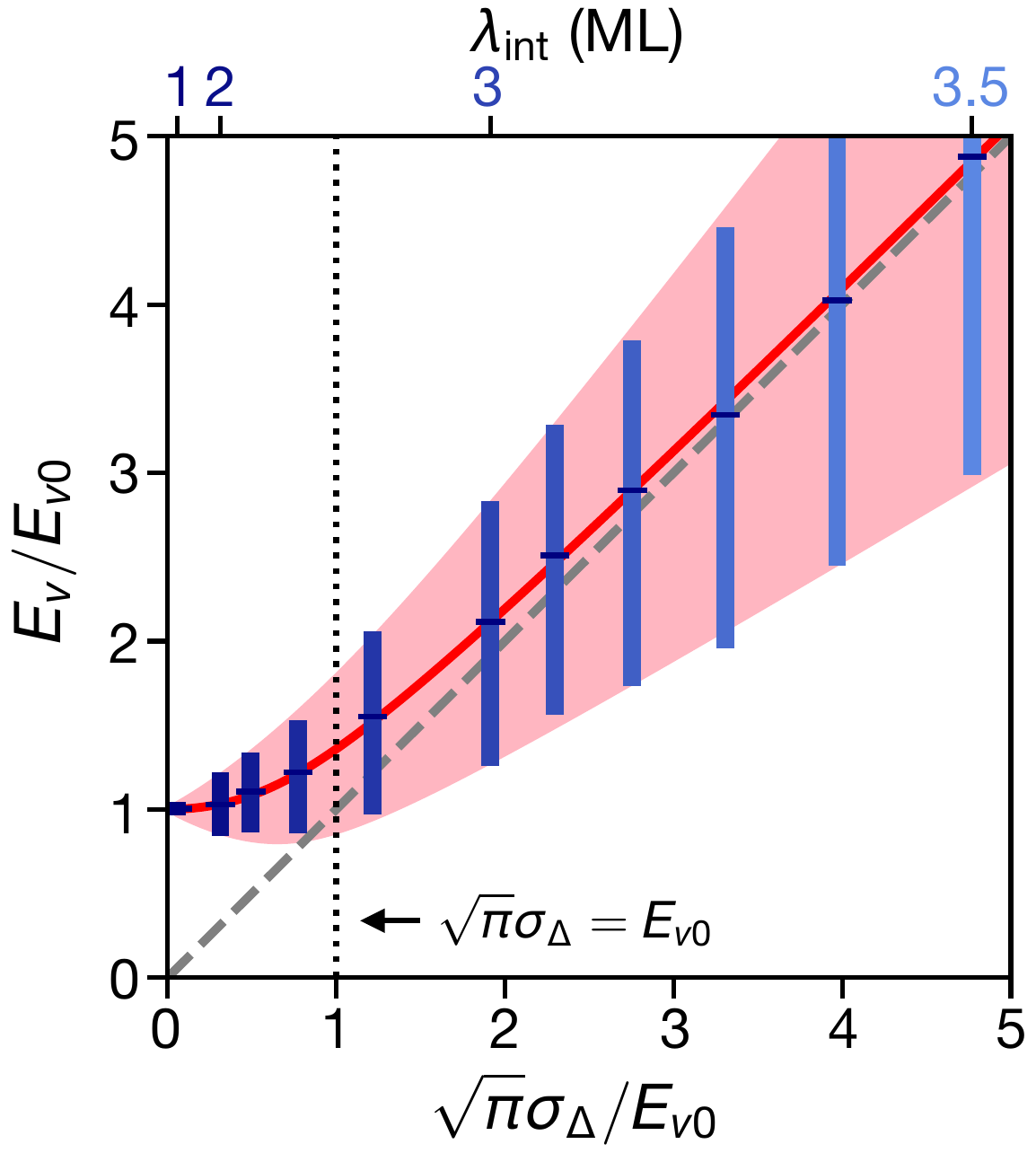}
	\centering
	\caption{Universal crossover between deterministically and randomly dominated valley splittings, in the presence of alloy disorder.
 1D effective-mass simulations are performed, using sigmoidal Ge profiles, as a function of interface width $\lambda_\text{int}$ (top axis).
 500 simulations are performed at each $\lambda_\text{int}$ value, with 25-75 percentiles shown as vertical bars.
 The mean valley splittings $\bar E_v$ are shown as crossbars and corresponding standard deviations $\sigma_\Delta$ (bottom axis) are obtained from Eq.~(\ref{varDelta}).
 The two energy axes are normalized by the deterministic valley splitting $E_{v0}$, resulting in an asymptote of $E_v/E_{v0}=1$ in the low-$\sigma_\Delta$ limit, a slope of 1 in the high-$\sigma_\Delta$ limit, and a well-defined crossover between the two regimes at $\sqrt{\pi}\sigma_\Delta=E_{v0}$ (vertical dotted line).
The red curve shows the theoretical estimate for $\bar E_v$, obtained from the Rician distribution in Eq.~(\ref{riceMean}), taking $E_{v0}$ and $\sigma_\Delta$ as inputs from the effective-mass simulations. The pink shaded region shows the corresponding 25-75 percentile range for the Rician distribution, which closely matches the simulation results. 
For all results, we assume dots with orbital splittings of $\hbar \omega_\text{orb} = 2$~meV, vertical electric fields of $E_z = 5$~mV/nm, and well widths of $W = 80$~ML.}
	\label{fig:deterministicVsDisorder}
\end{figure}

To characterize the statistical properties of $\Delta$ in the presence of alloy disorder, we first consider the case with no interface steps.
From Eq.~(\ref{deltaDelta}) we see that, since $\bar \delta_l=\text{E}[\delta_l]=0$, we must have $\text{E}[\Delta]=\Delta_0$.
Following \cite{Wuetz:2022p7730} to compute the variance of Eq.~(\ref{deltaDelta}), we then obtain 
\begin{multline} \label{varDelta}
\sigma_\Delta^2=\mathrm{Var} \left[ \Delta \right] = \mathrm{Var} \left[ \delta \Delta \right] = \\
 {1 \over \pi}\left[ {a_0^2 \Delta E_c \over 8 a_\mathrm{dot} (X_w - X_s)} \right]^2 
 \sum_l  |\psi_\text{env} (z_l)|^4 \bar X_l (1 - \bar X_l) ,
\end{multline}
where $\sigma_\Delta$ is the standard deviation of the $\Delta$ distribution, and we have used the fact that concentrations fluctuations in different layers are statistically independent. 

If many atomic layers contribute to the sum in Eq.~(\ref{discreteDelta}), then according to the central limit theorem, we can approximate the intervalley coupling $\Delta$ as a circular gaussian random variable centered at $\Delta_0$, for which the variances of the real and imaginary parts of $\Delta$ are both $(1/2) \mathrm{Var} \left[\Delta\right]$. 
This approximation should be accurate for quantum wells with wide interfaces. 
For structures with very sharp interfaces, the sum in Eq.~(\ref{discreteDelta}) may be dominated by just a few layers, as in Fig.~\ref{fig:introFourier}(e).
In this case, the central limit theorem is less accurate, and $\Delta$ may have a non-circular distribution. 
Nonetheless, the approximation provides reasonable estimates, even in cases where it is not well justified, and we adopt it in all cases below.

The valley splitting $E_v=2|\Delta|$ is real.
For a circular gaussian distribution of $\Delta$ values, the corresponding $E_v$ probability distribution is Rician~\cite{Aja-Fernandez:2016}, defined as 
\begin{equation} \label{riceDistrib}
f_\mathrm{Rice}(z | \nu, \sigma) = {z \over \sigma^2} \exp \left( - {z^2 + \nu^2 \over 2 \sigma^2} \right) I_0 \left( z \nu \over \sigma^2 \right) ,
\end{equation}
where $I_0 (y)$ is a modified Bessel function of the first kind.
Here, the `center' parameter is given by $\nu=E_{v0}=2|\Delta_0|$, and the `width' parameter is given by $\sigma = \sqrt{2}\sigma_\Delta$. 
To illustrate valley splitting distributions in different limiting regimes, the tight-binding results from Fig.~\ref{fig:introFourier}(h) are replotted in histogram form in Fig.~\ref{fig:introFourier}(i).
Here we also plot the corresponding Rician distributions, using the parameters $E_{v0}=2|\Delta_0|$ and $\sigma_\Delta$ computed in Eqs.~(\ref{eq:D0}) and (\ref{varDelta}). 
For quantum wells with wide interfaces (purple data), the predicted distributions show excellent agreement with the simulations. 
For wells with narrow interfaces (gold data), the Rician distribution is somewhat skewed, since the distribution of $\Delta$ is no longer perfectly circular, and the central limit theorem is less-well-satisfied. However, the Rician model still provides a reasonable estimate of the results. 

We may therefore use the known properties of the Rician distribution to characterize the statistical properties of the valley splitting in the presence of disorder.
The mean valley splitting is thus given by
\begin{equation} \label{riceMean}
    \bar E_v = \sigma \sqrt{\pi/2}L_{1/2}(-\nu^2/2\sigma^2) ,
\end{equation}
where $L_{1/2}(x)$ is a Laguerre polynomial~\cite{Aja-Fernandez:2016}. 
In the randomly dominated regime, corresponding to $\nu\ll\sigma$ (or $|\Delta_0|\ll \sigma_\Delta$), the Rice distribution reduces to a Rayleigh distribution with $\Delta \approx \delta \Delta$ and $\bar E_v \approx \sqrt{\pi}\sigma_\Delta(1+E_{v0}^2/8\sigma_\Delta^2)$.
Since $\text{Var}[\Delta]=\sigma_\Delta^2$ is approximately proportional to the average Ge in the quantum well, given by $Y_l$ in layer $l$, we see that the mean valley splitting can be increased by simply exposing the wavefunction to more Ge.
This is an important result.
In contrast, the deterministic correction to $\bar E_v$ (the second term in the Rayleigh expression) is proportional to $(E_{v0}/\sigma_\Delta)^2$, which has almost no effect on the valley splitting.
In the opposite limit, $\nu\gg\sigma$ (or $|\Delta_0|\gg \sigma_\Delta$), the mean valley splitting is simply given by $\bar E_v\approx E_{v0}$.
This is the expected result in the deterministically enhanced regime.

\begin{figure*}[t] 
	\includegraphics[width=16cm]{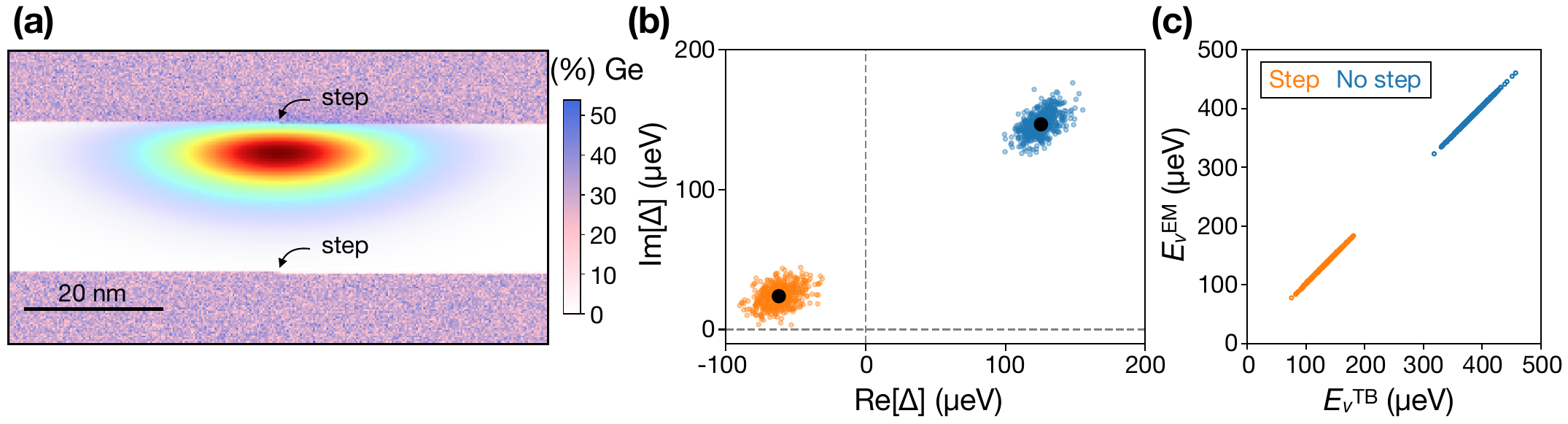}
	\centering
	\caption{ 
Effects of interface steps on the valley splitting for a super-sharp interface.
(a) A typical 2D cell geometry and corresponding 2D envelope function, obtained in a quantum well with a single-atom step located at the center of the quantum dot confinement potential along $\hat x$.
(b) $\Delta$ distributions obtained from Eq.~(\ref{discreteDelta2D}), plotted on the complex plane, for 500 random-alloy realizations of the geometry shown in (a) (orange). 
We adopt the sigmoidal model of Eq.~(\ref{eq:sigmoid}) for the quantum well Ge profile, with $X_s=0.7$ and $X_w=1$, and an interface width of $\lambda_\text{int} = 4\tau=1$~ML.
The 2D envelope function is recomputed for each simulation, taking into account the concentration fluctuations, and we assume a vertical electric field of $E_z = 5$~mV/nm and an isotropic harmonic confinement potential of strength $\hbar \omega_\text{orb} = 2$~meV.
Blue points show results for the same geometry, but in the absence of steps.
The central black points correspond to $\Delta_0$ for each distribution.
For such sharp interfaces, the presence of a step suppresses the average valley splitting by a factor of 3-4, although it has little effect on the standard deviation of the distribution.
It is interesting to note that, despite being strongly suppressed, the valley splitting remains in the deterministically enhanced regime, even in the presence of a step, for the case of a super-sharp interface.
(c) Correlation plot of the effective-mass valley splitting results, $E_v^\text{EM}=2|\Delta|$, taken from (b), vs.\ 2D tight-binding results obtained for the same 2D cell geometries and disorder realizations, demonstrating nearly perfect agreement.
The results show that the $2k_0$ theory of valley splitting also holds for step disorder.}
	\label{fig:fourierStep}
\end{figure*}

We now examine the crossover between deterministic and randomly dominated behaviors. 
In Fig.~\ref{fig:deterministicVsDisorder}, we plot the total valley splitting $E_v$, including both deterministic and random contributions (normalized by the deterministic value, $E_{v0}$) as a function of the standard deviation, $\sqrt{\pi} \sigma_\Delta$ (also normalized by $E_{v0}$).
Results are obtained from effective-mass calculations, including 500 realizations of alloy disorder.
Here, the blue bars represent the 25-75 percentile range, and the crossbars indicate the mean value.
The solid red line shows the Rician estimate, and the pink shading represents the corresponding 25-75 percentiles of the Rice distribution.
The asymptotic behaviors derived above are then shown as the dashed gray line (randomly dominated regime) and the constant value $E_v/E_{v0}=1$ (deterministically enhanced regime).
We can identify the crossover between these two regimes as 
\begin{equation} 
\label{cuttoffPoint}
E_{v0} \approx \sqrt{\pi}\sigma_\Delta  .
\end{equation}
In practice, the crossover appears abrupt, with the deterministic valley splitting quickly becoming overwhelmed by disorder as the wavefunction is exposed to more Ge.

\subsection{Device failure analysis} 

For qubit applications, a key outcome of our analysis is that valley splitting distributions exhibit qualitatively different behaviors, depending on whether they fall into the deterministically enhanced vs.\ disorder-dominated regimes.
In general, robust qubit operations require valley splittings that are reliably large, so that excited valley states do not compete with spins as qubits, and do not interfere with spin-qubit dynamics. 
To form large and uniform arrays of qubits, we must therefore ensure that all quantum dots have valley splittings above a specified threshold energy value, $E_v^\text{min}$. 
Throughout this work, we adopt the threshold value of $E_v^\text{min} = 100$ $\mu$eV, which corresponds to 1.2~K, or about 10$\times$ the typical electron temperature in a dilution refrigerator. 

We may then ask the question, what fraction of dots fail according to this criterion?
For the Rice distribution, this fraction is given by
\begin{equation} \label{pFailureDetApprox}
    P_\text{fail} = \int_{0}^{E_v^\text{min}} \!\!\! dE \, f_\text{Rice} ( E \;|\; E_{v0}, \sqrt{2}\sigma_\Delta) .
\end{equation}
In the disorder-dominated regime ($\sigma_\Delta\gg E_{v0}$), we find that 
\begin{equation}
    P_\text{fail} \approx 1 - \exp\left( -{E_v^\text{min}}^2 / 4 \sigma_\Delta^2 \right) \quad\quad\text{(disordered)} .
    \label{eq:Pfaildisordered}
\end{equation}
If we further assume that $E_v^\text{min} > \sigma_\Delta$, as is often true for wide-interface heterostructures, we find $P_\text{fail} \sim O(1)$. 
On the other hand, if we assume that $E_v^\text{min} \lesssim \sigma_\Delta$, as found in some high-disorder heterostructures, we obtain the power law behavior
\begin{equation}
P_\text{fail} \approx \left( E_v^\text{min}/ 2 \sigma_\Delta\right)^2 \quad\quad\text{(disordered)} .
\end{equation}
In either case, the failure rate is found to be unacceptably high.
In the deterministically enhanced regime ($\sigma_\Delta\ll E_{v0}$), on the other hand, the failure rate is exponentially suppressed:
\begin{equation}
P_\text{fail} \propto \exp(-E_{v0}^2/4\sigma_\Delta^2) \quad\quad\text{(deterministic)} .
\end{equation}
In this case, it is possible that no qubits have unacceptably low valley splittings, even in large arrays. 
Taking an example: for a dot with the same parameters as Fig.~\ref{fig:introFourier}, and a 1~ML interface width, we obtain $E_{v0} \approx 386$~$\mu$eV, $\sigma_\Delta \approx 12$~$\mu$eV, and $P_\text{fail} \approx 10^{-61}$, which is extremely small. 
However, $P_\text{fail}$ also increases extremely quickly with interface width. 
For example, for a 2~ML interface with all other parameters unchanged, we obtain $E_{v0} \approx 79$~$\mu$eV, $\sigma_\Delta \approx 14$~$\mu$eV, and $P_\text{fail} \approx 0.82$. 
So even though 2~ML interfaces fall into the deterministically enhanced regime ($E_{v0}\gtrsim\sigma_\Delta$), since $E_{v0} < E_v^\text{min}$, $P_\text{fail}$ can still be large. 
In recent experiments where quantum wells were found to have sigmoidal interfaces of width $\lambda_\text{int} = 0.8$~nm~\cite{Wuetz:2022p7730}, the measured 100~$\mu$eV failure rate was found to be $\sim 50$\%. 
For the $E_z$ and $\hbar \omega_\text{orb}$ values reported in that work, we predict a similar value of 62\%, while for the $E_z = 5$~mV/nm, $\hbar \omega_\text{orb} = 2$~meV parameters used elsewhere in this work, we predict $P_\text{fail} = 99$\%.
Finally we note that valley splitting distributions are not perfectly Rician when interfaces are very narrow, so the estimates given above are rough. 
However, these results highlight the fact that $E_v$ can be consistently large in the deterministically enhanced regime, although this requires extremely sharp heterostructure features.

\subsection{Interface steps} \label{sec:stepsTheory}

Si/SiGe heterostructures are grown on surfaces that may be intentionally miscut away from the [001] crystallographic axis, resulting in single-atom steps at all device interfaces. 
Steps may also arise as a consequence of strain or other natural fluctuations, which are very difficult to control at the single-atom level.
Such steps are known to significantly effect the valley splitting, and are therefore very well-studied~\cite{Friesen:2006p202106,Kharche:2007p092109, Tariq:2019p125309, Hosseinkhani:2020p043180, Ercan:2021p235302, Hosseinkhani:2021p085309, Dodson:2022p146802, Tariq:2022p53}.
Reductions in $E_v$ of up to 75\% for a single step have been reported theoretically, depending on the particular step geometry, location, and other dot parameters. 

The effective-mass description of valley splitting, developed in preceding sections, also applies to devices with steps.
In this case, we use the 2D intervalley matrix element, Eq.~(\ref{discreteDelta2D}). 
This requires first calculating the 2D envelope function $\psi(x,z)$, which we do by solving a discretized Schr\"odinger equation on a grid of cells, in the absence of valleys, while including alloy disorder.
Figure~\ref{fig:fourierStep}(a) shows a typical envelope solution for the case where a step is located at the center of the quantum dot confinement potential.
In Fig.~\ref{fig:fourierStep}(b), we show the complex $\Delta$ distributions obtained for cases with (orange) and without (blue) single-atom steps, again located at the center of the quantum dot confinement potential.
In each case, the effective-mass results are obtained for 500 different disorder realizations. 
As we might expect, the step significantly reduces the central value of the distribution, $|\Delta_0|$, in this case, by a factor of 3-4.
This particular geometry, with the step centered on the dot, is found to be a worst-case scenario for suppressing the valley splitting, although we do not consider other step geometries here.
It is interesting to note however that the variances of the two distributions in Fig.~\ref{fig:fourierStep}(a) are nearly identical.
This can be understood from the fact that the variance in Eq.~(\ref{varDelta}) depends on quantities that vary slowly in space, such as $\psi_\text{env}(\mathbf{r})$ and $\bar X_l$.
In other words, the statistical properties of the valley splitting depend on the local Ge concentration, and are not particularly sensitive to the presence of steps.
It is also interesting to note that the valley splitting remains in the deterministically enhanced regime, $|\Delta_0|\gtrsim \sigma_\Delta$, even in the presence of a step, when the interface is very sharp.
Below, we will show that this is no longer true for wider interfaces.

Figure~\ref{fig:fourierStep}(c) shows a correlation plot comparing effective-mass and tight-binding results for cell geometries and confinement parameters identical to those used in Fig.~\ref{fig:fourierStep}(b).
Similar to Fig.~{\ref{fig:introFourier}}(h), we observe nearly perfect correlations, including cases with and without a step.
This demonstrates that the universal $2k_0$ theory of valley splitting also captures the effects of interface steps.
In the following section, we explore the interplay between steps and alloy disorder more thoroughly.

\section{Conventional Si/SiGe heterostructures} \label{sec:conventional}

In this section, we use the theoretical tools developed above to analyze conventional Si/SiGe heterostructures. 
First, we more thoroughly explore the interplay between alloy disorder, interface width, and interface steps.
In particular, we show that for devices with realistically broadened interfaces, step disorder is less important than alloy disorder. 
We then show how alloy disorder and step disorder impact the variability of valley splitting across a device. 
Finally, we use theoretical and numerical approaches to study how the specific profile of an interface affects its valley splitting.

\subsection{Interplay between interface steps and interface width}
\label{sec:interplay}

\begin{figure*}[] 
	\includegraphics[width=15cm]{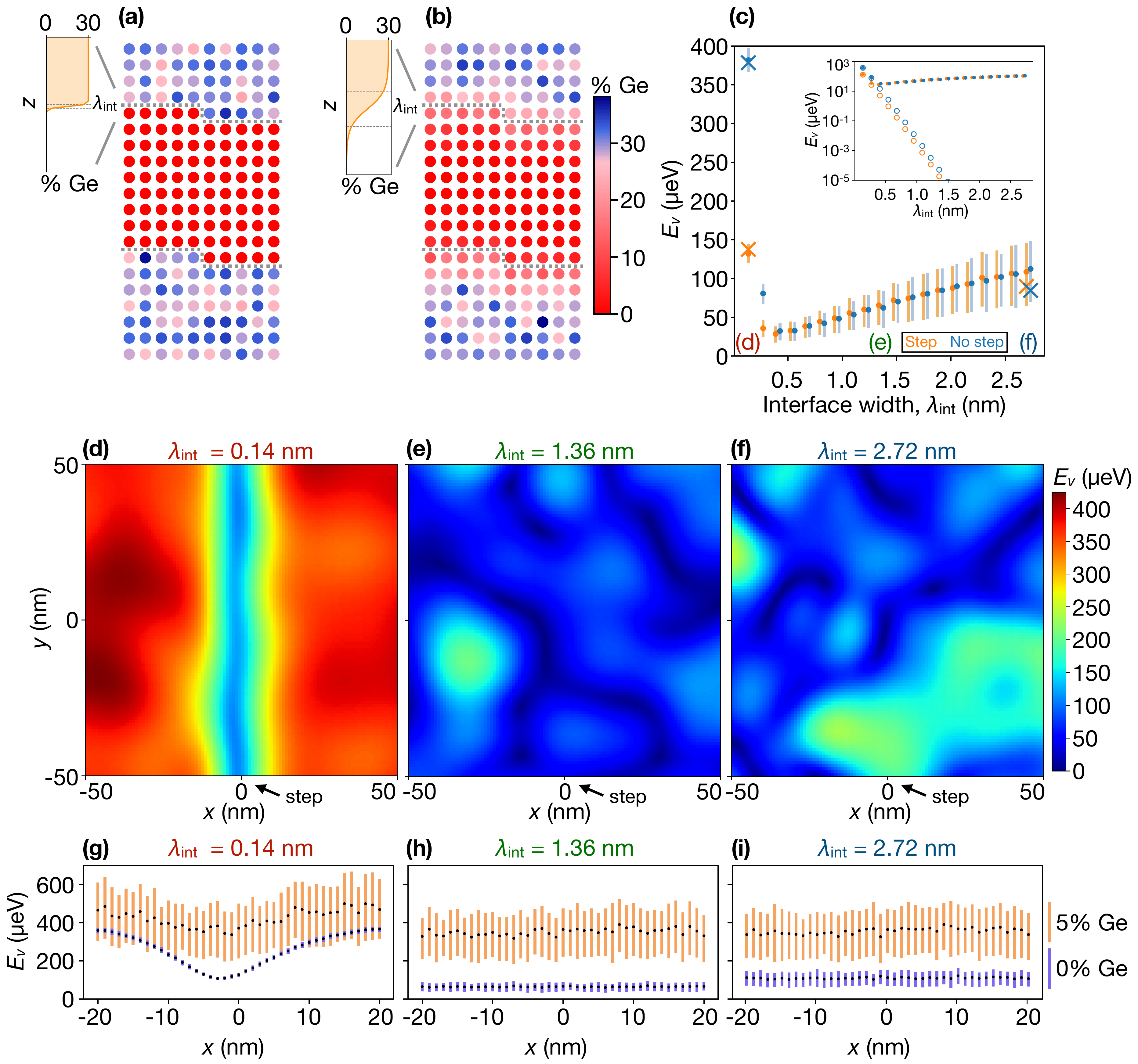}
	\centering
	\caption{
Interplay between interface step disorder and random alloy disorder. 
(a), (b) Schematic illustrations of 2D tight-binding models for quantum wells with (a) sharp interfaces, or (b) wider interfaces, in the presence of a single-atom step through the center of the device. 
Here interfaces are modelled as sigmoid functions [Eq.~(\ref{eq:sigmoid})] with a characteristic width of $\lambda_\text{int}$, as indicated by dotted lines in the insets. 
Quantum dots are modelled as in Eq.~(\ref{Uqw}), with orbital energy splittings of $\hbar \omega_\text{orb} = 2$~meV, and a vertical electric field of $E_z = 5$~mV/nm.
(c) Valley-splitting results are shown for the case of no steps at the interface (blue), or a single-atom step going through the center of the quantum dot (orange), as a function of the interface width.
Circular markers show the mean values for 1,000 different alloy-disorder realizations, while error bars show the 25-75 percentile range. 
Inset: filled circles shown the same mean values plotted in the main panel.
Open circles show the results of similar simulations, performed without alloy disorder.
(d)-(f) Valley splittings as a function of dot position, in quantum wells with alloy disorder and a single-atom step at the quantum well interface, located at $x=0$. 
The only difference between the three maps is the interface thickness, as indicated at the top of each panel.
(g)-(i) Valley splitting results as a function of dot position, as the dot is moved across a step located at $x = 0$.
Here, interface widths are the same as in (d)-(f), but the simulations are repeated for 200 realizations of alloy disorder, with circular markers showing the mean values, and error bars showing the 25-75 percentiles.
Blue data: the same interface/step geometries as (d)-(f), with no additional Ge in the quantum well.
Orange data: the same interface/step geometries as (d)-(f), with an average uniform concentration of 5\% Ge added to the quantum well.
For the orange data, we use Ge barrier concentrations of 35\%, to maintain a 30\% concentration offset between the barriers and the quantum well. 
(All other simulations, without Ge in the quantum well, use Ge barrier concentrations of 30\%.)}
	\label{fig:wideInt}
\end{figure*}

Here we use the 2D minimal tight-binding model to explore the interplay between interface steps and interface widths on the valley splitting. 
We choose a smooth quantum well confinement potential defined in terms of sigmoid functions, as
\begin{multline}
\bar X(z)= X_w \\
+\frac{X_s-X_w}{1+\exp[(z-z_t)/\tau]}+\frac{X_s-X_w}{1+\exp[(z_b-z)/\tau]} , 
\label{eq:sigmoid}
\end{multline}
where we adopt the convention that $z=0$ at the top surface of the sample, and $z>0$ inside the sample, including in the quantum well.
Here, $z_t$ and $z_b$ are the positions of the top and bottom interfaces of the quantum well, with $z_b-z_t=W$, and the interface width is given by $\lambda_\text{int}=4\tau$.
In cases with narrow interfaces, we choose $z_t$ and $z_b$ to lie halfway between atom sites.
Steps may be included by inserting Eq.~(\ref{eq:sigmoid}) into Eq.~(\ref{SiConcStep}), with $\bar X_l(x,y)=\bar X(x,y,z_l)$.
Some typical narrow and wide-interface geometries are illustrated schematically in Figs.~\ref{fig:wideInt}(a) and \ref{fig:wideInt}(b).
The quantum dots are confined laterally using Eq.~(\ref{eq:Uiso}).
Here and throughout this section, we choose the orbital excitation energy to be $\hbar\omega_\text{orb}=2$~meV.
To reduce the 3D cell geometry to 2D, we assume the step is oriented along $\hat y$.
As described in Sec.~\ref{sec:qwPotential}, we are then able consider a separable wavefunction, with $\psi_y(y)$ being the ground state of a harmonic oscillator along $\hat y$, also adopting a confinement potential with $\hbar\omega_\text{orb}=2$~meV.
We finally take a weighted average of the Si concentration fluctuations along $\hat y$ for each element of the 2D cell, oriented in the $x$-$z$ plane. 
Note that the wavefunction $\psi_y(y)$ is used only in the averaging procedure; the remainder of the 2D simulation is performed using the tight-binding model.

\subsubsection{Narrow interfaces}

Although it is extremely difficult to grow ultra-sharp interfaces of width 1~ML, or $\lambda_\text{int}=0.14$~nm, this limit is often considered in theoretical calculations.
For example, in this limit, the valley coupling can be treated as a $\delta$-function in effective-mass theory~\cite{Friesen:2007p115318, Tariq:2019p125309, Hosseinkhani:2020p043180, Hosseinkhani:2021p085309}. 
We therefore also begin by considering the ultra-sharp limit here, as illustrated in Fig.~\ref{fig:wideInt}(a).
We further consider a 100$\times$100~$\text{nm}^2$ section of heterostructure in the $x$-$y$ plane, with a linear step running down the middle of the geometry. 
Si and Ge atoms are assigned to a 3D cell geometry after determining the average concentration for each cell.
We then raster the center position of the dot across the $x$-$y$ plane, apply the 3D-to-2D cell reduction procedure described above at each location (eliminating the $y$ coordinate), and compute the valley splitting in the $x$-$z$ plane using the tight-binding theory.

Figure~\ref{fig:wideInt}(d) shows the resulting valley splitting as a function of dot position. 
Away from the step, because of the sharpness of the interface, we find that $E_v$ can be quite high, typically on the order of 350~$\mu$eV or more. 
Near the step however, $E_v$ is reduced by about 71\%. 
Although there is some variability due to alloy disorder, the step is, without question, the dominant feature in the data.
This is consistent with the fact that the ultra-sharp interface falls within the deterministically enhanced valley-splitting regime. 
Indeed, using effective-mass theory to solve the the same geometry, we find that $E_{v0} \approx 386$~$\mu$eV (away from the step), while $\sigma_\Delta \approx 12$ $\mu$eV, confirming that $E_{v0}\gg \sigma_\Delta$.

\subsubsection{Wide interfaces}

Theoretically, it is well known that $E_v$ depends sensitively on the width of the interface and decays quickly for wider interfaces \cite{Chen:2021p044033}.
Figures~\ref{fig:wideInt}(e) and \ref{fig:wideInt}(f) show results for calculations similar to the previous section, but with interface widths of $\lambda_\text{int} = 1.36$~nm or 10~ML in panel (e), and 2.72~nm or 20~ML in panel (f). 
In contrast to Fig.~\ref{fig:wideInt}(d), the step feature is no longer visible in either of these maps, and the valley splitting variability is fully consistent with alloy disorder. 
Indeed, for the 10~ML interface, we find that $E_{v0} \approx 5 \times 10^{-5}$~$\mu$eV while $\sqrt{\pi} \sigma_\Delta \approx 64$ $\mu$eV, indicating that this quantum well lies deep within the disordered regime: $E_{v0}\ll \sqrt{\pi} \sigma_\Delta$.

We study the crossover between deterministic and disordered behavior in more detail in Fig.~\ref{fig:wideInt}(c).
Here in the main panel, we plot the valley splitting as a function of interface width for the case of no step (blue), and when the dot is centered at a step (orange). 
The markers indicate the mean values obtained from
1,000 minimal-model tight-binding simulations, with different disorder realizations, and the error bars show the 25-75 percentile ranges.
For comparison, NEMO-3D simulation results ($\times$ markers) are also shown for several interface widths, averaged over 10 disorder realizations.
The data show a distinctive \emph{minimum} in the valley splitting, which occurs at the interface width of 3~ML.
The crossover between deterministic and disordered behavior is abrupt, occurring at interface widths of 2-3~ML.
The crossover is observed more clearly in the inset, where the filled circles show the same mean values as the main panel, while the open circles show $E_{v0}$ computed in the virtual crystal approximation, where the Ge concentration in a given layer is given by $\bar Y_l$.
The abrupt divergence of the two data sets between 2-3~ML confirms the crossover location, and clearly demonstrates that alloy disorder has essentially no effect in the deterministically enhanced regime.

The valley splitting behavior on either side of the crossover is also distinctive.
For narrow interfaces, $E_v$ is initially large, dropping precipitously with interface width.
As in Fig.~\ref{fig:fourierStep}, the step is seen to significantly reduce the valley splitting in both the two-band and NEMO-3D simulations. 
In the wide-interface regime, the valley splitting is seen to \emph{increase} with interface width, while the error bars also grow.
These effects can both be attributed to the increasing exposure to Ge.
In this regime, we further note that the difference in $E_v$ for stepped vs.\ non-stepped heterostructures essentially disappears. 

Finally we note that the magnitude and details of the valley splitting depend on the precise shape of the interface.
We also explore the relationship between interface widths and shapes in more detail in Sec.~\ref{sec:interfaces}.

\subsubsection{Effect of steps for very strong alloy disorder}\label{sec:disorderedsteps}

In Fig.~\ref{fig:wideInt}(c), the effects of alloy disorder were found to overwhelm step disorder for increasing levels of Ge in the quantum well.
It is possible to explore the effects of even larger Ge concentrations by introducing Ge directly into the quantum well.
In Figs.~\ref{fig:wideInt}(g), \ref{fig:wideInt}(h), and \ref{fig:wideInt}(i), we show results for geometries similar to Figs.~\ref{fig:wideInt}(d), \ref{fig:wideInt}(e), and \ref{fig:wideInt}(f).
Here the blue data correspond to the same geometries as panels (d)-(f), with the same interface widths.
The orange data correspond to the same geometries, but with an (average) uniform 5\% concentration of Ge added to the quantum wells.
In both cases, the markers show the mean valley-splitting values, averaged over 200 disorder realizations, and the error bars show the corresponding 25-75 percentile ranges, as the dot is moved across a step located at $x=0$. 
For quantum wells with 5\% Ge, the random component of the valley splitting is greatly enhanced, as revealed by the size of the error bars.
For the narrow interfaces of Fig.~\ref{fig:wideInt}(g), the effect of the step is still (barely) visible for the quantum well with 5\% Ge, although it is much less prominent than in the quantum well without Ge. 
In Figs.~\ref{fig:wideInt}(h) and \ref{fig:wideInt}(i), the interface seems to have no effect on the valley splitting, while the random fluctuations dominate. 
Importantly, we see that adding 5\% Ge to the quantum well significantly increases the average valley splitting in all cases. 
This enhancement represents one of the main results of the present work, and we explore it in further detail in Sec.~\ref{sec:highGe}.

\begin{figure*}[] 
	\includegraphics[width=16cm]{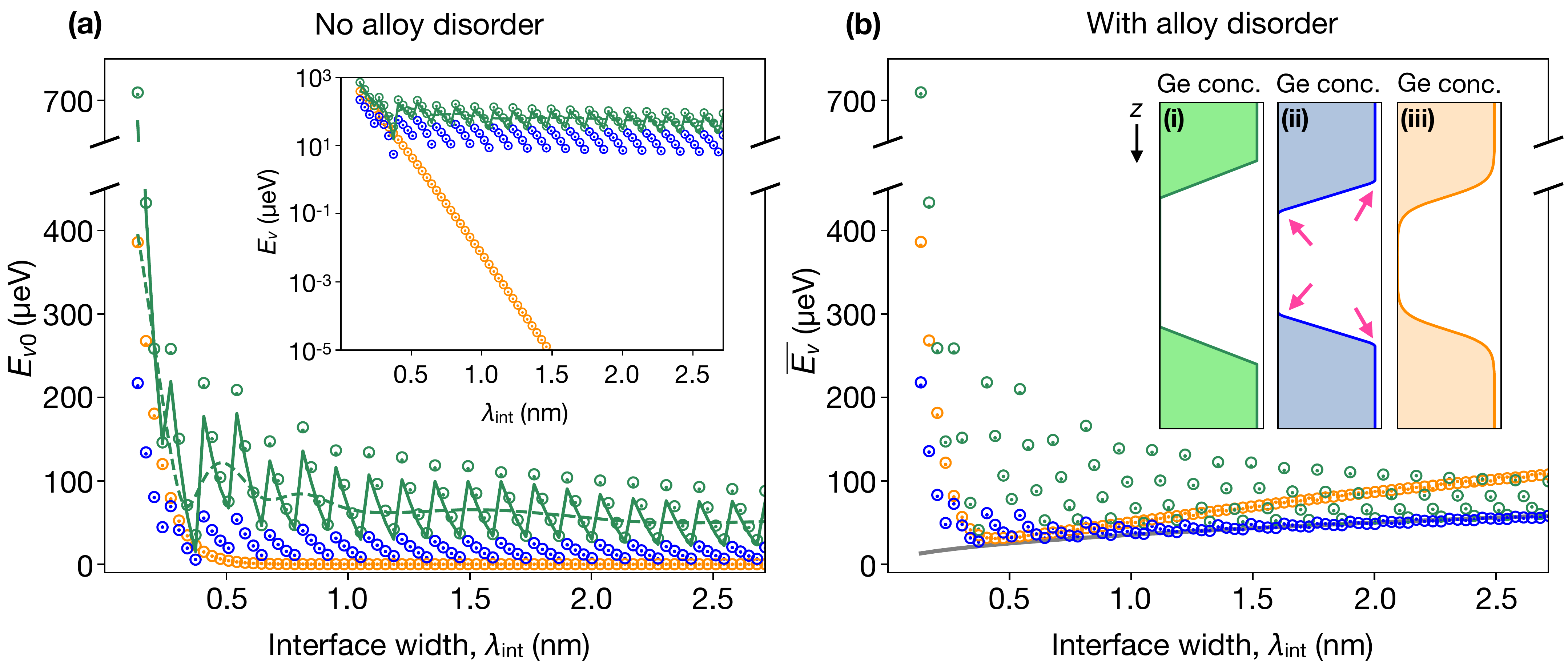}
	\centering
	\caption{
     Valley splitting ($E_v$) depends sensitively on the width ($\lambda_\text{int}$) and shape of the interface, and requires alloy disorder to attain realistic results. 
     (a) The deterministic valley splitting $E_{v0}$ as a function of $\lambda_\text{int}$, obtained using the virtual crystal approximation to remove alloy disorder.
     Results are shown for the three Ge concentration profiles illustrated as insets in (b), with corresponding color coding.
     Dot markers: minimal tight-binding model.
     Open circles: effective-mass model.
     Dashed and solid green lines correspond, respectively, to a continuum variational calculation, Eq.~(\ref{delta0narrow}), and the discretized version of the same calculation.
     Inset: the same data plotted on a semilog scale. 
     (b) Valley splittings computed for the same geometries as (a), in the presence of alloy disorder.
     Dot markers: averaged value of 1,000 minimal tight-binding simulations with different disorder realizations. 
     Open circles: averaged effective-mass model, Eq.~(\ref{riceMean}).
     Insets: (i) linear interface of width $\lambda_\text{int}$, (ii) same linear interface with smoothed corners (see main text), (iii) sigmoidal interface of width $4\tau=\lambda_\text{int}$.
     All simulations here assume a vertical electric field of $E = 5$~mV/nm, a wide 200~ML quantum well to ensure that the wavefunction only feels the top interface, and a quantum well Ge concentration offset of $\Delta Y=30$\%.
	}
	\label{fig:narrowInt}
\end{figure*}

\subsection{Valley splitting variability}

One of the key feature of the valley splitting, apparent in the color maps of Figs.~\ref{fig:wideInt}(d)-\ref{fig:wideInt}(f), is the variability of the valley splitting across a device. 
In the narrow-interface regime, these variations are dominated by the presence of a step, while for wider interfaces, they are mainly caused by alloy disorder.
In both cases, the variations have a characteristic length scale arising from the self-averaging of the concentration fluctuations by the lateral extent of the quantum dot. 
The characteristic length scale of the variations is therefore proportional to the dot size, $a_\text{dot} = \sqrt{\hbar / m_t \omega_\text{orb}}$, which is of order 10-20~nm for typical excitation energies $\hbar\omega_\text{orb}\sim2$~meV assumed here.
We therefore expect to observe significant changes in the valley splitting as the dot is moved over these length scales.
Such tunability is of great practical interest, and we study it in further detail in Sec.~\ref{sec:highGe}.
The ranges of tunability and variability of the valley splitting observed in recent experiments can be quantitatively explained by taking into account the alloy disorder, as we show below.

\subsection{Effect of interface shape} \label{sec:interfaces}

In Sec.~\ref{sec:interplay}, we compared 1~ML interfaces to wider interfaces, finding that the interface shape significantly affects the valley splitting.
Although it is not possible to study all possible interface profiles, we consider three representative cases here, to gain intuition.

The first geometry we consider is the only one used in our simulations so far -- the sigmoidal quantum well, defined in Eq.~(\ref{eq:sigmoid}).
[See Fig.~\ref{fig:narrowInt}(b)iii.
Note that the ultra-sharp interfaces considered in Sec.~\ref{sec:interplay} were also sigmoidal, with characteristic widths of $\lambda_\text{int}=1$~ML.]
The second geometry we consider is the linearly graded interface [Fig.~\ref{fig:narrowInt}(b)i].
For this profile, in the ultra-sharp limit, the interface Ge concentration jumps from its minimum value to its maximum value over a single cell width.
Below, we also explore a range of linear interfaces with smaller slopes.
While such interfaces are more realistic than ultra-sharp interfaces, they possess unphysically sharp corners that induce $2k_0$ components in the Fourier spectrum of $U_\text{qw}(z)$, which artificially enhances the valley splitting.
To correct this problem, we consider a third geometry [Fig.~\ref{fig:narrowInt}(b)ii], which is similar to the linear geometry, but includes a slight rounding of the corners, obtained by averaging the Ge concentration over three successive cell layers: $X_l \rightarrow X_l'=(X_{l-1} + X_l + X_{l+1})/3$.

\subsubsection{Without alloy disorder} \label{sec:narrowInts}

To provide a baseline for analyzing the three model geometries, we first consider the virtual crystal approximation, which does not include alloy disorder (by definition).
In the next section, we solve the same geometries while including alloy disorder.

Figure~\ref{fig:narrowInt}(a) shows the results of valley-splitting simulations as a function of interface width, in the absence of alloy disorder.
We compare the three quantum well profiles, which are color-coded to match the insets of Fig.~\ref{fig:narrowInt}(b).
We also compare two different calculation methods: the minimal tight-binding model (dots) and effective-mass theory of Eq.~(\ref{discreteDelta}) (open circles).
In the latter case, the envelope function $\psi_\text{env}$ is computed numerically using a Schr\"{o}dinger equation.
The excellent theoretical agreement again demonstrates the validity of the effective-mass approach.
For all three geometries, the valley splittings are found to be larger for narrow interfaces, while quickly decreasing for wider interfaces.

Figure~\ref{fig:narrowInt}(a) also illustrates the strong dependence of $E_v$ on the shape of the interface. 
We note that perfectly linear Ge profiles (green data) yield valley splittings that are deterministically enhanced, compared to the other two methods, due to the sharp corners.
It is interesting to note that even minimal smoothing of the sharp corners causes a significant reduction of the valley splittings (blue data), compared to the sharp-corner geometry. 
The more realistic sigmoidal geometry has even lower valley splitting over most of its range (orange data).

The two linear geometries (blue and green) exhibit periodic oscillations, which can be explained as sampling effects, or discreteness of the atoms at the interface. 
We may confirm this hypothesis analytically as follows.
The dashed green line in Fig.~\ref{fig:narrowInt}(a) shows the results for a continuum-model variational approximation for $E_v$, which does not take into account the discreteness of the atoms, and does not correctly reproduce the tight-binding oscillations in Fig.~\ref{fig:narrowInt}(a). 
In contrast, a numerical, but discrete, solution of the same variational model (solid green line), exhibits the same oscillations as the tight-binding results.
Details of these calculations are presented in Appendix~\ref{appendix:variational}. 

Finally, to make contact with experiments, we consider a sigmoidal interface of width of $\lambda_\text{int} \approx 0.8$~nm, as consistent with recent atom probe tomography (APT) measurements~\cite{Wuetz:2022p7730}.
The corresponding valley-splitting estimate, from Fig.~\ref{fig:narrowInt}(a), is given by $E_v\approx 0.1$~$\mu$eV.
This predicted value is much smaller than the average measured value of $\bar E_v\approx 42$~$\mu$eV, indicating that random-alloy disorder is a key ingredient for understanding the experimental results.
A corollary to this statement is that the quantum wells studied in Ref.~\cite{Wuetz:2022p7730} fall into the disorder-dominated regime, despite having interface widths below 1~nm.

\subsubsection{With alloy disorder} \label{sec:wideInts}

Figure~\ref{fig:narrowInt}(b) shows the same type of valley-splitting results as Fig.~\ref{fig:narrowInt}(a), for the same three quantum well geometries, but now including alloy disorder.
Here the sigmoidal results (orange) are the same as in Fig.~\ref{fig:wideInt}(c), where we used the same quantum well geometry.
In Fig.~\ref{fig:narrowInt}(b), the dot markers show the average of 1,000 minimal tight-binding model simulations, while open circles show effective-mass results from Eq.~(\ref{riceMean}), where $\Delta_0$ is taken from Fig.~\ref{fig:narrowInt}(a) and $\sigma_\Delta$ is computed in Eq.~(\ref{varDelta}).
Here again we observe excellent agreement between the two theoretical approaches.

Figures~\ref{fig:narrowInt}(a) and \ref{fig:narrowInt}(b) are nearly identical in the deterministically enhanced regime (very low interface widths).
This is expected since alloy disorder plays a weak role in this case.
In the randomly dominated regime ($\lambda_\text{int} \gtrsim 0.4$~nm), the mean valley splitting is enhanced above its deterministic value, similar to results obtained in previous sections.
In particular, for the interface width $\lambda_\text{int}=0.8$~nm, we obtain $\bar E_v\approx 42$~$\mu$eV in the presence of alloy disorder, which is \emph{more than two orders of magnitude higher} than the disorder-free result, and much more in line with experimental measurements~\cite{Wuetz:2022p7730}.

While the valley splitting generally trends downward for large $\lambda_\text{int}$ in Fig.~\ref{fig:narrowInt}(a), it is interesting to note that the opposite is true in Fig.~\ref{fig:narrowInt}(b).
This is due to the wavefunction being exposed to higher Ge concentrations, and more disorder, for larger $\lambda_\text{int}$.
The effect is especially prominent for the sigmoidal Ge profile.
We also observe that the deterministic enhancement of the valley splitting, due to unphysically sharp corners, persists into the large-$\lambda$ regime.
In this case, the valley splittings closely match those shown in Fig.~\ref{fig:narrowInt}(a), when $E_{v0} \approx \bar E_v$.
Finally, we note that it is possible to derive an analytical estimate for $\bar E_v$ in the randomly dominated regime using a variational calculation, as shown by the gray line in Fig.~\ref{fig:narrowInt}(b) and explained in Appendix~\ref{appendix:variational}.

\section{Alternative heterostructures} \label{sec:specialized}

Moving beyond conventional SiGe/Si/SiGe heterostructures, several alternative schemes have been proposed to boost the valley splitting.  
In this section, we analyze the performance of such proposals, focusing on the effects of alloy disorder. 
In Sec.~\ref{sec:highGe}, we consider quantum wells containing a uniform concentration of Ge, as proposed in Ref.~\cite{Wuetz:2022p7730}.
Here we study the dependence of the valley splitting on Ge content and electric field, finding that the extra Ge greatly enhances the valley splitting on average, but also increases the variability.
Taking this a step further, we explore how such variability allows for enhanced tuning of the valley splitting in these structures. 
In Sec.~\ref{sec:narrowWells}, we study narrow quantum wells and compare our simulation results to the experimental results of Ref.~\cite{Chen:2021p044033}, obtaining very good agreement. 
In Sec.~\ref{sec:spike}, we explore the effects of a narrow Ge spike centered inside a quantum well~\cite{McJunkin:2021p085406}, focusing on how the spike width affects the valley splitting in the presence of alloy disorder. 
In Sec.~\ref{sec:wigglewell}, we comment on the Wiggle Well heterostructure, which contains intentional Ge concentration oscillations with a carefully chosen period~\cite{McJunkin:2022p7777}. 
Finally in Sec.~\ref{algorithm}, we develop a procedure for determining the optimal Ge profile for a quantum well, to maximize $E_v$ in either the deterministic or disorder-dominated regime.

\subsection{Uniform Ge in the quantum well}  \label{sec:highGe}

\subsubsection{Valley-splitting distributions}
\begin{figure}[t] 
	\includegraphics[width=6.5cm]{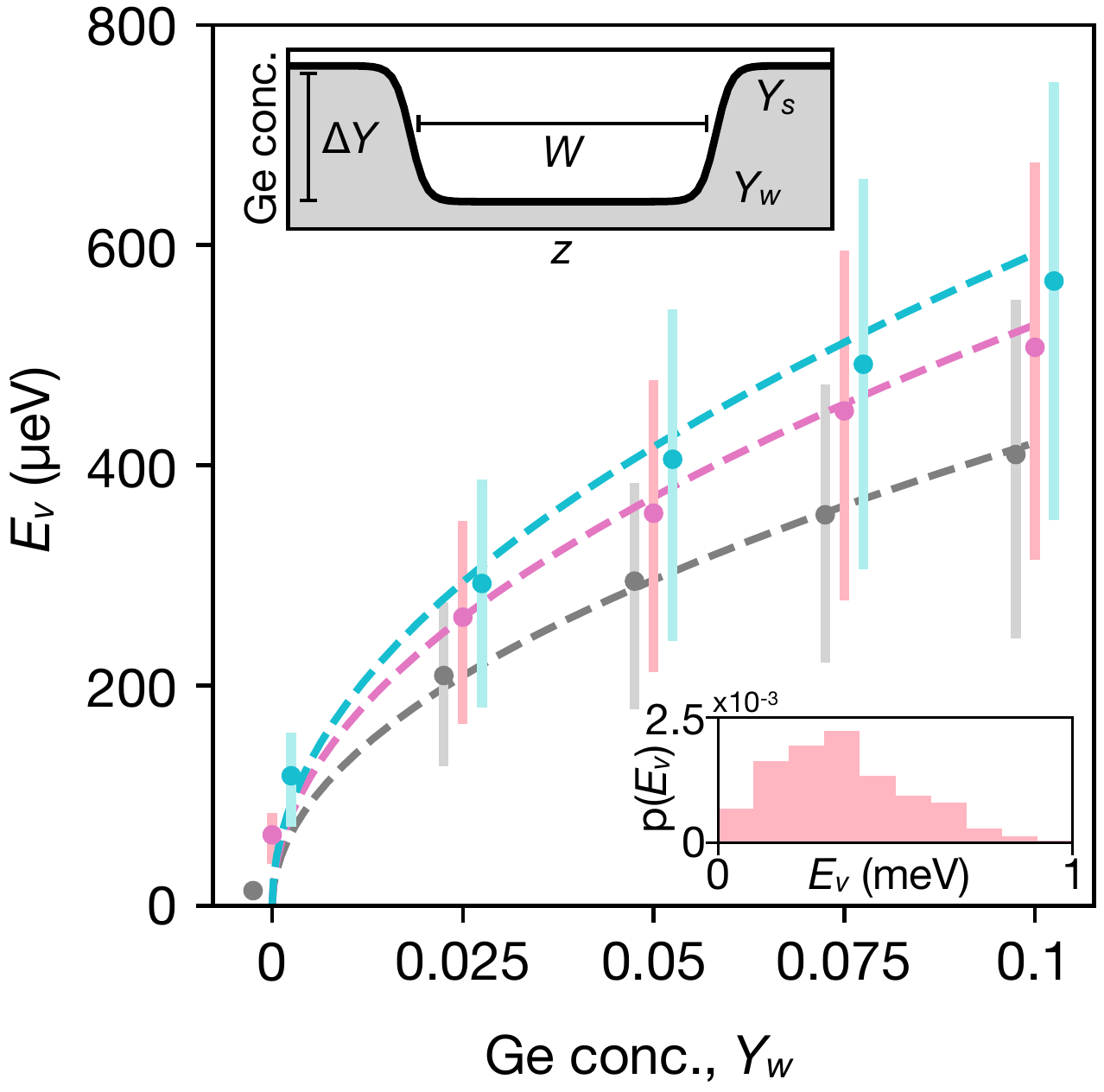}
	\centering
	\caption{Valley splittings as a function of uniform Ge concentration $Y_w$ inside the quantum well for three vertical electric fields: $E_z=0$ (gray), 5~mV/nm (pink), and 10~mV/nm (blue). 
 Here, the dots indicate mean values, and error bars indicate the 25-75 percentile range, for 1D minimal tight-binding simulations based on 1,000 different disorder realizations. 
 (Blue and gray dots are offset to the left and right, respectively, for clarity.)
 Dashed lines show theoretical predictions from Eq.~(\ref{uniformGeNoF}). 
 The simulations assume a dot with an orbital energy splitting of $\hbar \omega_\text{orb} = 2$~meV and are performed in a quantum well of width $W = 80$~ML, concentration offset $\Delta Y = 30$\%, and sigmoidal interface of width $\lambda_\text{int} = 10$~ML. 
 Top inset: schematic illustration of the quantum well simulation geometry. 
 Bottom inset: valley-splitting distribution results, corresponding to $Y_w = 5$\% and $E_z = 5$~mV/nm in the main plot.}
	\label{fig:uniform}
\end{figure}

In Ref.~\cite{Wuetz:2022p7730}, it was proposed to add a uniform concentration of Ge to the quantum well to significantly increase the random component of the intervalley coupling $\delta \Delta$ and the average valley splitting $\bar E_v$. 
The resulting valley splittings fall deep within the disorder-dominated regime.

Figure~\ref{fig:uniform} shows the results of minimal-model tight-binding simulations of the valley splitting, as a function of the uniform Ge concentration $Y_w$, for vertical electric fields $E_z=0$ (gray), 5~mV/nm (pink), and 10~mV/nm (blue), in quantum wells of width $W$ as defined in the upper inset.
Here we plot the average of 1,000 disorder realization (closed circles), and the corresponding 25-75 percentiles (bars). 
We see that even a small amount of Ge produces a large enhancement of the valley splitting, as compared to a conventional, Ge-free quantum well ($Y_w=0$). 

We can approximate the scaling form for the mean valley splitting using $\bar E_v\approx \sqrt{\pi}\sigma_\Delta$ in the disorder-dominated regime, where $\sigma_\Delta$ is given in Eq.~(\ref{varDelta}), and we adopt a very simple approximation for the envelope function:
\begin{equation} \label{eq:varform}
\psi_\text{env}(z)= 
\begin{cases}
\sqrt{2/L_z} \sin \left( \pi z /  L_z \right), & (0\leq z\leq L_z), \\
0, & (\text{otherwise}).
\end{cases}
\end{equation}
Approximating the sum in Eq.~(\ref{varDelta}) as an integral, we obtain the useful scaling relation
\begin{equation} \label{uniformGeNoF}
\bar E_v \approx \sqrt{3 \over 32} {a_0^{3/2} \over a_\mathrm{dot} L_z^{1/2}} {|\Delta E_c| \over X_w - X_s} \sqrt{X_w(1-X_w) } ,
\end{equation}
where $\Delta E_c$ is determined by the quantum well concentration offset, taken to be $\Delta Y=30\%$ in this section.
For the case of no vertical electric field, $L_z$ is given by the physical width of the quantum well, $W=z_t-z_b$, as defined in Eq.~(\ref{eq:sigmoid}).
When the electric field is large enough that the wavefunction does not feel the bottom of the well, the well can be treated as a triangle potential.
In this case, a variational calculation using Eq.~(\ref{eq:varform}) gives
\begin{equation} \label{sigmaVariational}
    L_z \approx \left( {2 \hbar^2 \pi^2 \over e E_z m_l} \right)^{1/3},
\end{equation}
as described in Appendix~\ref{appendix:variational}.
These analytical estimates are shown as dashed lines in Fig.~\ref{fig:uniform}.

The resulting distribution of tight-binding results is plotted in the lower inset of Fig.~\ref{fig:uniform}, for the case $Y_w=0.05$.
We note that the distribution takes the characteristic Rician form expected in the disorder-dominated regime, $\sigma_\Delta\gg E_{v0}$, for which the density of states near $E_v=0$ is nonvanishing.
As the Ge concentration increases, the whole distribution moves towards higher energies, and fewer samples have low energies. 
In the following section, we explain how to leverage this important result.

\subsubsection{Spatial variability and tunability of the valley splitting} 

\begin{figure*}[] 
	\includegraphics[width=6.4in]{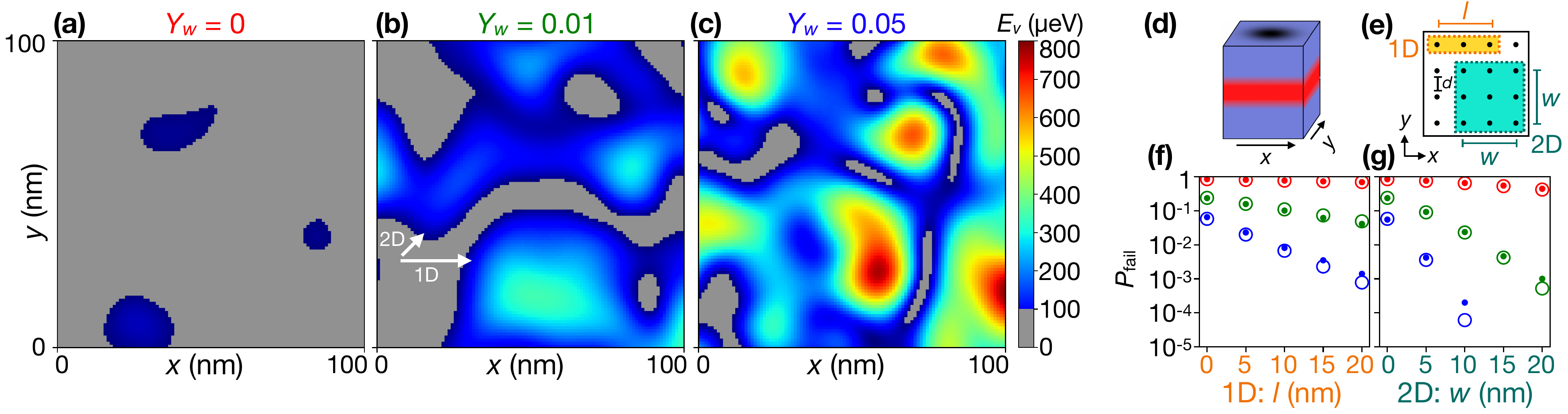}
	\centering
	\caption{
 Tuning the valley splitting in quantum wells with uniform Ge concentrations.
 (a)-(c) The valley splitting $E_v$ is computed as a function of the quantum dot center position $(x,y)$, in the presence of alloy disorder. 
 Results are shown for three uniform Ge concentrations inside the quantum well: (a) $Y_w=0$, (b)  $Y_w=0.01$, and (c) $Y_w=0.05$.
We assume sigmoidal quantum wells with interface widths of $\lambda_\text{int} = 10$~ML, quantum well widths of $W = 80$~ML, quantum well offsets $\Delta Y=Y_s-Y_w$ = 30\%, and vertical electric fields $E_z=5$~mV/nm. 
In the color maps, valley splittings below 100~$\mu$eV are considered dangerous for qubit operation, and are shaded gray. 
(d) Quantum well simulation geometry, with $x$ and $y$ directions labeled. 
The quantum dot is located in the quantum well, with the density profile indicated by the shading on the top of the device. 
(e) Schematic illustration of the simulation procedure used to compute the failure probability $P_\text{fail}$ of finding an acceptable valley splitting (i.e., with $E_v\geq 100$~$\mu$eV) when a quantum dot is allowed to explore different locations inside a 1D or 2D search box of size $l$ or $w$, respectively. (See main text for details.) 
(f), (g) $P_\text{fail}$ as a function of search-box size. 
For simplicity, we only consider quantum dots centered at grid points, with spacing $d = 5$~nm. 
Red, green, and blue markers show results for heterostructures with average Ge contents of $Y_w = 0$, 0.01, and 0.05, respectively. 
Solid dots correspond to numerical averages, computed from 10,000 disorder realizations. 
Open circles show corresponding theoretical results, as computed in Appendix~\ref{appendix:movingDot}.
	 }
	\label{fig:movingDot}
\end{figure*}

As the mean valley splitting increases with Ge concentration, the variability also increases.
This can be seen in Figs.~\ref{fig:movingDot}(a)-\ref{fig:movingDot}(c), where the valley splitting is plotted as a function quantum dot position, for three different Ge concentrations.
Here, the calculations are performed similarly to Fig.~\ref{fig:wideInt}, although the effective simulation geometry is now reduced to 1D, since there are no interface steps.
Regions with dangerously low valley splittings (here defined as $E_v<100$~$\mu$eV) are shaded gray.
These gray regions decrease in size as the Ge concentration increases, as consistent with the previous section, while the variability is seen to increase significantly.

We can take advantage of this behavior by proposing that, when a quantum dot is initially centered at a dangerous location, the gate voltages should be adjusted to change its location.
For larger Ge concentrations, such desirable locations are typically found in close proximity.
Indeed, dot motion of up to 20~nm has been reported in recent experiments~\cite{Dodson:2022p146802, McJunkin:2022p7777}, which has in turn been used to explain the large observed variations in valley splitting. 

We now study the likelihood of being able to find a nearby `safe' alloy-disorder configuration, for which $E_v \geq 100$~$\mu$eV.
We consider two scenarios, illustrated schematically in Figs.~\ref{fig:movingDot}(d) and \ref{fig:movingDot}(e).
In the first case, the dot can be shifted in a single, fixed direction. In the second, the dot can be shifted in either of two directions. 
To begin, we generate valley-splitting maps, similar to Figs.~\ref{fig:movingDot}(a)-\ref{fig:movingDot}(c), for the same three Ge concentrations.
To simplify the search procedure, we divide each map into a 2D grid of points separated by 5~nm, as shown in Fig.~\ref{fig:movingDot}(e).
For every grid point initially characterized as `dangerous' (i.e., with $E_v< 100$~$\mu$eV), we search for at least one non-dangerous grid point within a 1D or 2D search box of size $l$ or $w$, as illustrated in  Fig.~\ref{fig:movingDot}(e).
Defining $P_\text{fail}$ as the probability of failure, we repeat this procedure for 10,000 disorder realization and plot the average $P_\text{fail}$ values in Figs.~\ref{fig:movingDot}(f) and \ref{fig:movingDot}(g) (dots) as a function of $l$ or $w$.
Analytical results for the same disorder realizations are also shown as open circles, by accounting for the correlations between valley splittings in neighboring sites, as described in Appendix~\ref{appendix:movingDot}.
We see that the ability to search over larger regions greatly enhances the success rate of these procedures, particularly for 2D searches.
For larger Ge concentrations, many orders of magnitude of improvement can be achieved in this way.

\subsection{Narrow quantum wells} \label{sec:narrowWells}

\begin{figure}[t] 
	\includegraphics[width=6.5cm]{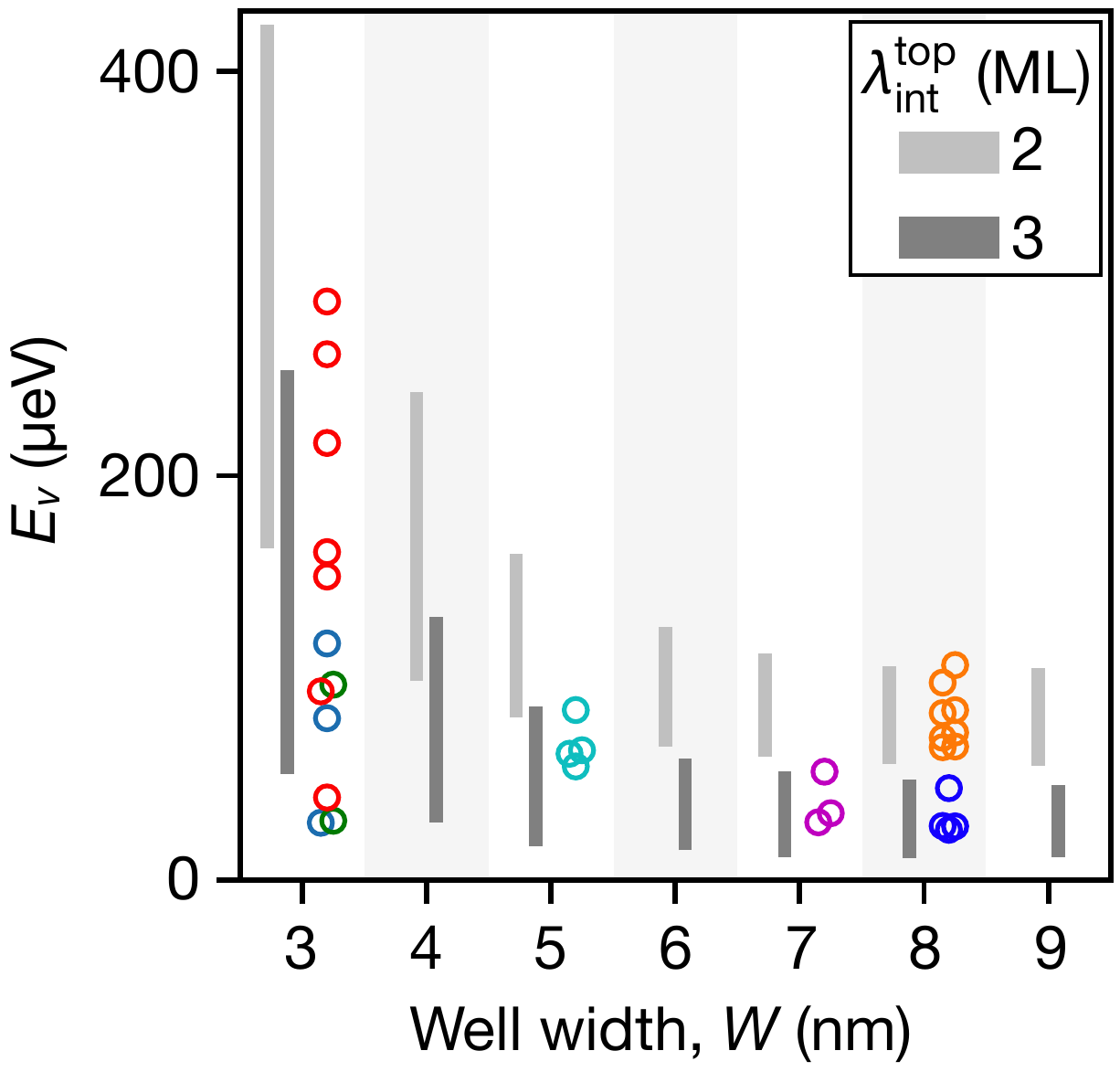}
	\centering
	\caption{Comparison of experimental valley splitting measurements~\cite{Chen:2021p044033} (open circles) to tight-binding  simulations with alloy disorder, as a function of quantum well width.
 Experimental data points with the same color represent quantum dots fabricated on the same chip.
Bars indicate the 10-90 percentile ranges of 1,000 1D minimal tight-binding simulations, assuming a relatively sharp top interface of width $\lambda_\text{int}^\text{top}=2$~ML (light-gray bars) or 3~ML (dark-gray bars). 
As in Ref.~\cite{Chen:2021p044033}, the bottom interface width is taken to be $\lambda_\text{int}^\text{bot} = 8$~ML, and we assume an electric field of $E_z = 5$~mV/nm, and an orbital splitting of $\hbar \omega_\mathrm{orb} = 1.5$~meV.  }
	\label{fig:hrlTwoInterface}
\end{figure}

It is known from theory and experiment that narrowing a quantum well while keeping other growth and confinement parameters fixed should enhance its valley splitting~\cite{Boykin:2004p115, Friesen:2007p115318}. 
For example, in Fig.~\ref{fig:hrlTwoInterface} we reproduce the valley-splitting results reported in Ref.~\cite{Chen:2021p044033} (open circles), where it was found that 3~nm wells have higher valley splittings on average than wider wells.
However, significant variability is also observed, and some 3~nm quantum wells are still found to have valley splittings that are dangerously low.
Such behavior is in contrast with the deterministic enhancement predicted for narrow wells, but can be fully explained by alloy disorder, noting that the electron wavefunctions are forced to overlap with Ge in the barriers when the quantum wells are very narrow.

To study this behavior, we perform tight-binding simulations of the valley splitting as a function of well width, assuming sigmoidal barriers and a vertical electric field of $E_z=5$~mV/nm, for the same well widths as Ref.~\cite{Chen:2021p044033}.
Similar to theoretical calculations reported in that work, we consider a range of widths for the top interface, while keeping the bottom interface width fixed at 8~ML.
In contrast with that work, we include the effects of alloy disorder by performing simulations with 1,000 different disorder realizations for each geometry.
We plot our results as 10-90 percentile ranges (gray bars) in Fig.~\ref{fig:hrlTwoInterface}.
Here we only show the results for interface widths of 2 or 3~ML and  quantum dot confinement potentials $\hbar\omega_\text{orb}=1.5$~meV, since those values provide excellent agreement with the data, for both the mean valley splitting values and the variability.
These results lend strong support for the role of alloy disorder in determining the variability of the valley splitting.

We conclude that valley splittings can be enhanced on average by using narrow quantum wells, due to increased overlap with Ge in the barrier regions.
Considering the trends observed in Fig.~\ref{fig:hrlTwoInterface}, it is interesting to ask whether deterministically enhanced behavior (e.g., exponential suppression of small valley splittings) could potentially be achieved in ultra narrow quantum wells.
We can answer this question using the crossover criterion of Eq.~(\ref{cuttoffPoint}), finding that, for all the results shown in Fig.~\ref{fig:hrlTwoInterface}, only wells with interface widths of $\lambda_\text{int}^\text{top}=2$~ML fall into the deterministically enhanced regime.
This is consistent with our more general results for wider quantum wells, showing that deterministic enhancement of the valley splitting still requires super sharp interfaces, and indicates that there is no deterministic advantage in using narrow quantum wells.
Finally, by comparing the experimental and theoretical results in Fig.~\ref{fig:hrlTwoInterface}, for $W=3$~nm, we see that the experiments are most consistent with $\lambda_\text{int}^\text{top}=3$~ML, whose behavior is randomly dominated.
This again emphasizes the difficulty of achieving deterministically enhanced valley splittings.


\subsection{Ge spike} \label{sec:spike}

\begin{figure*}[] 
	\includegraphics[width=6.4in]{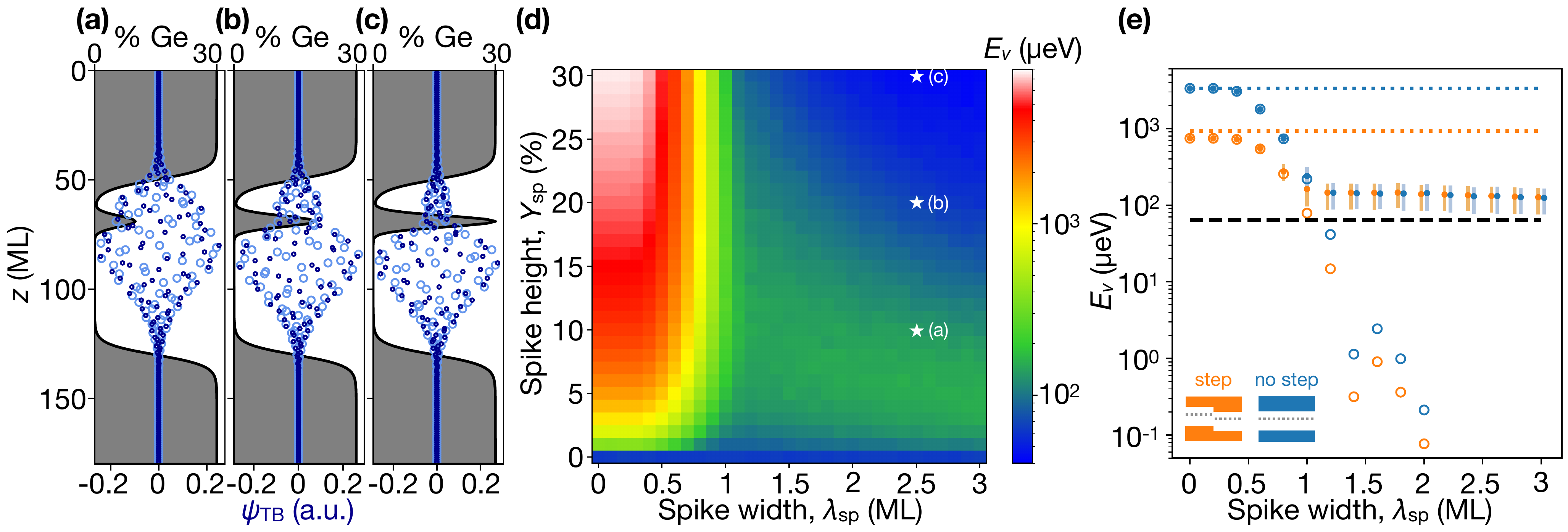}
	\centering
	\caption{Interplay between deterministic and random contributions to the valley splitting, for the Ge spike geometry.
 (a)-(c) Ge concentration profiles (gray) for spike geometries of width $\lambda_\text{sp}=2.5$~ML and heights $Y_\text{sp}=10$, 20, and 30\%, respectively.
 The resulting ground-state (closed circles) and excited-valley state (open circles) tight-binding wavefunctions are also shown.
 For increasing spike heights, we see that the wavefunction is pushed away from the spike, resulting in lower Ge overlap and lower disorder-enhanced valley splittings.
(d) Mean valley splittings (color scale) as a function of spike width $\lambda_\text{sp}$ and spike height $Y_\text{sp}$. 
Here, each pixel represents an average of 1,000 1D minimal tight-binding simulations, and the results corresponding to panels (a)-(c) are indicated with stars. 
(e) Valley splitting as a function of spike width, for a fixed 10\% Ge spike height, and a variety of device geometries, including a virtual crystal approximation without alloy disorder ($E_{v0}$, open circles), and averaged results from 1,000 random-alloy simulations ($\bar E_v$, closed circles, with 25-75 percentile error bars).
In both cases, we consider both step-free geometries (blue), and geometries with a single step through the center of the dot (orange). 
Here, deterministically enhanced behavior occurs only for very narrow spikes of width $<1$~ML, corresponding to a single atom in our model geometry.
The black dashed line shows the average valley splitting for the same quantum well as the other simulations, including random alloy, but without the Ge spike.
The blue dotted line is the maximum valley splitting, corresponding to a perfect single-atom spike geometry, computed as in Ref.~\cite{McJunkin:2021p085406} for a spike height of $Y_\text{sp}=10$\%.
The orange dotted line shows the same result for the case of a single-atom step running through the center of the dot.
Here we assume the step is present on both top and bottom interfaces, as well as the single-atom spike, resulting in a suppression of the valley splitting by a factor of $\sim 0.28$, as explained in the main text.
In all calculations reported here, we assume an isotropic quantum dot with an orbital energy splitting of $\hbar \omega_\text{orb} = 2$~meV and a vertical electric field of $E_z = 5$~mV/nm.
	}
	\label{fig:geSpike}
\end{figure*}

In Ref.~\cite{McJunkin:2021p085406}, it was shown that a narrow spike of Ge in the quantum well can increase the valley splitting by a factor of two, and theoretical calculations indicate that much larger enhancements are possible.
However, alloy disorder was not considered in that work.
In this section, we explore the interplay between optimal geometries (single monolayer spikes, which are difficult to grow), realistically diffused geometries, varying spike heights (i.e., the Ge concentrations at the top of the spike), and interface steps, and we include the effects of alloy disorder.

We first consider the case without interface steps.
Figures~\ref{fig:geSpike}(a)-\ref{fig:geSpike}(c) show heterostructures with Ge spikes of height $Y_\text{sp}=10$, 20, and 30\%, respectively, and their corresponding tight-binding wavefunctions.
Here we define the Ge spike profile as $Y(z) = Y_\text{sp} \exp[-(z-z_\text{sp})^2 / 2 \lambda_\text{sp}^2]$, where $z_\text{sp}$ is the position of the center of the spike, and $\lambda_\text{sp}$ is the spike width.
An important effect can be observed by comparing panels (a)-(c): for increasing spike heights, the wavefunction envelope is suppressed, at and above the spike location, with consequences for disorder-induced valley coupling.
We explore this dynamic in Fig.~\ref{fig:geSpike}(d), where we plot the average valley splitting for varying spike widths and heights.
Here, the stars labelled (a)-(c) correspond to Figs.~\ref{fig:geSpike}(a)-\ref{fig:geSpike}(c).
We also include the singular case of a monolayer spike with no Ge outside this layer, which is defined as $\lambda_\text{sp}=0$ in the figure, and is of interest because it allows for analytical estimates, as obtained in Ref.~\cite{McJunkin:2021p085406}.

Three types of limiting behavior are observed in Fig.~\ref{fig:geSpike}(d), which we also indicate with horizontal lines in Fig.~\ref{fig:geSpike}(e):
(i) In the limit of vanishing spike height, $Y_\text{sp}\rightarrow 0$, we recover results for a conventional quantum well without a spike [dashed black line in Fig.~\ref{fig:geSpike}(e)],  which falls into the disorder-dominated regime for the quantum well considered in Fig.~\ref{fig:geSpike};
(ii) In the limit of ultra-narrow spikes, $\lambda_\text{sp}\rightarrow 0$, we recover the analytical predictions of Ref.~\cite{McJunkin:2021p085406} [dotted blue line in Fig.~\ref{fig:geSpike}(e)], which fall into the deterministically enhanced regime;
(iii) For larger $Y_\text{sp}$ and $\lambda_\text{sp}$, we observe disorder-dominated behavior, characterized by larger valley splittings when the electron overlaps significantly with the Ge (lower spike heights), and smaller valley splittings otherwise (larger spike heights). 

In Fig.~\ref{fig:geSpike}(e), we compare several types of spike behavior, for spikes of height $Y_\text{sp}=10$\%, including the case when the dot is centered at a step.
For simulations with (orange) and without (blue) a step, we plot the average valley splitting values (closed circles) and the corresponding 25-75\% quartiles (error bars).
We also plot the corresponding deterministic, disorder-free results (open circles), obtained using the virtual-crystal approximation.
These results indicate a well-defined crossover from deterministic to disorder-dominated behavior when $\lambda_\text{sp}\approx 1$~ML, suggesting that deterministic enhancement of the valley splitting should be difficult to achieve in this system.
The theoretical maximum $E_v$ due to a single-layer Ge spike (blue dotted line) is computed following Ref.~\cite{McJunkin:2021p085406}, assuming a vertical electric field of $E_z = 5$~mV/nm, obtaining $E_v \approx 3.3$~meV.
We also note that, in the absence of a step, the valley splitting abruptly transitions from its deterministically enhanced upper bound (blue dotted line) to an asymptote that is slightly larger than the valley splitting in the absence of a spike (black dashed line).
Such enhancement is expected in the disorder-dominated regime, as discussed in Sec.~\ref{sec:conventional}.
In the presence of a step, the valley splitting asymptotes to the same value, as anticipated in Sec.~\ref{sec:disorderedsteps}, since the steps do not have a strong effect in this regime.
In the deterministically enhanced regime however, the valley splitting approaches a value suppressed below the step-free result by a factor of ${1 \over 2}| 1 + \exp(- i k_0 a_0/2)| \approx 0.28$ (orange dotted line), for the case where the step runs through the center of the dot.

\subsection{Wiggle Well} \label{sec:wigglewell}

\begin{figure}[h] 
	\includegraphics[width=6cm]{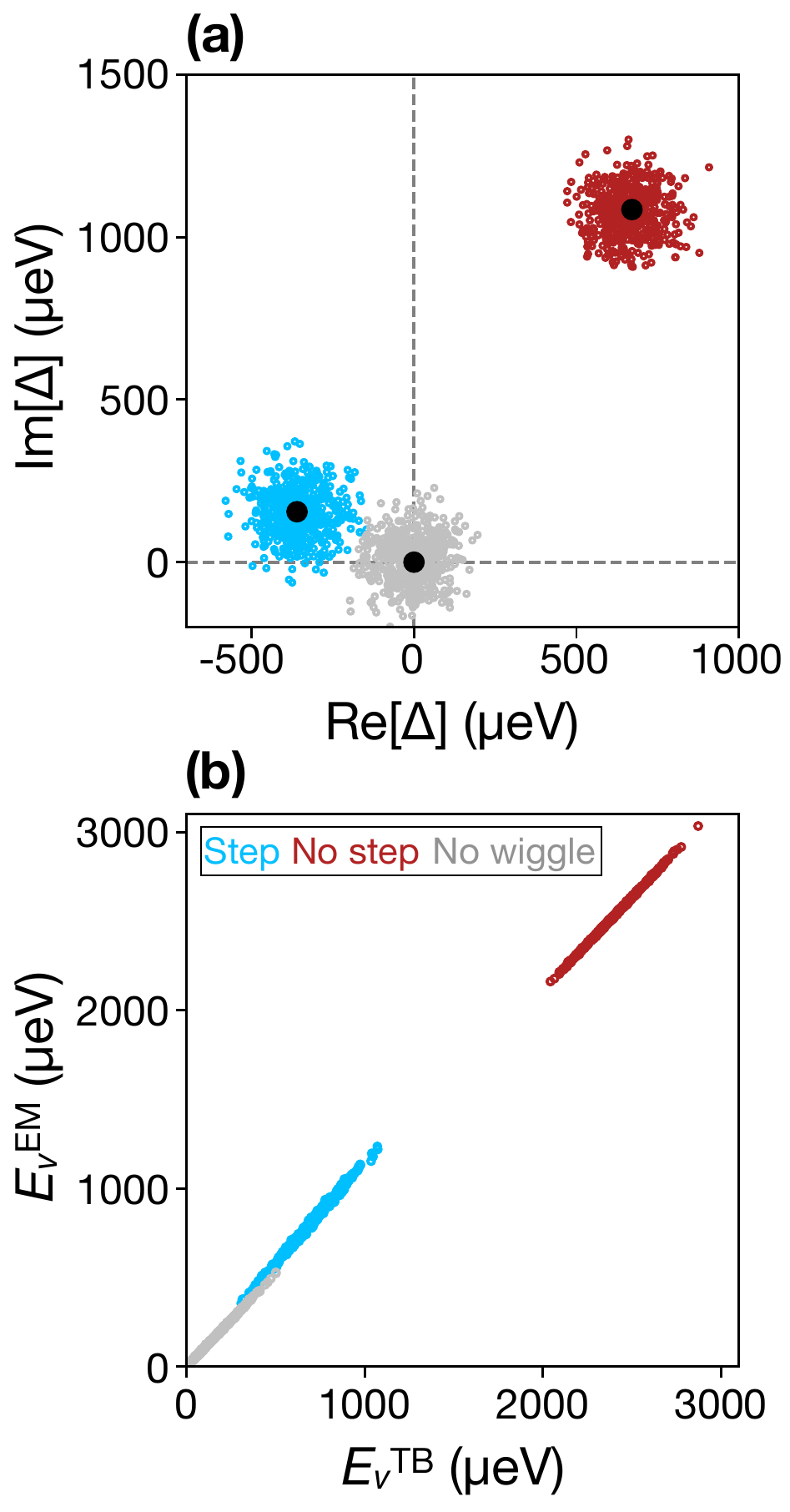}
	\centering
	\caption{Short-period Wiggle Wells provide deterministic enhancement of the valley splitting. 
 (a) Intervalley matrix elements $\Delta$ are computed for short-period Wiggle Wells using the 1D effective-mass approach, and are plotted in the complex plane.
 Results are shown for geometries with (blue dots) and without (red dots) a step passing through the center of the quantum dot confinement potential. 
 Additionally we show results for quantum wells with uniform Ge in the quantum well, but no concentration oscillations, with the same average Ge concentration as the Wiggle Wells, $Y_w=A_w=0.5$\% (gray dots).
All three geometries include random-alloy disorder, and each simulation set is comprised of 500 disorder realizations.
Additionally, we choose a quantum well offset of $\Delta Y =Y_s-\bar Y_w= 30$\%, a vertical electric field of $E_z = 5$~mV/nm, a quantum well width of $W=80$~ML, and an interface width of $\lambda_\text{int} = 10$~ML.
To ensure that $E_v < E_\text{orb}$ in each case, we choose $\hbar \omega_\text{orb} = 4$~meV for all geometries. 
(b) A correlation plot comparing tight-binding and effective-mass simulations of the valley splitting, for the same disorder realizations shown in (a).
Nearly perfect correlations confirm the importance of the $2k_0$ wavevector for determining the valley splitting.
	}
	\label{fig:wiggleWell}
\end{figure}

The most effective method for deterministically enhancing the $2k_0$ component of the confinement potential $U_\text{qw}(z)$ in Eq.~(\ref{eq:EMDelta}) is to add it directly to the quantum well, where it overlaps strongly with the electron wavefunction. 
This is accomplished by introducing Ge concentration oscillations of the form 
\begin{equation}
Y_w(z)=A_w[1-\cos(qz+\phi)] ,
\label{eq:YWW}
\end{equation}
where $A_w$ is the average Ge concentration and $q$ is the oscillation wavevector.
Several wavelengths were proposed to enhance the valley splitting in Refs.~\cite{McJunkin:2022p7777} and \cite{Feng:2022p085304}, including $\lambda=2\pi/q =1.8$~nm and 0.32~nm.
The latter corresponds to $q=2k_0$, which we refer to as the short-period Wiggle Well. 
Below, we make use of the Wiggle Well's large valley splitting to further characterize the transition between deterministic and random-dominated behavior.

In Fig.~\ref{fig:wiggleWell}(a), we compare three closely related calculations of the valley-coupling matrix element $\Delta$, each of which shows the averaged results of 500 alloy disorder realizations.
The simulation geometries include (i) a short-period Wiggle Well with no interface steps (red dots), (ii) a short-period Wiggle Well with a single-atom step passing through the center of the dot (blue dots), and (iii) uniform Ge in the quantum well with no concentration oscillations, but with the same average Ge concentration as the Wiggle Wells, $Y_w=A_w=0.5$\% (gray dots).
The results reveal several interesting features.
First, the red data display a striking deterministic enhancement of the valley splitting, for which the number of solutions with $\Delta$ values near zero is exponentially suppressed.
This is particularly impressive given the small amplitude of the concentration oscillations.
The valley splitting is strongly reduced for the blue data, due to the step; however, a significant deterministic enhancement is still apparent, attesting to the potency of Wiggle Well approach.
Finally, as anticipated in Sec.~\ref{sec:highGe}, we see that devices formed in quantum wells with uniform Ge have $\Delta$ values centered near zero, as consistent with randomly dominated behavior.
A second important observation in Fig.~\ref{fig:wiggleWell}(a) is that the standard deviations of the three distributions about their mean values are nearly identical for the three distributions.
This is consistent with the fact that the mean Ge concentration, and therefore the alloy disorder, is the same in all three cases.
Thus, by moving from the Wiggle Well to the uniform-Ge geometry, we observe a clear crossover from deterministically enhanced to disorder-dominated behavior.

Finally, in Fig.~\ref{fig:wiggleWell}(b) we show a correlation plot comparing effective-mass and tight-binding calculations, similar to Fig.~\ref{fig:heterostructureCartoon}(f), that includes the three data sets of Fig.~\ref{fig:wiggleWell}(a).
Here again we observe nearly perfect correlations, demonstrating that the $2k_0$ theory explains the full range of valley splitting behavior, from extreme deterministic enhancement to totally disorder-dominated.

\subsection{Optimizing the Ge distribution} \label{algorithm}

\begin{figure*}[] 
	\includegraphics[width=12.5cm]{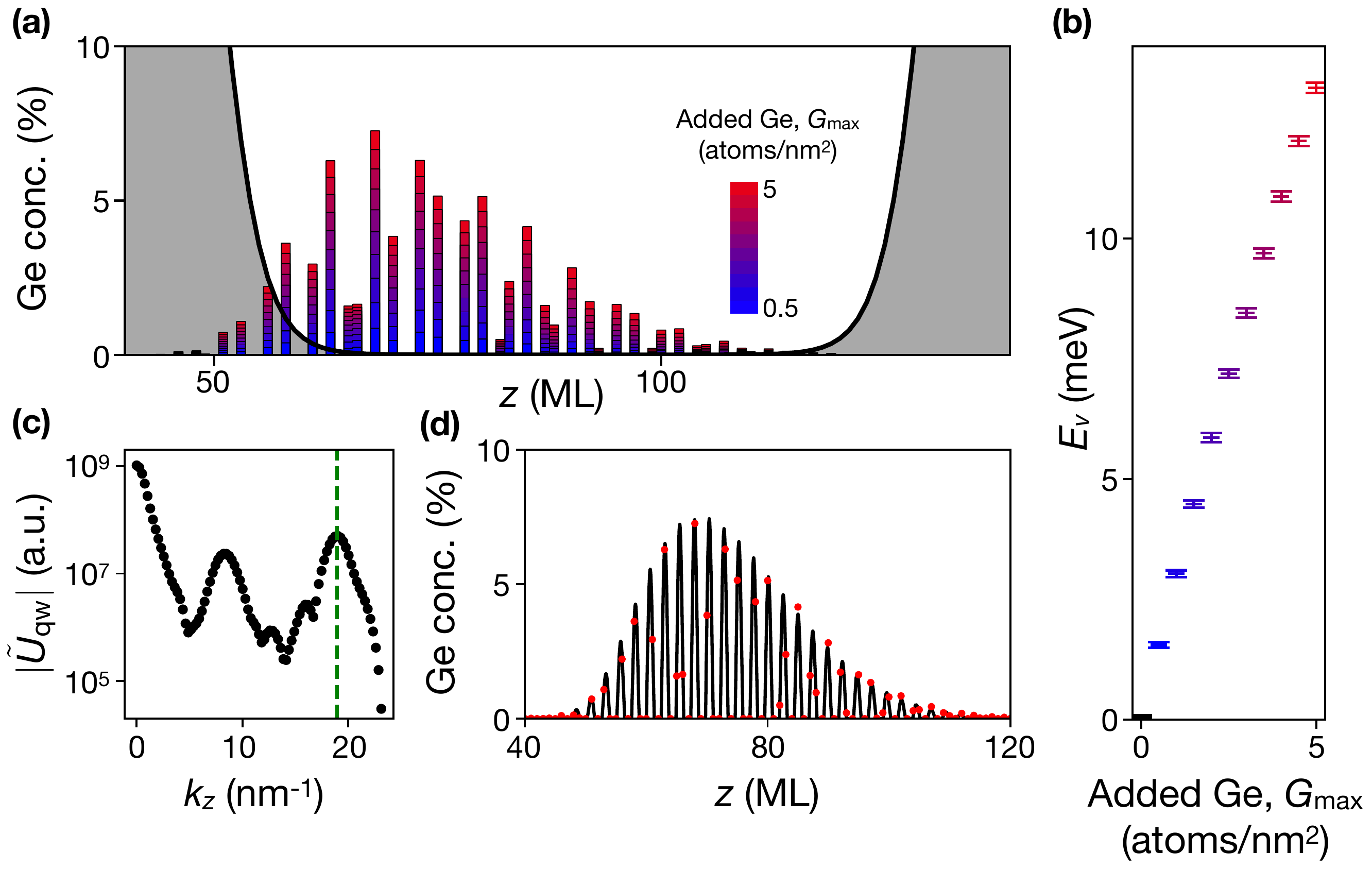}
	\centering
	\caption{Optimizing the Ge concentration profile to provide a large deterministic enhancement of the valley splitting. 
 (a) The initial Ge profile (gray) and added Ge content (vertical bars centered on individual atomic layers) corresponding to the $G_\text{max}$ value indicated by the colorbar.
 Optimization is performed as described in the main text, in the absence of random-alloy disorder. 
The initial sigmoidal profile is defined in Eq.~(\ref{eq:sigmoid}), with $\lambda_\text{int}=10$ ML and $W=80$ ML, and with a vertical electric field of $E_z = 5$~mV/nm and quantum dot orbital splitting of $\hbar \omega_\text{orb} = 2$~meV. 
(b) Optimized valley splitting distributions, including random-alloy disorder, as a function of the added Ge. 
Results are shown for the mean values and 25-75 percentile range of 1,000 1D tight-binding simulations at each $G_\text{max}$ value. 
(c) Discrete Fourier transform of the weighted barrier potential $|\tilde U_\text{qw}|$, as a function of the reciprocal wavevector, for the optimized concentration profile corresponding to $G_\text{max}=5$~atoms/nm$^2$. 
The green dashed line identifies the wavevector $q = 2 k_0$ responsible for valley splitting. 
(d) Best fit of Eq.~(\ref{eq:Yfit}) (solid curve) to the optimized Ge concentration profile (red dots) shown in (a), for the case of $G_\text{max}=5$~atoms/nm$^2$. 
}
	\label{fig:optimizePureVS}
\end{figure*}

\begin{figure}[] 
	\includegraphics[width=7cm]{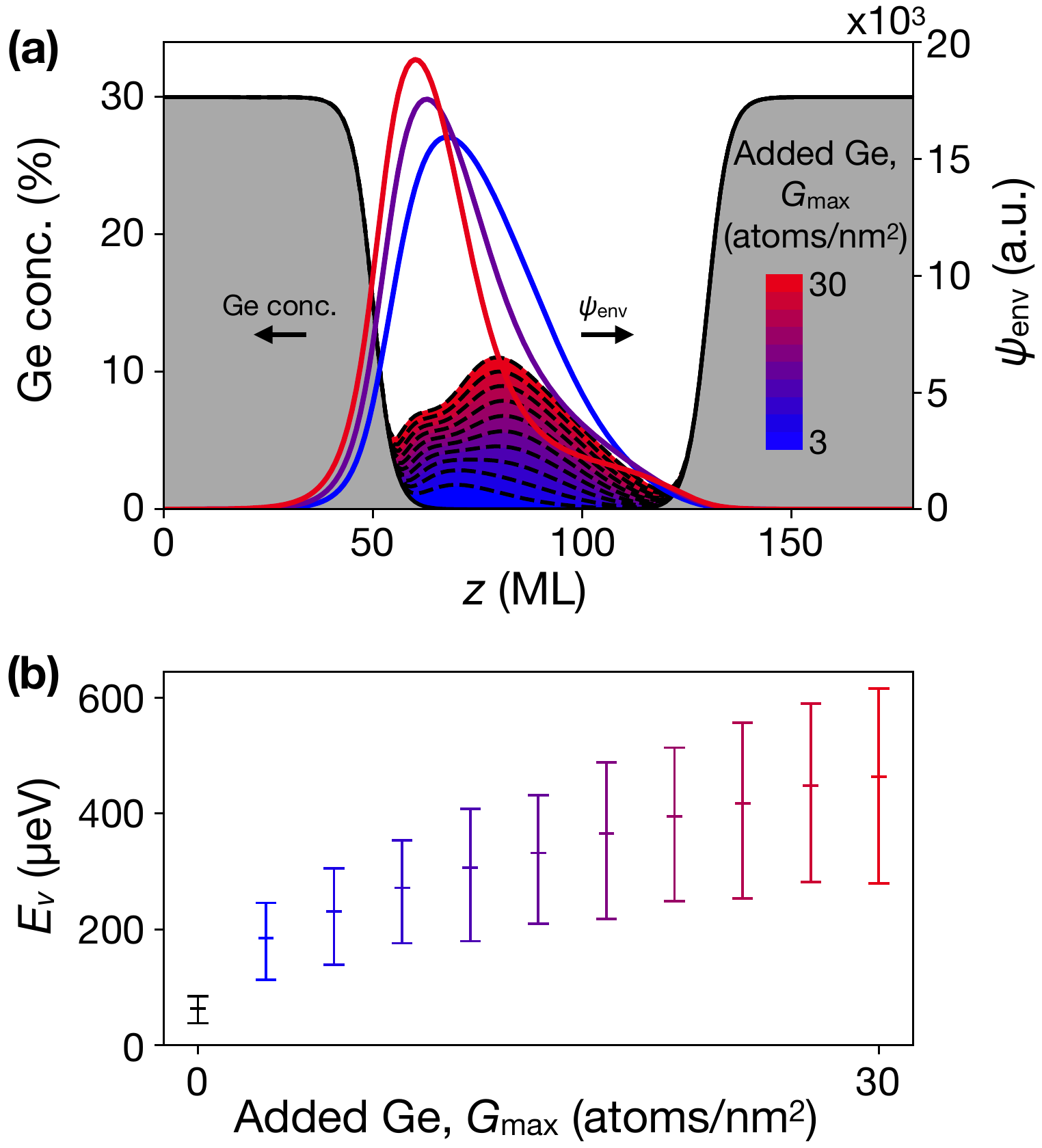}
	\centering
	\caption{
 Optimized Ge profiles for enhancing the valley splitting in the disorder-dominated regime.
(a) The initial Ge concentration profile (gray) and added Ge content (colored region), using the procedure described in the main text and the same quantum well as Fig.~\ref{fig:optimizePureVS}.
The colorbar indicates the total added Ge concentration, $G_\text{max}$, and the resulting envelope functions $\psi_\text{env}(z)$ are shown for the cases of $G_\text{max}=3$, 15, and 30~atoms/$\text{nm}^2$.
(b) Optimized valley splitting distributions as a function of $G_\text{max}$, showing the mean values and 25-75 percentile ranges of 1,000 1D tight-binding simulations at each $G_\text{max}$ value. }
	\label{fig:optimize}
\end{figure}

We have shown that the valley splitting can be deterministically enhanced in structures such as the Wiggle Well, or enhanced on average in quantum wells with uniform Ge.
However, most schemes considered here require increasing the contact with random Ge alloy, which has the undesired side effect of decreasing the mobility~\cite{McJunkin:2022p7777}.
It is therefore interesting to search for Ge concentration profiles that maximize the valley splitting while reducing the total amount of Ge in the quantum well. 
In this section, we use a projected-gradient-descent routine to maximize the valley splitting for a specified amount of Ge.
In its simplest form, this algorithm tends to remove all Ge inside or outside the quantum well, which is not the desired solution.
We therefore constrain the procedure to only \emph{add} Ge to an initial sigmoidal quantum well.
The following steps are then repeated until a steady-state solution is achieved: (i) estimate the gradient of the reward function, which we take to be the valley splitting computed using the 1D minimal tight-binding method, (ii) update the Ge concentration in each layer in the direction of the gradient, (iii) renormalize the Ge concentration in all layers so that the resulting concentration profile has a fixed density of additional Ge atoms, $G_\mathrm{max}$. 
(With out this renormalization step, the algorithm would continually increase the amount of Ge in the quantum well, which is also not a desired solution.) 
Full details of the optimization procedure can be found in Appendix~\ref{appendix:algorithm}.
Note that the total Ge content is computed by summing the contributions from individual layers in the $z$ direction; $G_\mathrm{max}$ is therefore reported in units of atoms/$\text{nm}^2$.
Below, we apply the routine separately for optimizations in the deterministic vs.\ random regimes.

\subsubsection{Deterministically enhanced regime}

Deterministically enhanced valley splittings are achieved by allowing the routine to optimize both the short-wavelength oscillations and the large-scale Ge concentration envelope that determines the shape of the wavefunction envelope.
To focus on deterministic effects, we perform the optimization in the virtual crystal approximation (i.e., without including random-alloy fluctuations).
Results for the added Ge concentration are shown in color in Fig.~\ref{fig:optimizePureVS}(a), for the initial sigmoidal concentration profile shown in gray.
Here the color scale represents the total added Ge concentration, where $G_\text{max}\in (0.5,5)$~atoms/nm$^2$, for a fixed electric field of $E_z=5$~mV/nm.
The resulting behavior is reminiscent of the Wiggle Well.

To analyze this behavior, we Fourier transform the weighted confinement potential defined in Eq.~(\ref{eq:EMDelta}), $\tilde U_\text{qw}(z)=U_\text{qw}(z)|\psi_\text{env}(z)|^2$.
Results are shown in Fig.~\ref{fig:optimizePureVS}(c) as a function of the reciprocal wavevector $k_z$, for the case of $G_\text{max}=5$ atoms/nm$^2$.
Here as usual, the $2k_0$ component, $\tilde U_\text{qw}(k_z=2k_0)$, determines the valley splitting.
We therefore expect to find an enhancement of $|\tilde U_\text{qw}(k_z)|$ at this wavevector (dashed-green line).
Indeed this is the observed behavior, indicating that our optimization routine naturally reproduces key features of the short-period Wiggle Well.

The results also differ from the Wiggle Well in interesting ways. 
First, we note that approximately half the atomic layers in Fig.~\ref{fig:optimizePureVS}(a) contain no added Ge.
Indeed, the secondary peak observed at $k_z\approx 8.3$~nm$^{-1}$ in Fig.~\ref{fig:optimizePureVS}(c) corresponds to the first harmonic of $2k_0$, shifted to lower $k_z$ values due to aliasing effects on a discrete lattice.
Such harmonics are a hallmark of \emph{truncated} sinusoids. 
For example, the following `truncated Wiggle Well' yields such harmonics:
\begin{equation}
Y_w(z)=(\pi A_w)\,\text{max}[\cos(2k_0z+\phi),0] .
\label{eq:Ytrunc}
\end{equation}
As in Eq.~(\ref{eq:YWW}), $A_w$ is defined here as the average Ge concentration.
However, the Fourier component of $U_\text{qw}(2k_0)$ for this concentration profile is $\pi/2$ times larger than for the conventional Wiggle Well, for the same value of $A_w$.
Therefore, the truncated Wiggle Well found by our optimization procedure should improve the valley splitting of the Wiggle Well by a factor of $\pi/2$, for the same total Ge content.
We confirm this prediction through simulations in Appendix~\ref{app:truncWiggleWell}.

A second difference between Fig.~\ref{fig:optimizePureVS}(a) and a Wiggle Well is in the nonuniform envelope of the concentration profile, which mimics the density profile of the wavefunction $|\psi_\text{env}(z)|^2$.
This behavior enhances the valley splitting by increasing the wavefunction overlap with Ge.
(We note that the precise shape of the Ge concentration envelope depends on the quantum well profile [gray region in Fig.~\ref{fig:optimizePureVS}(a)] and the electric field.)
Based on these observations, we hypothesize that the optimal Ge distribution observed in Fig.~\ref{fig:optimizePureVS}(a) should be well approximated as a truncated Wiggle Well weighted by the envelope probability, $|\psi_\text{env}(z)|^2$. 
To test this hypothesis, we fit the added Ge profile in Fig.~\ref{fig:optimizePureVS}(a), for the case of $G_\text{max}=5$~atoms/nm$^2$, to the form
\begin{equation}
    Y_\text{fit}(z) = a_\text{fit} |\psi_\text{env}(z)|^2 \text{max}[0,\cos(2k_0 z+\phi_\text{fit})] ,
    \label{eq:Yfit}
\end{equation}
where $a_\text{fit}$ and $\phi_\text{fit}$ are fitting parameters, and $\psi_\text{env}(z)$ is obtained from effective-mass theory by solving a Schr\"odinger equation.
The resulting fits are excellent, as shown in Fig.~\ref{fig:optimizePureVS}(d), where the solid line is $Y_\text{fit}(z)$ and the red dots are the optimized simulation results.

Finally in Fig.~\ref{fig:optimizePureVS}(b), we study the statistical effects of alloy disorder by performing 1,000 valley splitting simulations for the optimized Ge profiles obtained at each $G_\text{max}$ value considered in Fig.~\ref{fig:optimizePureVS}(a).
Here, the markers indicate mean values and the error bars show the corresponding 25-75 percentile range.
The very small standard deviations are indicative of very strong deterministic enhancements, which are robust in the presence of alloy disorder.

\subsubsection{Randomly dominated regime}

Concentration profiles like Fig.~\ref{fig:optimizePureVS}(a) are challenging to grow in the laboratory (just like the Wiggle Well), due to their short-period features.
We therefore also apply a concentration-optimizing procedure in the randomly dominated valley splitting regime, where the Ge profiles are more slowly varying (analogous to quantum wells with uniform Ge).
Here, to avoid obtaining a deterministically enhanced profile, we choose $\sigma_\Delta^2$ as the reward function, as defined in Eq.~(\ref{varDelta}).
This has the effect of maximizing the mean valley splitting as well as the variance, since $\sigma_\Delta^2\approx\bar E_v^2/\pi$ in the randomly dominated regime. 

Figure~\ref{fig:optimize}(a) shows concentration profiles obtained from this procedure, where the color scale indicates the added Ge content in the range of $G_\text{max}\in (3,30)$~atoms/$\text{nm}^2$.
Here we adopt the same initial quantum well and electric field as Fig.~\ref{fig:optimizePureVS}(a). 
The corresponding envelope functions $\psi_\text{env}(z)$ are shown for the cases $G_\text{max}=3$, 15, and 30~atoms/$\text{nm}^2$. 
For low $G_\text{max}$ values, the resulting Ge profiles are roughly uniform, with the Ge shifted slightly towards the top of the quantum well where $|\psi_\text{env}(z)|^2$ is large. 
For high $G_\text{max}$ values, the peak of added Ge shifts towards the center of the quantum well, squeezing the electron more tightly against the top interface. 
(Note again that the final Ge profile depends on the precise shape of the quantum well and the electric field.)
This has a two-fold effect of (i) exposing the wavefunction to more Ge in the barrier region, and (ii) causing the electron to overlap with fewer atomic layers, which also enhances the concentration fluctuations.
We interpret these results as a preference for narrower quantum wells.
 Figure~\ref{fig:optimize}(b) shows the resulting mean values and 25-75 percentile distributions of the valley splitting, as a function of $G_\text{max}$.
 The valley splitting enhancements here are slightly larger than for the case of uniform Ge, if we compare the same total Ge content.
Overall, optimized valley splittings in the disorder-dominated regime (Fig.~\ref{fig:optimize}) are found to be much smaller than in the deterministically enhanced regime (Fig.~\ref{fig:optimizePureVS}), although the devices are much easier to grow.

\section{Summary} 
\label{sec:summary}

In this paper, we derived a universal effective-mass theory of valley splitting in Si/SiGe heterostructures, based on the $2k_0$ reciprocal wavevector of the Fourier transform of the weighted confinement potential, $\tilde U_\text{qw}(z)=U_\text{qw}(z)|\psi_\text{env}(z)|^2$ \red{(Sec.~III~D)}.
By comparing our results to those of tight-binding simulations, we showed that this theory accurately predicts the valley splitting across a diverse set of heterostructures and disorder models. 
We then used the $2k_0$ theory to identify two valley splitting regimes \red{(Sec.~III~F)}: (i) the deterministic regime, in which the valley splitting is determined by atomistic details of the quantum well, such as the sharpness of the interface or the location of an atomic step at the interface, and (ii) the disorder-dominated regime, in which the valley splitting is fully determined by SiGe random alloy disorder. 
In the deterministic regime, the valley splitting is reliably large and independent of the alloy disorder, and the probability of finding a low valley splitting is exponentially suppressed \red{(Sec.~III~G)}. 
In the disordered regime, valley splittings can be large on average, but there is still a good chance of finding a small valley splitting.
\red{The crossover between these two regimes was shown to be sharp and universal (Fig.~4), since it depends only on the integrated overlap of the electron with Ge in the quantum well or the quantum well interface. 
However it was also shown that the crossover occurs in a regime where heterostructure features are very sharp (e.g., sharp interfaces or single-atom spikes).
Such sharp features are difficult to achieve in the laboratory; therefore, deterministically enhanced valley splittings are difficult to achieve in physical devices.}

\red{
Several conventional heterostructure geometries were investigated by means of simulations (Sec.~IV). 
Sharp interfaces were found to give a large deterministic enhancement of the valley splitting (Sec.~IV~A~1), but only in the ultra-sharp limit ($\lambda_\text{int}<$3~ML), which is difficult to achieve in the laboratory.
Single-atom steps at the interface were shown to suppress the valley splitting by up to 71\% at ultra-sharp interfaces, but were found to have almost no effect for interfaces with $\lambda_\text{int}>3$~ML (Sec.~IV~A~2).
Indeed, the precise shape of the interface was found to be important only for the case of ultra-sharp interfaces (Sec.~IV~C).
Wider interfaces were also shown to enhance the \emph{variability} of the valley splitting, due to greater exposure of the electron to the SiGe random alloy (Sec.~IV~B). 
This enhanced variability has the interesting side-effect of \emph{increasing} the average valley splitting as a function of $\lambda_\text{int}$ in the disorder-dominated regime (Figs.~6 and 7).}

\red{
Several unconventional heterostructures were also investigated (Sec.~V).
Uniform Ge in the quantum well was found to enhance both the average valley splitting and its standard deviation, due to significant overlap of the electron with Ge inside the quantum well, where the wavefunction is largest (Sec.~V~A).
Similar effects occur in other geometries, including the sharp Ge spike, where the Ge concentration is maximized where the wavefunction is largest (Sec.~V~C),
and narrow quantum wells, for which the wavefunction is squeezed into the quantum well barrier region where the Ge concentration is high (Sec.~V~B). 
The Wiggle Well geometry provides a very effective enhancement of the valley splitting by engineering the $2k_0$ wavevector directly into Ge concentration oscillations inside the quantum well.
This geometry also experiences enhanced valley splitting variability due to high Ge exposure (Sec.~V~D).}

\red{
Such unconventional geometries are found to be \emph{optimal}, in the following sense.
When the Ge concentration profile is optimized numerically, to obtain the maximum valley splitting while allowing for short-wavelength concentration oscillations, it naturally converges to a Wiggle Well-like geometry (Sec.~V~E~1).
This procedure can be understood as optimizing the valley splitting, \emph{ deterministically}.
Alternatively, in the disorder-dominated regime, the average valley splitting is proportional to its standard deviation.
Maximizing this quantity yields results similar to the geometry with uniform Ge in the quantum well (Sec.~V~E~2).}

\section{Conclusions: Best Strategies for Enhancing $E_v$}
\label{sec:conclusions}

We now conclude by describing the best strategies for enhancing the valley splitting in Si/SiGe heterostructures. 
Just as there are two types of valley splitting behavior, there are also two approaches for obtaining large valley splittings.
The first is to establish layer-by-layer growth control, which would allow for the implementation of structures like short-period Wiggle Wells, single-atom spikes, and super-sharp interfaces.
We emphasize however, that if 1-2 monolayer growth accuracy cannot be achieved, then deterministic attempts to enhance the valley splitting will be overwhelmed by random-alloy disorder, and will fail. 
In this case, there is no benefit to striving for deterministic enhancement.

An alternative strategy for enhancing the valley splitting is to intentionally add Ge to the quantum well, to increase the exposure to disorder.
This has the effect of increasing the mean value as well as the standard deviation of the valley splitting.
Additionally, and just as importantly, one should arrange for electrostatic control of the dot position. 
The most straightforward approach for adding Ge is to choose a simple, smooth Ge profile, such as a broadened interface or uniform Ge in the quantum well, because such structures are easy to grow. 
Very narrow quantum wells are also effective for increasing the exposure to Ge.
Finally, we note that even low Ge concentrations and modest tunability of the dot's position can increase the probability of achieving useful valley splittings by many orders of magnitude.

\section*{Acknowledgements}

We acknowledge helpful discussions with D.\ Savage, N.\ Hollman, C.\ Raach, E.\ Ercan, L.\ Tom, B.\ Woods, A.\ Vivrekar, P.\ Gagrani, and D.\ DiVincenzo. 
This research was sponsored in part by the Army Research Office (ARO) under Awards No.\ W911NF-17-1-0274 and No.\ W911NF-22-1-0090. 
The work was performed using the computing resources and assistance of the UW-Madison Center For High Throughput Computing (CHTC) in the Department of Computer Sciences. The CHTC is supported by UW-Madison, the Advanced Computing Initiative, the Wisconsin Alumni Research Foundation, the Wisconsin Institutes for Discovery, and the National Science Foundation, and is an active member of the OSG Consortium, which is supported by the National Science Foundation and the U.S. Department of Energy's Office of Science.
The views, conclusions, and recommendations contained in this document are those of the authors and are not necessarily endorsed nor should they be interpreted as representing the official policies, either expressed or implied, of the Army Research Office (ARO) or the U.S. Government. The U.S. Government is authorized to reproduce and distribute reprints for Government purposes notwithstanding any copyright notation herein.


\appendix

\section{Theoretical treatment of alloy disorder in 1D and 2D models} \label{appendix:2Dalloy}

In this Appendix, we describe how to account for alloy disorder in the coarse-grained 1D and 2D models we use in both tight-binding and effective-mass simulations. 
First, we consider how to compute the weighted average Si concentration in each cell, starting from a fully atomistic, 3D model of the heterostructure. 
Then, we derive effective probability distributions for the Si concentrations in each cell. 
These effective distributions allow us to sample many realizations of alloy disorder without generating fully atomistic, 3D models, thus greatly improving our computational efficiency.

\subsection{Averaging method for obtaining Si and Ge concentrations}

As described in Section~\ref{sec:qwPotential}, for 1D models, the Si concentration in each cell $X_\text{cell}= X_l$, where $X_l$ is the average Si concentration at layer $l$, weighted by the dot probability density. 
We can define this quantity as follows:
\begin{equation} \label{1dSiconc}
    X_l = \sum_{a \in A_l} \mathds{1}\left[ a = \text{Si} \right] w(a)
\end{equation}
where $A_l$ is the set of atoms in layer $l$, $a = \{ \text{Si}, \text{Ge} \}$ is a Si or Ge atom, and $\mathds{1}\left[ a = \text{Si} \right]$ is the indicator function which returns 1 if $a = \text{Si}$ and $0$ otherwise. 
The weight function $w(a)$ is proportional to the dot probability density, which we assume to be the ground state of a 2D isotropic harmonic oscillator. 
Following Ref.~\cite{Wuetz:2022p7730}, $w(a)$ should be normalized such that $\sum_{a \in A_l} w(a) = 1$. 
Because there are two atoms in each layer of the Si cubic unit cell, spread out over an area of size $a_0^2$, we can approximate the sum over atoms as an integral using the transformation
\begin{equation} \label{eq:sum2int1D}
    \sum_{a \in A_l} \rightarrow {2 \over a_0^2} \int_{-\infty}^{+\infty} dx \; dy.
\end{equation}
Hence, the correctly normalized weight function is
\begin{equation} \label{weight1d}
    w(a) = {a_0^2 \over 2 \pi a_\text{dot}^2} e^{-r_a^2 / a_\text{dot}^2}
\end{equation}
where $r_a$ is the distance of atom $a$ from the center of a dot of radius $a_\text{dot} = \sqrt{\hbar / m_t \omega_\text{orb}}$, $m_t = 0.19 m_e$ is the transverse effective mass in Si, $m_e$ is the bare electron mass, and $\hbar \omega_\text{orb}$ is the characteristic energy level spacing of the confinement potential. 
So, starting from a fully atomistic model of a heterostructure, we can incorporate concentration fluctuations into a 1D cell model by taking the weighted average Si concentration in each cell, where the weight function is given by Eq.~(\ref{weight1d}).
As noted in the main text, we assume a cell size of $\Delta z=a_0/4$ here.

For 2D models, we divide the atoms on layer $l$ into cells of width $\Delta x$. 
We assume the dot envelope function is separable, $\Psi(x,y,z) = \psi_{xz}(x,z)\psi_y(y)$, and we take $\psi_y(y)$ to be the ground state of a parabolic confinement potential with characteristic orbital splitting $\hbar \omega_y$. 
The Si concentration in each cell, weighted by the dot wavefunction, is given by
\begin{equation} \label{2Dconcdist}
X_{j,l} = \sum_{a \in A_{j,l}} \mathds{1} \left[ a = \mathrm{Si} \right] w_\text{2D}(a)
\end{equation}
where $A_{j,l}$ is the set of atoms in the $j$th cell along $\hat x$, in the $l$th layer, $\mathds{1}$ is the indicator function, and $w_\text{2D}(a)$ is proportional to the dot orbital wavefunction in the $y$ direction. 
Proper normalization should ensure that
\begin{equation}
\sum_{a \in A_{j,l}} w_\text{2D}(a) = 1.
\end{equation}
Because there are 2 atoms per unit cell per layer, we can approximate the sum as an integral using
\begin{equation}
\sum_{a \in A_{j,l}} \rightarrow {2 \Delta x \over a_0^2} \int_{-\infty}^\infty  dy .
\end{equation}
Thus, we find 
\begin{equation} \label{weight2d}
w_\text{2D}(a) = {a_0^2 \over 2 \Delta x a_y \sqrt{\pi} } e^{-y_a^2 / a_y^2},
\end{equation}
where $y_a$ is measured from the center of the dot, and $a_y = \sqrt{\hbar / m_t \omega_y}$. 
In this case, we adopt the same vertical cell dimension $\Delta z$ as the 1D model, and a lateral cell width of $\Delta x = a_0/2$ for the 2D cell model. 
Thus, we account for concentration fluctuations in each cell of a 2D model by taking the weighted average Si concentration in each cell of size $\Delta x\times \Delta z$, where the weight function is given by Eq.~(\ref{weight2d}).

\subsection{Generating probability distributions for Si and Ge concentrations}

In the previous section, we took alloy disorder into account by generating fully 3D heterostructures atomistically, then populating each cell of a coarse-grained model by computing the  weighted average Si concentration in each 1D or 2D cell. 
However, it is slow and computationally expensive to repeat this procedure for every simulation. 
Here, we show that we can generate valley splitting distributions by randomly sampling the Si concentration in each cell from known probability distributions based on the Si/Ge concentration profile.

\begin{figure}[t] 
	\includegraphics[width=3.1in]{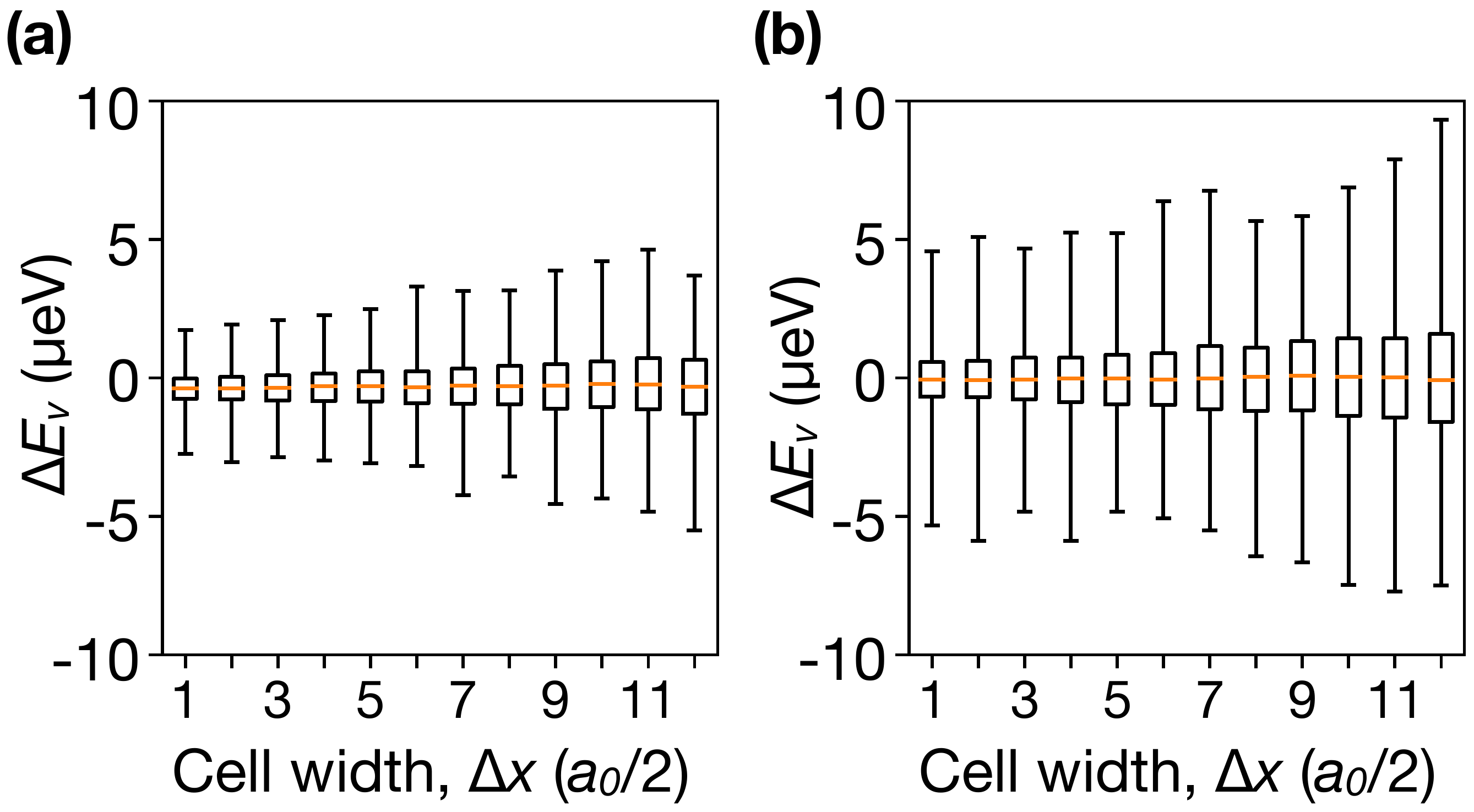}
	\centering
	\caption{
Discrepancy between 1D and 2D tight-binding models in systems without steps. 
 (a), (b) Plots show the median values (red lines), quartile ranges (boxes), and maximum ranges (whiskers) of $\Delta E_v = E_v^\text{2D} - E_v^\text{1D}$, for the same disorder realizations in 1D and 2D cell geometries (see main text and Appendix~\ref{appendix:2Dalloy}), as a function of the cell size $\Delta x$. 
 Here we consider quantum wells with smoothed linear interface profiles, as described in Section~\ref{sec:interfaces}, with interface widths of (a) $\lambda_\text{int} = 1$~ML, and (b) $\lambda_\text{int} = 10$~ML, and well widths of $W=80$~ML. 
For all simulations we assume a vertical electric field of $E_z = 5$~mV/nm and an isotropic orbital energy splitting of $\hbar \omega_\text{orb} = 2$~meV.}
	\label{fig:1d2dDifference}
\end{figure}

In Ref.~\cite{Wuetz:2022p7730}, by examining the statistical properties of Eq.~(\ref{1dSiconc}), 
we showed that the Si concentration in each cell of a 1D model can be described as a binomial random variable: 
\begin{equation} \label{1Dsampling}
    X_l \sim {1 \over N_\mathrm{eff}} \text{Binom}( N_\mathrm{eff}, \bar X_l) ,
\end{equation}
where $N_\mathrm{eff} = 4 \pi a_\mathrm{dot}^2 / a_0^2$ is the number of Si atoms per layer inside a 2D dot of radius $\sqrt{2}\, a_\text{dot}$~\cite{Wuetz:2022p7730} and $\bar X_l$ is the average Si concentration in layer $l$.
Here, we derive a similar sampling rule for cells in a 2D model. 
Taking the variance of Eq.~(\ref{2Dconcdist}), we find 
\begin{equation}
\begin{split}
\mathrm{Var} \left[X_{j,l}\right] &= \bar X_{j,l} (1 - \bar X_{j,l}) \sum_{a \in A_{j,l}} w_\text{2D}^2(a) \\
& =  \bar X_{j,l} (1 - \bar X_{j,l}) {1 \over 2 \sqrt{2 \pi} } {a_0^2 \over a_y \Delta x} \\
& = {\bar X_{j,l} (1 - \bar X_{j,l})  \over N_\mathrm{eff}^\mathrm{2D} },
\end{split}
\end{equation}
where $N_\mathrm{eff}^\mathrm{2D} = 2 \sqrt{2 \pi} a_y \Delta x / a_0^2$ and $\bar X_{j,l}$ is the expected Si concentration of the 2D cell with indices $(j,l)$, obtained by averaging uniformly over the entire layer ($\bar X_{j,l}=\bar X_l$, for the case of no step), or by using Eq.~(\ref{SiConcStep}) (for the case of a step).
Comparing these relations to the known properties of a binomial distribution, we can identify
\begin{equation} \label{2Dsampling}
X_{j,l} \sim {1 \over N_\mathrm{eff}^\mathrm{2D}} \mathrm{Binom}(N_\mathrm{eff}^\mathrm{2D}, \bar X_{j,l}) .
\end{equation}
In this way, we can account for alloy disorder in the 1D and 2D models by sampling each cell according to Eqs.~(\ref{1Dsampling}) and (\ref{2Dsampling}), rather than generating a full 3D lattice geometry and explicitly averaging the Ge concentration in every cell. 

\section{Characterizing tight-binding models} \label{appendix:tb}
\subsection{Comparing NEMO-3D, two-band tight-binding model, and effective-mass theory}

\begin{figure}[] 
	\includegraphics[width=3.1in]{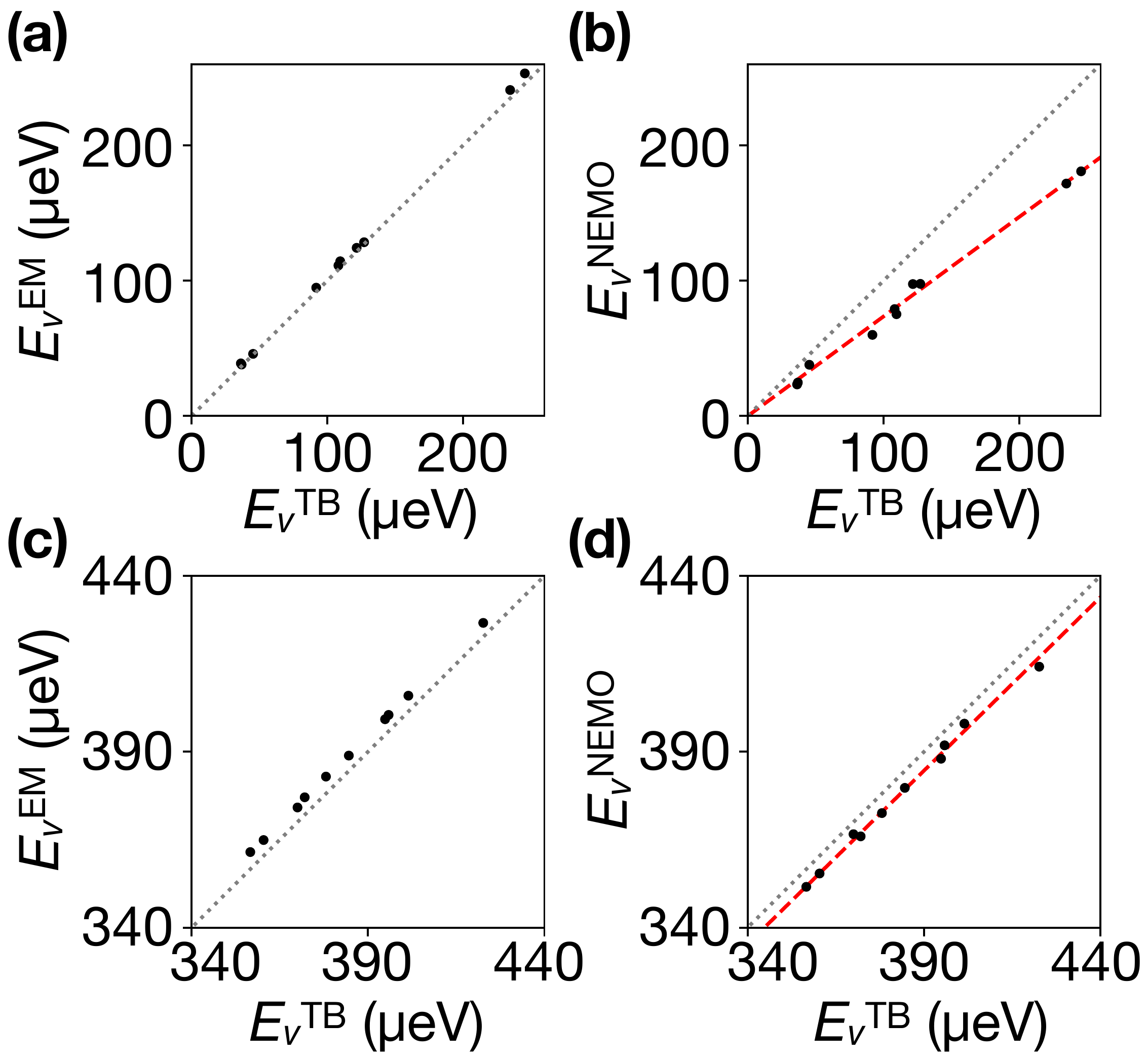}
	\centering
	\caption{
 Valley splitting comparisons between effective-mass theory, two-band tight-binding, and NEMO-3D models. 
 (a), (b) Using the same 10 alloy disorder realizations, we calculate valley splittings in three ways. 
 In (a), we compare valley splittings computed from the minimal tight-binding model ($E_v^\text{TB}$) and effective-mass theory [$E_v^\text{EM}$, from Eq.~(\ref{discreteDelta})], obtaining nearly perfect agreement. 
 In (b), we compare the minimal tight-binding model to NEMO-3D ($E_v^\text{NEMO}$).
 The red dashed line indicates the best fit to $E_v^\text{NEMO} = \alpha E_v^\text{TB}$ with $\alpha = 0.74$. 
 Gray dotted lines indicate $y = x$. 
 For both (a) and (b), we use quantum wells of width $W=80$~ML and sigmoid interfaces of width $\lambda_\text{int} = 20$~ML. 
 (c), (d) show the same data as (a), (b) for 10 heterostructures with $\lambda_\text{int} =1$~ML. 
 In (d), the red line indicates the best fit, with $\alpha = 0.99$. 
 In all simulations, we assume a vertical electric field of $E_z = 5$~mV/nm and an orbital energy splitting of $\hbar\omega_\text{orb}=2$~meV. 
 The NEMO-3D valley splitting results shown here are the same used in Fig.~\ref{fig:wideInt}(c).}
	\label{fig:NemoTbModel}
\end{figure}

\begin{figure}[] 
	\includegraphics[width=6.5cm]{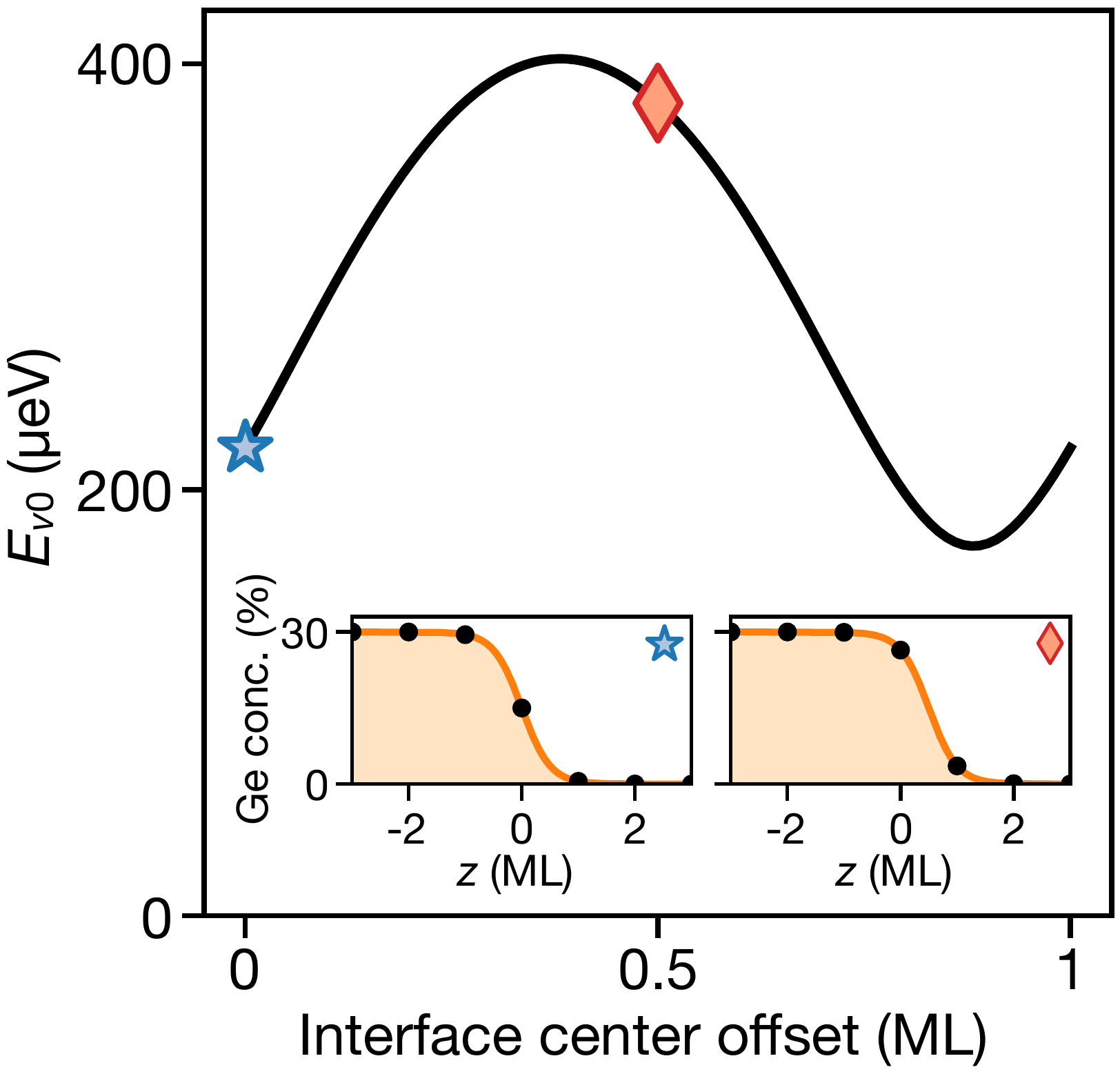}
	\centering
	\caption{
 For quantum wells with sharp interfaces, the deterministic valley splitting $E_{v0}$ depends sensitively on the interface position. 
 Here we compute $E_{v0}$ for quantum wells with sigmoid interfaces, and interface widths of $\lambda_\text{int} = 1$~ML, as the center of the interface is varied between two adjacent atomic layers.
 Insets: (blue star) an interface centered exactly on an atomic layer; 
 (red diamond) an interface centered halfway between atomic layers.}
	\label{fig:interfacePosition}
\end{figure}

In this Appendix, we compare results obtained from NEMO-3D, the minimal two-band tight-binding model, and effective-mass theory in heterostructures without steps.
We first construct a 3D crystal lattice atom by atom, including alloy disorder.
Taking the dot to be in the orbital ground state of a harmonic confinement potential with a characteristic strength of $\hbar \omega_\text{orb} = 2$~meV, we then reduce the 3D model to an effective 1D cell geometry as described in Appendix~\ref{appendix:2Dalloy}.
This geometry can be solved immediately, using the 1D minimal tight-binding model to obtain the valley splitting.
For the effective-mass model, we use the same 1D concentration profile to numerically solve for $\psi_\text{env}(z)$, and then compute $E_v^\text{EM}$ from Eq.~(\ref{discreteDelta}).
Figures~\ref{fig:NemoTbModel}(a) and \ref{fig:NemoTbModel}(c) show correlation plots for valley splittings computed this way for 10 disorder realizations, obtaining nearly perfect agreement between the two methods, for both (a) wide and (c) sharp interfaces. 

The full 3D crystal lattice geometries used in these simulations are then used to obtain the valley splitting in NEMO-3D.
Figures~\ref{fig:NemoTbModel}(b) and \ref{fig:NemoTbModel}(d) show correlation plots for the same 10 heterostructures, now comparing the minimal tight-binding model to the NEMO-3D model, for (b) wide and (d) sharp interfaces. 
The NEMO-3D valley splitting results consistently fall slightly below the minimal tight-binding values; however the results are very strongly correlated, and well-approximated by a linear scaling relation, $E_v^\text{NEMO} \approx \alpha E_v^\text{(TB)}$, where $\alpha = 0.74$ for wide interfaces and $\alpha=0.99$ for sharp interfaces. 
Thus, the effect of the higher bands ignored in the two-band model are effectively captured by a modest linear scaling of the valley splitting, which depends on interface width.

\subsection{Comparing 1D and 2D tight-binding models}

In this paper, we employ the 1D tight-binding model in systems without steps, and the 2D tight-binding model in systems with steps, assuming a cell width of $\Delta x = a_0/2$. 
In this section, we show that these choices yield consistent results. 

We first choose a value for the cell width. 
We generate a full 3D crystal lattice atom-by-atom, including alloy disorder. 
We then create a coarse-grained 2D cell geometry using the methods described in Appendix~\ref{appendix:2Dalloy}, and solve this using the 2D minimal tight-binding model.
Next, we coarse grain the model a second time, as described in Appendix~\ref{appendix:2Dalloy}, to obtain a 1D cell geometry, and solve this using the 1D minimal tight-binding model.
We then compute the corresponding differences $\Delta E_v = E_v^\text{2D} - E_v^\text{1D}$. 
This procedure is repeated for 1,000 realizations of random alloy disorder. 
$\Delta x$ is then modified, choosing values that are integer multiples of $a_0/2$.
The results are plotted in Fig.~\ref{fig:1d2dDifference} for a range of $\Delta x$ values, showing
the resulting median values (red lines), 25-75\% quartiles (boxes), and maximum ranges (whiskers).
The very small values obtained for $\Delta E_v$ indicate excellent consistency between the 1D and 2D models for systems without steps.
Since the smallest cell width, $\Delta x=a_0/2$, is found to provide the best agreement, we adopt this as the cell width for our 2D model.

\section{Choosing the center of the quantum well interface} \label{appendix:interface}

In quantum wells with very sharp interfaces, the exact location of the interface [e.g., $z_b$ or $z_t$ in Eq.~(\ref{eq:sigmoid})] strongly affects the deterministic valley splitting $E_{v0}$. 
Figure~\ref{fig:interfacePosition} shows this variation in $E_{v0}$ for a $\lambda_\text{int} = 1$~ML sigmoidal interface, as the center of the sigmoid is moved between two adjacent atomic monolayers. 
While all interfaces in Fig.~\ref{fig:interfacePosition} are drawn from the same sigmoid profile, $E_{v0}$ nonetheless varies by a factor of 2. 
In this work, whenever narrow interfaces are considered, we choose the center of the interface to be exactly halfway between two adjacent monolayers, as shown in the inset labeled with a red diamond.

\begin{figure}[] 
	\includegraphics[width=8cm]{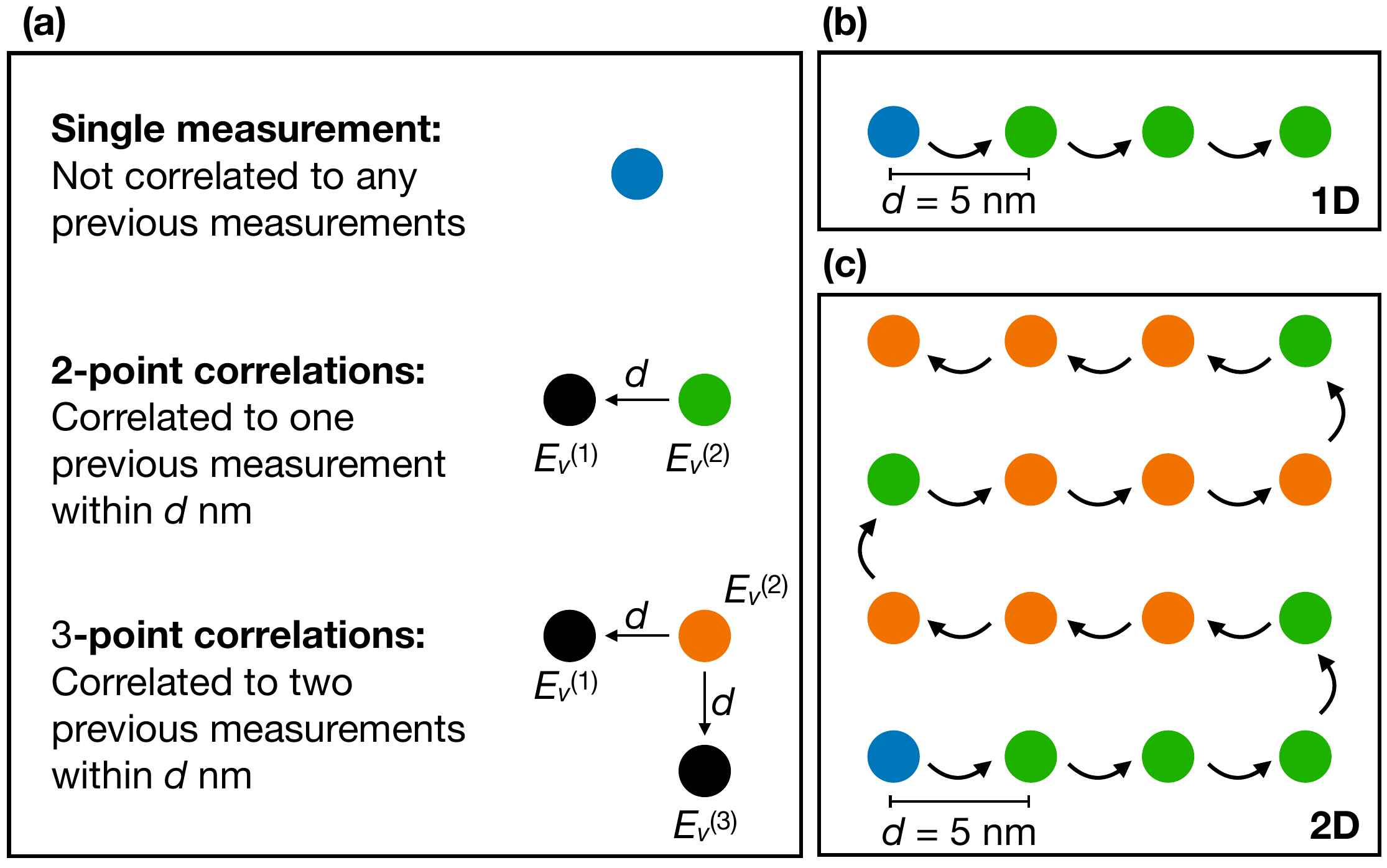}
	\centering
	\caption{ 
 An explanation of the types of correlations taken into account when computing $P_\text{fail}$. 
 (a) The difference between a single $E_v$ measurement, two-point correlations between neighboring measurements, and three-point correlations between neighboring measurements. 
 We only consider these three types of correlations in our model. 
 (b) Illustration of a series of $E_v$ measurements, in increments of $d$, along a 1D line. 
 The first measurement is uncorrelated to anything prior, and the next 3 are correlated to one prior measurement. 
 (c) Illustration of a series of $E_v$ measurements along a square grid with spacing $d$. 
 We see one initial measurement, six measurements correlated to one prior measurement within distance $d$, and 9 measurements correlated to two prior measurements within distance $d$. }
	\label{fig:movingDotCorr}
\end{figure}

\section{Valley splitting spatial correlations} \label{appendix:movingDot}

In this Appendix, we theoretically compute the probability $P_\text{fail}$ of measuring a valley splitting lower than a nominal threshold value $E_v^\text{min}$, as defined in Sec.~\ref{sec:highGe} of the main text.
In that Section, we considered geometries with uniform Ge in the quantum well, which are always disorder-dominated, with $|\Delta_{0}|\ll \sigma_\Delta$.
The probability of failure for a single valley splitting measurement is therefore given by Eq.~(\ref{eq:Pfaildisordered}):
\begin{equation} \label{initProb}
	p_1 := P(E_v < E_v^\text{min}) \approx 1 - \exp \left[-{(E_v^\text{min})^2 / 4 \sigma_{\Delta}^2} \right] .
\end{equation}

In the protocol described in Sec.~\ref{sec:highGe}, a dot is allowed to sample either a 1D or 2D region of a device, along a set of grid points with spacing 5~nm. 
If $n$ sites are sampled, a naive estimate for $P_\text{fail}$ would be $p_1^n$.
However, this estimate is inaccurate because the quantum dot is larger than the grid spacing, so the valley splittings measured when the dot is centered on nearby grid points are not independent. 
A more accurate estimate of $P_\text{fail}$ should therefore account for correlations between nearby grid points. 
Empirically, for the setup described in Sec.~\ref{sec:highGe} and Fig.~\ref{fig:movingDot}(e), we find that it is sufficient to account for nearest-neighbor correlations between sites separated by $\leq 5$~nm. 

Figure~\ref{fig:movingDotCorr} illustrates the different types of correlations relevant to the 1D and 2D simulation schemes shown in Fig.~\ref{fig:movingDot}(e). 
In Fig.~\ref{fig:movingDotCorr}(b), we consider four neighboring sites in a 1D geometry.
To explain the correlation analysis, we can think of the valley splitting simulations at each of these sites as `measurements' performed in a particular order, as indicated by the arrows.
The first measurement (blue dot) clearly has no prior measurements to correlate with, so its probability of failure is simply given by $p_1$ in Eq.~(\ref{initProb}).
However, the next three measurements (green dots) are all correlated with the prior measurement. 
If we define the conditional probability of failure at site (2), given failure at site (1), as
\begin{equation}
    p_2 := P(E_v^{(2)} < E_v^\text{min} | E_v^{(1)} < E_v^\text{min})  .
\end{equation}
then the total estimated probability of failure for the linear geometry shown in Fig.~\ref{fig:movingDotCorr}(b) is given by $P_\text{fail} \approx p_1 p_2^3$. 
In general, for a 1D chain of $n$ sites, we have
\begin{equation}
    P_\text{fail} \approx p_1 \; p_2^{n-1}.
\end{equation}

Figure~\ref{fig:movingDotCorr}(c) illustrates a series of valley splitting measurements exploring a 2D region, on a 4$\times$4 grid. 
If we imagine performing these measurements by snaking across the lattice as indicated by the arrows, we see there is only one measurement (blue dot) having no prior correlations. 
There are six measurements (green dots) correlated with one prior measurement, and nine measurements (orange dots) correlated with two prior measurements. 
If we define the conditional probability function
\begin{equation} \label{p3Intro}
    p_3 := P(E_v^{(2)} < E_v^\text{min} | E_v^{(1)} < E_v^\text{min}, E_v^{(3)} < E_v^\text{min})
\end{equation}
for measurements at sites (1), (2), and (3) in Fig.~\ref{fig:movingDotCorr}(a), then we can estimate $P_\text{fail} \approx p_1 p_2^6 p_3^9$. 
In general, for an $n \times n$ lattice, we have
\begin{equation} 
    P_\text{fail} \approx p_1 \; p_2^{2n-2} \; p_3^{(n-1)^2}.
\end{equation}

In Appendix~\ref{sec:2pointCorr} (below) we compute the probability $p_2$, and in Appendix~\ref{sec:3pointCorr} we compute $p_3$. 
Numerical values for $p_1$, $p_2$, and $p_3$ are given in Table~\ref{table:1}, and the resulting $P_\text{fail}$ values are given in Tables~\ref{table:2} and~\ref{table:3}.

\subsection{Two-point correlations}  \label{sec:2pointCorr}

We consider quantum dots centered at two neighboring grid sites, (1) and (2).
The corresponding dot positions in the $x$-$y$ plane are $\mathbf{r}_1$ and $\mathbf{r}_2$, and we assume the dots are separated by the grid spacing $d=|\mathbf{r}_2-\mathbf{r}_1|$. 
We want to compute $p_2$, the conditional probability that $E_v^{(2)} < E_v^\text{min}$, given that $E_v^{(1)} < E_v^\text{min}$, where as usual, the valley splittings are related to the intervalley couplings through $E_v^{(1)}=2|\Delta_1|$ and $E_v^{(2)}=2|\Delta_2|$.
In the disorder-dominated regime, we simply assume the deterministic intervalley couplings are zero, such that $\Delta_1$ and $\Delta_2$ become complex random variables centered at zero.
The probability distributions of $\Delta_1$ and $\Delta_2$ are assumed to be circular gaussian distributions in the complex plane.
To account for the correlations between $\Delta_1$ and $\Delta_2$, we need to compute the covariances between real and imaginary components of $\Delta_1 $ and $\Delta_2$, defined as $\Delta_1^R$, $\Delta_1^I$, $\Delta_2^R$, and $\Delta_2^I$. 

We begin with the following identity for covariances:
\begin{equation} \label{covEq}
\mathrm{Cov}\left[X,Y\right] = {1\over2} \left( \mathrm{Var}\left[X\right] + \mathrm{Var}\left[Y\right] - \mathrm{Var}\left[X-Y\right] \right) .
\end{equation}
Previously, we have found that $\text{Var}[\Delta^R]=\text{Var}[\Delta^I]=\sigma_\Delta^2/2$, or $\text{Var}[\Delta]=\sigma_\Delta^2$.
Following Ref.~\cite{Wuetz:2022p7730}, we also find that 
\begin{multline} \label{eq:varD1D2}
    \mathrm{Var} \left[ \Delta_2 -  \Delta_1 \right] = \left( {a_0 \over 4} {\Delta E_c \over X_w - x_s} \right)^2 \\ \sum_l |\psi_\text{env}(z_l)|^4  \mathrm{Var} \left[ \delta_l^{(2)} - \delta_l^{(1)} \right],
\end{multline}
where from Eq.~(\ref{1dSiconc}) we have
\begin{equation} \label{eq:deltalSupp}
\delta_l^{(j)} =  {X_l^{(j)}} -\bar X_l^{(j)} = \sum_{a \in A_l} \mathds{1} \left[ a = \text{Si} \right] w_j(a) - \bar X_l^{(j)} .
\end{equation} 
Here, $X_l^{(j)}$ are the weighted Si concentrations at sites $j=1,2$, and the properly normalized probability density for site $j$ is given by
\begin{equation} \label{eq:weightFunc12}
w_{j}(a) = {a_0^2 \over 2 \pi a_\mathrm{dot}^2} e^{-|\mathbf{r}_a-\mathbf{r}_{j}|^2 /a_\mathrm{dot}^2} ,
\end{equation}
where $\mathbf{r}_a$ is the position of atom $a$ in layer $l$. 
Using Eqs.~(\ref{eq:varD1D2}), (\ref{eq:deltalSupp}), (\ref{eq:weightFunc12}), and the sum-to-integral transformation Eq.~(\ref{eq:sum2int1D}), we can evaluate
\begin{equation}
    \mathrm{Var} \left[ \Delta_2 - \Delta_{1} \right] = 2 \left( 1 - e^{-d^2 / 2 a_\mathrm{dot}^2} \right) \sigma_\Delta^2 .
\end{equation}
Using Eq.~(\ref{covEq}), we then have
\begin{equation}
    \mathrm{Cov}\left[\Delta_2^R,\Delta_{1}^R\right] = \mathrm{Cov}\left[\Delta_2^I,\Delta_{1}^I\right] = {1 \over 2} e^{-d^2 / 2 a_\mathrm{dot}^2} \sigma_\Delta^2 .
\end{equation}

We can now construct the joint probability density function for $\Delta_1$ and $\Delta_2$. 
For the basis ordering $\left\{ \Delta_1^R, \Delta_1^I, \Delta_2^R, \Delta_2^I \right\}$ the covariance matrix is given by
\begin{equation} \label{eq:2ptCov}
    \mathbf{\Sigma} = 
    {\sigma_\Delta^2 \over 2}
    \begin{pmatrix}
        1 & 0 & A & 0 \\
        0 & 1 & 0 & A \\
        A & 0 & 1 & 0 \\
        0 & A & 0 & 1
    \end{pmatrix},
    \end{equation}
    where
    \begin{equation}
    A = \exp ( -d^2 / 2 a_\mathrm{dot}^2) . \label{eq:Adef}
\end{equation} 
Using the standard definition of conditional probability, we now have
\begin{equation} \label{p2}
    p_2 = {P(E_v^\mathrm{(2)} < E_v^\text{min} , E_v^\mathrm{(1)} < E_v^\text{min}) \over p_1}.
\end{equation}
We can evaluate the numerator using the joint probability density function, giving 
\begin{multline} \label{2pointFunc}
    P(E_v^\mathrm{(2)} < T, E_v^\mathrm{(1)} < T) = {1 \over \sqrt{(2 \pi)^4 |\mathbf{\Sigma}|}} \\ \int\limits_{\substack{|\Delta_1| < E_v^\text{min}/2 \\ |\Delta_2| < E_v^\text{min}/2}} 
    \hspace{-.3in} d \Delta_1 d \Delta_2 \exp \left( -{1\over 2}\mathbf{v}^T \mathbf{\Sigma}^{-1} \mathbf{v} \right)
\end{multline}
where we define $\mathbf{v} = \left(\Delta_1^R, \Delta_1^I, \Delta_2^R, \Delta_2^I\right)^T $, and $d\Delta_j$ is shorthand for $d\Delta_j^R d\Delta_j^I$. 
This integral can be evaluated numerically for a given set of parameters $\sigma_\Delta$, $d$, and $a_\text{dot}$.
Numerical results for $p_2$, for a typical set of parameters, are presented in Table~\ref{table:1}, and the corresponding results for $P_\text{fail}$ are reported in Table~\ref{table:2} for a 1D grid geometry.

\subsection{Three-point correlations}  \label{sec:3pointCorr}
The conditional probability $p_3$ is computed similarly to $p_2$. 
In this case, there are three intervalley couplings, so the covariance matrix becomes
\begin{equation}
    \mathbf{\Sigma} = 
   {\sigma_\Delta^2 \over 2} \begin{pmatrix}
        1 & 0 & A & 0 & B & 0\\
        0 & 1 & 0 & A & 0 & B \\
        A & 0 & 1 & 0 & A & 0 \\
        0 & A & 0 & 1 & 0 & A \\
        B & 0 & A & 0 & 1 & 0 \\
        0 & B & 0 & A & 0 & 1
    \end{pmatrix} ,
\end{equation}
in the basis $\left\{\Delta_{1}^R, \Delta_{1}^I, \Delta_{2}^R,\Delta_{2}^I, \Delta_{3}^R, \Delta_{3}^I   \right\}$, where $A$ is given in Eq.~(\ref{eq:Adef}) and $B = \exp(-d^2 / a_\mathrm{dot}^2)$, which differs from $A$ because the distance between sites (1) and (3) is given by $\sqrt{2}d$. 
Similar to Eq.~(\ref{p2}), we apply the standard definition of conditional probability,
\begin{equation} \label{condProb3point}
\begin{split}
    p_3  = {P(E_v^{(1)} < E_v^\text{min}, E_v^{(2)} < E_v^\text{min},  E_v^{(3)} < E_v^\text{min} ) \over P( E_v^{(1)} < E_v^\text{min}, E_v^{(3)} < E_v^\text{min} ) }
\end{split} .
\end{equation}
Here, the denominator can be evaluated using Eq.~(\ref{2pointFunc}), while the numerator is given by
\begin{equation}
\begin{split}
    &P(E_v^{(1)} < E_v^\text{min}, E_v^{(2)} < E_v^\text{min}, E_v^{(3)} < E_v^\text{min}) =\\
    &\int\limits_{\substack{|\Delta_1| < E_v^\text{min}/2 \\ |\Delta_2| < E_v^\text{min}/2 \\ |\Delta_3| < E_v^\text{min}/2}} 
    \hspace{-.3in} d\Delta_1 d\Delta_2 d\Delta_3 {1 \over \sqrt{(2\pi)^6 |\mathbf{\Sigma}|}} \exp \left( -{1\over 2} \mathbf{v}^T \mathbf{\Sigma}^{-1} \mathbf{v} \right)
\end{split} ,
\end{equation}
where now, $
\mathbf{v} = \begin{pmatrix}
    \Delta_1^R, \Delta_1^I, \Delta_2^R, \Delta_2^I, \Delta_3^R, \Delta_3^I
\end{pmatrix}^T
$.
Numerical results for $p_3$ are presented in Table~\ref{table:1}, and the corresponding results for $P_\text{fail}$ are reported in Table~\ref{table:3} for a 2D grid geometry.

\begin{table}[]
\begin{center}
\begin{tabular}{c S[table-format=3.2] S[table-format=2.4] S[table-format=2.4] S[table-format=2.2]}
 \hline \hline
  & {$\sigma_\Delta$ ($\mu$eV)} & {$p_1$} & {$p_2$}  & {$p_3$}  \\ [0.5ex] 
 \hline
 0\% Ge & 36.21 & 0.8515 & 0.9531  &  0.98 \\ [1ex] 
 1\% Ge & 96.43 & 0.2357 & 0.6763  &  0.83 \\ [1ex] 
 5\% Ge & 203.82 & 0.0584 & 0.3397 &  0.53 \\
 \hline \hline
\end{tabular}
\caption{Numerical parameters used to calculate $P_\text{fail}$. 
To compute $\sigma_\Delta$, we use Eq.~(\ref{varDelta}), and we assume quantum wells with a sigmoidal profile defined by $\lambda_\text{int} = 10$~ML and $W = 80$~ML, an isotropic harmonic confinement potential of strength $\hbar \omega_\text{orb} = 2$~meV, corresponding to $a_\text{dot} \approx 14$~nm, and a vertical electric field of $E_z = 5$~mV/nm. 
To compute $p_2$ and $p_3$, we set $d = 5$~nm.}
\label{table:1}
\end{center}
\end{table}

\begin{table}[!htb]
\centering
\begin{tabular}{l S[table-format=2.5] S[table-format=2.5]} 
 \hline \hline
 &  {Calc. $P_\text{fail}$}  & {Sim. $P_\text{fail}$}   \\ [0.5ex] 
 \hline
 0\% Ge & & \\ [1ex] 
 { $n = 1$} & 0.85 & 0.8510 \\ [1ex] 
 { $n = 2$} & 0.81 & 0.8111 \\ [1ex] 
 { $n = 3$} & 0.77 & 0.7715 \\ [1ex] 
 { $n = 4$} & 0.74 & 0.7328 \\ [1ex]
 { $n = 5$} & 0.70 & 0.6936 \\ [1ex]
 \hline
 1\% Ge & & \\ [1ex] 
 { $n = 1$} & 0.24 & 0.2358 \\ [1ex] 
 { $n = 2$} & 0.16 & 0.1576 \\ [1ex] 
 { $n = 3$} & 0.11 & 0.1013 \\ [1ex] 
 { $n = 4$} & 0.073 & 0.0617 \\ [1ex]
 { $n = 5$} & 0.049 & 0.0411 \\ [1ex]
 \hline
 5\% Ge & & \\ [1ex] 
 { $n = 1$} & 0.058 & 0.0650 \\ [1ex] 
 { $n = 2$} & 0.020 & 0.0230 \\ [1ex] 
 { $n = 3$} & 0.0067 & 0.0081 \\ [1ex] 
 { $n = 4$} & 0.0023 & 0.0035 \\ [1ex]
 { $n = 5$} & 0.00078 & 0.0014 \\ [1ex]
 \hline \hline
\end{tabular}
\caption{Numerical and simulated values of $P_\text{fail}$ for a 1D grid geometry, using the same parameters as Table~\ref{table:1}.
These are the same values plotted in Fig.~\ref{fig:movingDot}(f) in the main text. 
Calculated values are computed using the methods described in Appendix~\ref{appendix:movingDot}. 
Simulated values are obtained by averaging 10,000 tight-binding simulations, as described in the main text.}
\label{table:2}
\end{table}

\begin{table}[!htb]
\centering
\begin{tabular}{l S[table-format = -1.5e1, table-align-exponent = false] S[table-format = -1.5e1, table-align-exponent = false]} 
 \hline \hline
  &  {Calc. $P_\text{fail}$}  & {Sim. $P_\text{fail}$}   \\ [0.5ex] 
 \hline
 0\% Ge & & \\ [1ex]
 { $n = 1$} & 0.85 & 0.8473 \\ [1ex] 
 { $n = 2$} & 0.76 & 0.7588 \\ [1ex] 
 { $n = 3$} & 0.65 & 0.6548 \\ [1ex] 
 { $n = 4$} & 0.53 & 0.5503 \\ [1ex]
 { $n = 5$} & 0.42 & 0.4441 \\ [1ex]
 \hline
 1\% Ge & & \\ [1ex]
 { $n = 1$} & 0.24 & 0.2359 \\ [1ex] 
 { $n = 2$} & 0.090 & 0.0925 \\ [1ex] 
 { $n = 3$} & 0.023 & 0.0237 \\ [1ex] 
 { $n = 4$} & 0.0042 & 0.0046 \\ [1ex]
 { $n = 5$} & 5.2e-4 & 0.0010 \\ [1ex]
 \hline
 5\% Ge & & \\ [1ex]
 { $n = 1$} & 0.058 & 0.0552 \\ [1ex] 
 { $n = 2$} & 0.0036 & 0.0042 \\ [1ex] 
 { $n = 3$} & 6.1e-5 & 2e-4 \\ [1ex] 
 { $n = 4$} & 3.0e-7 & 0 \\ [1ex]
 { $n = 5$} & 4.0e-10 & 0 \\ [1ex]
 \hline \hline
\end{tabular}
\caption{Numerical and simulated values of $P_\text{fail}$ for a 2D grid geometry, using the same parameters as Table~\ref{table:1}.
These are the same values plotted in Fig.~\ref{fig:movingDot}(f) in the main text. 
Calculated values are computed using the methods described in Appendix~\ref{appendix:movingDot}. 
Simulated values are obtained by averaging 10,000 tight-binding simulations, as described in the main text.}
\label{table:3} 
\end{table}

\begin{figure}[b] 
	\includegraphics[width=2.5in]{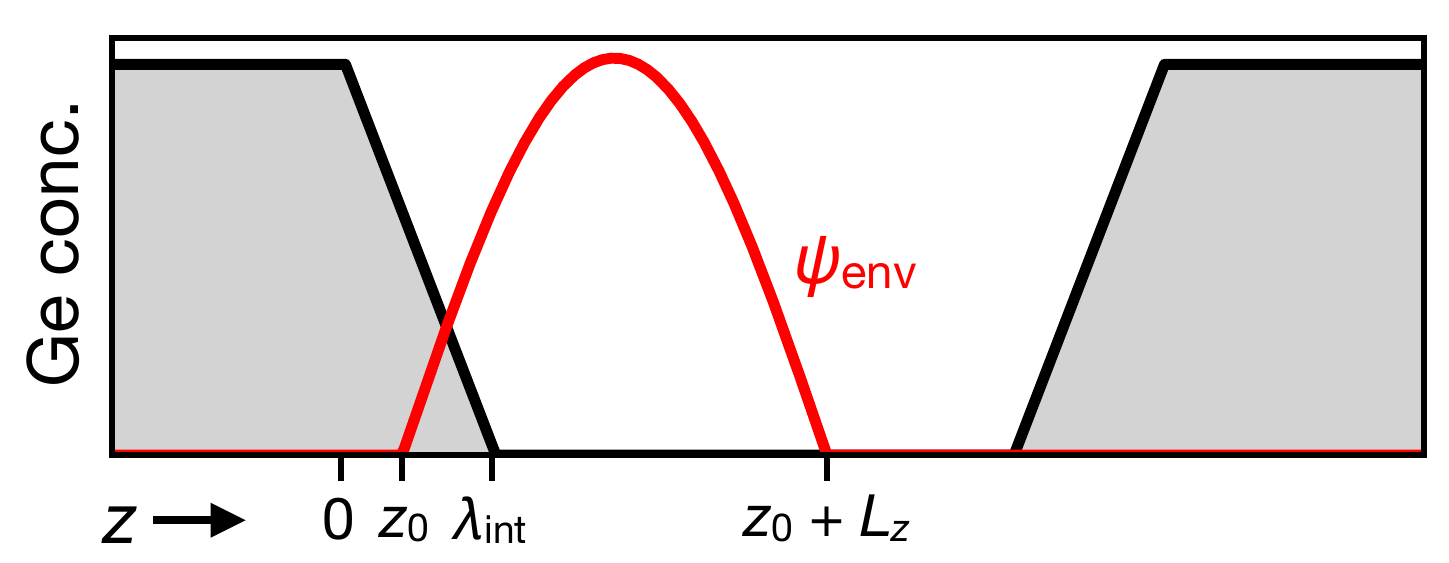}
	\centering
	\caption{Schematic illustration of the sinusoidal variational envelope function $\psi_\text{env}(z)$ and the variational parameters $z_0$, $\lambda_\text{int}$, and $L_z$ used to study linear interfaces.}
	\label{fig:variationalParams}
\end{figure}

\section{Variational approach for studying $E_v$ vs.\ interface width} \label{appendix:variational}

In this Appendix, we use a variational method to derive the valley splitting as a function of interface width, both with and without alloy disorder, as described in Sec.~\ref{sec:interfaces}. 
In both cases, we use the perfectly linear interface model described in Section~\ref{sec:interfaces} and Fig.~\ref{fig:narrowInt}(b)i. 

We consider the variational envelope function
\begin{equation} \label{varEnv}
\psi_\text{env} =
\begin{cases}
	\sqrt{2/L_z} \sin \left[ \pi (z-z_0) / L_z \right], &  z_0 \leq z \leq z_0 + L_z , \\
	0 &  \left( \mathrm{otherwise} \right) ,
\end{cases}
\end{equation}
with variational parameters $z_0$ and $L_z$, shown schematically in Fig.~\ref{fig:variationalParams}. 
Since the calculation only depends on the wavefunction near the top interface, a simple sinusoidal envelope suffices~\cite{Friesen:2007p115318}. 
The variational energy is given by $\langle H \rangle = \langle T \rangle + \langle \phi \rangle + \langle U_\text{qw} \rangle$, where the kinetic component $\langle T \rangle = \hbar^2 \pi^2 / 2 m_l L_z^2$ and the vertical field component $\langle \phi \rangle = (1/2) e E_z (L_z + 2 z_0 )$. 
The quantum well has barriers with Ge concentration $Y_s = 1-X_s$ and linear interfaces of width $\lambda_\text{int}$, as illustrated in Fig.~\ref{fig:variationalParams}. 
In the remainder of this section, we drop the subscript on $\lambda$ to avoid clutter.
In this section, to simplify the variational calculation, we set $z = 0$ at the top of the interface, as indicated in Fig.~\ref{fig:variationalParams}, and we use the quantum well potential 
\begin{equation}\label{eq:UqwVar}
    U_\text{qw}(z) = |\Delta E_c| \left( 1 - {X(z) - X_s \over 1 - X_s} \right). 
\end{equation}
Equation~(\ref{eq:UqwVar}) is equivalent to Eq.~(\ref{Uqw}), offset by a constant, so that $U_\text{qw} = 0$ in the middle of the quantum well.

\subsection{No alloy disorder}

\begin{figure*}[t] 
	\includegraphics[width=14cm]{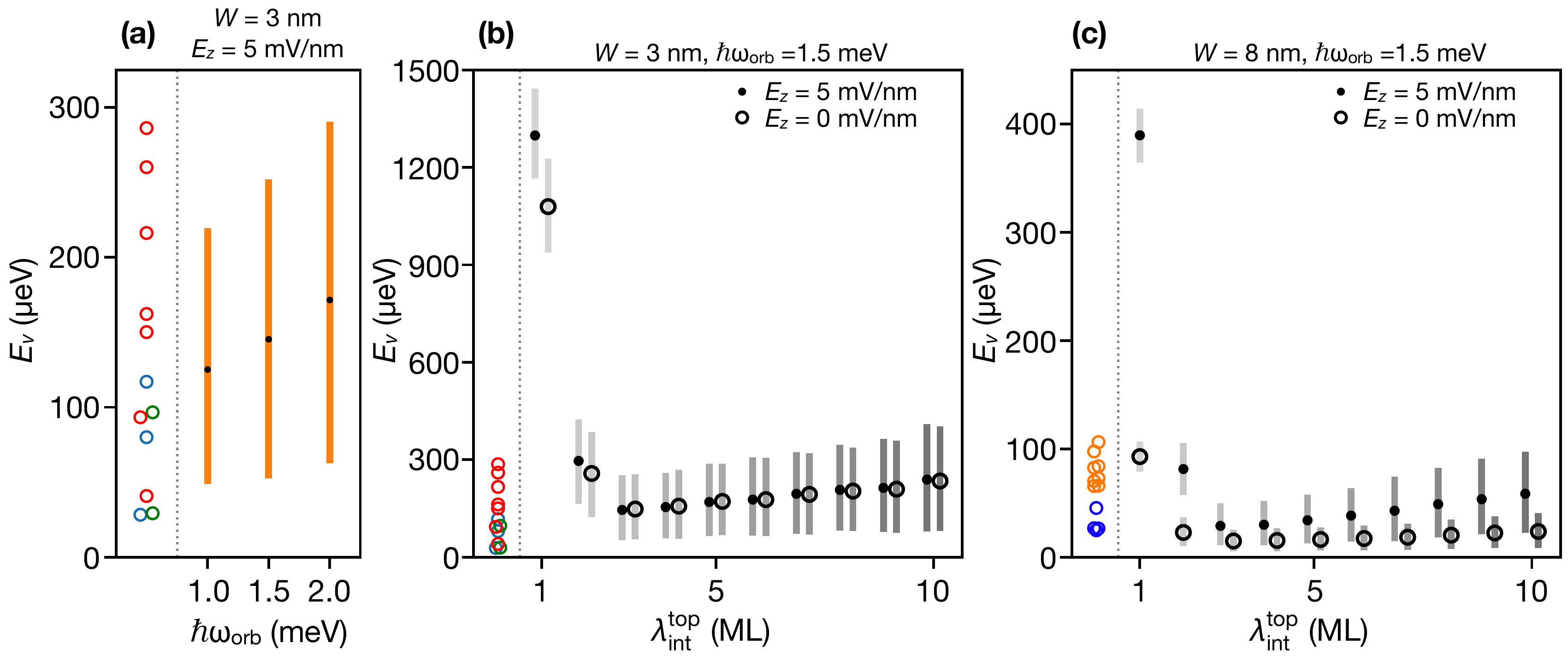}
	\centering
	\caption{
 Narrow quantum wells: our choice of vertical field $E_z$ and orbital splitting $\hbar \omega_\text{orb}$ yield simulation results consistent with experimental data considered in Sec.~\ref{sec:narrowWells}. 
 (a) The mean values (dots) and 10-90 percentile range (bars) obtained from 1,000 1D two-band tight-binding simulations of $E_v$, with orbital energies $\hbar \omega_\mathrm{orb}$ between 1 and 2~meV. 
 Here we use an electric field $E_z = 5$~mV/nm, a top-interface width of $\lambda_\text{int}^\text{top} = 3$~ML, and a well width of $W = 3$~nm. 
 $E_v$ data from Ref.`\cite{Chen:2021p044033}, for the 3~nm quantum wells, are included for comparison (open circles). 
 (b), (c) The 10-90 percentile ranges (bars) and mean values (dots and open circles) from 1,000 1D two-band tight-binding simulations of $E_v$, for different $\lambda_\text{int}^\text{top}$ and $E_z$. 
 Dots are used for $E_z = 5$~mV/nm, and open circles are used for $E_z = 0$~mV/nm. 
 (b) Results for a well width of $W=3$~nm, including $E_v$ data from Ref.~\cite{Chen:2021p044033}. 
 (c) Results for a well width of 8~nm, including experimental $E_v$ data from Ref.~\cite{Chen:2021p044033}. 
 Dots and circles have the same meaning as in (b).
 Both (b) and (c) assume an orbital splitting $\hbar \omega_\mathrm{orb} = 1.5$~meV.  
 }
	\label{fig:hrlSuppOther}
\end{figure*}

First, we examine the system without alloy disorder. 
This is accomplished by employing the virtual crystal approximation, $X(z)=\bar X_z$.
We separate the calculation into two cases: $z_0 < 0$ or $z_0 \geq 0$. 
The quantum well contribution to the variational energy is then given by
\begin{multline} \label{varUqw}
\langle U_\text{qw} \rangle = {|\Delta E_c | \over 4 \pi^2 L_z \lambda_\text{int} } \\
\times \begin{cases}
 2 \pi^2 \lambda_\text{int}^2  - 4 \pi^2 \lambda_\text{int} z_0  + L_z^2 \cos \left[ {2 \pi (\lambda_\text{int}-z_0) \over L_z }\right] \\
 \quad - L_z^2 \cos \left( {2 \pi z_0 \over L_z }\right), 
& z_0 < 0 , \\
 -L_z^2 + 2 \pi^2 (\lambda_\text{int} - z_0)^2 + L_z^2 \cos \left({2 \pi (\lambda_\text{int} - z_0) \over L_z } \right),
& z_0 \geq 0.
\end{cases}
\end{multline}
We then expand the cosine functions to fourth order and 
solve for the variational parameters by minimizing the variational energy, yielding
\begin{equation} \label{varParams}
\begin{split}
z_0 & \approx
\begin{cases}
{\lambda_\text{int} \over 2} - {1 \over 2 \pi} \left( {2 e E_z L_z^3 \over |\Delta E_c|} - {\pi^2 \lambda_\text{int}^2 \over 3} \right)^{1/2}, & z_0 < 0 ,  \\
\lambda_\text{int} - \left({3 e E_z L_z^3 \over  2 |\Delta E_c | \pi^2 } \right)^{1/3} \lambda_\text{int}^{1/3}, & z_0 \geq 0 ,
\end{cases} \\
L_z & \approx \left( { 2 \hbar^2 \pi^2 \over e E_z m_l} \right)^{1/3}.
\end{split}
\end{equation}

Applying these parameters to the envelope function, Eq.~(\ref{varEnv}), we can compute the intervalley coupling and the valley splitting, using Eq.~(\ref{discreteDelta}). 
The results are plotted as a solid green line in Fig.~\ref{fig:narrowInt}(a), showing very good agreement with numerical tight-binding and effective mass solutions. 
Thus, this simple variational model captures most of the interface physics. 

\begin{figure}[t] 
	\includegraphics[width=8cm]{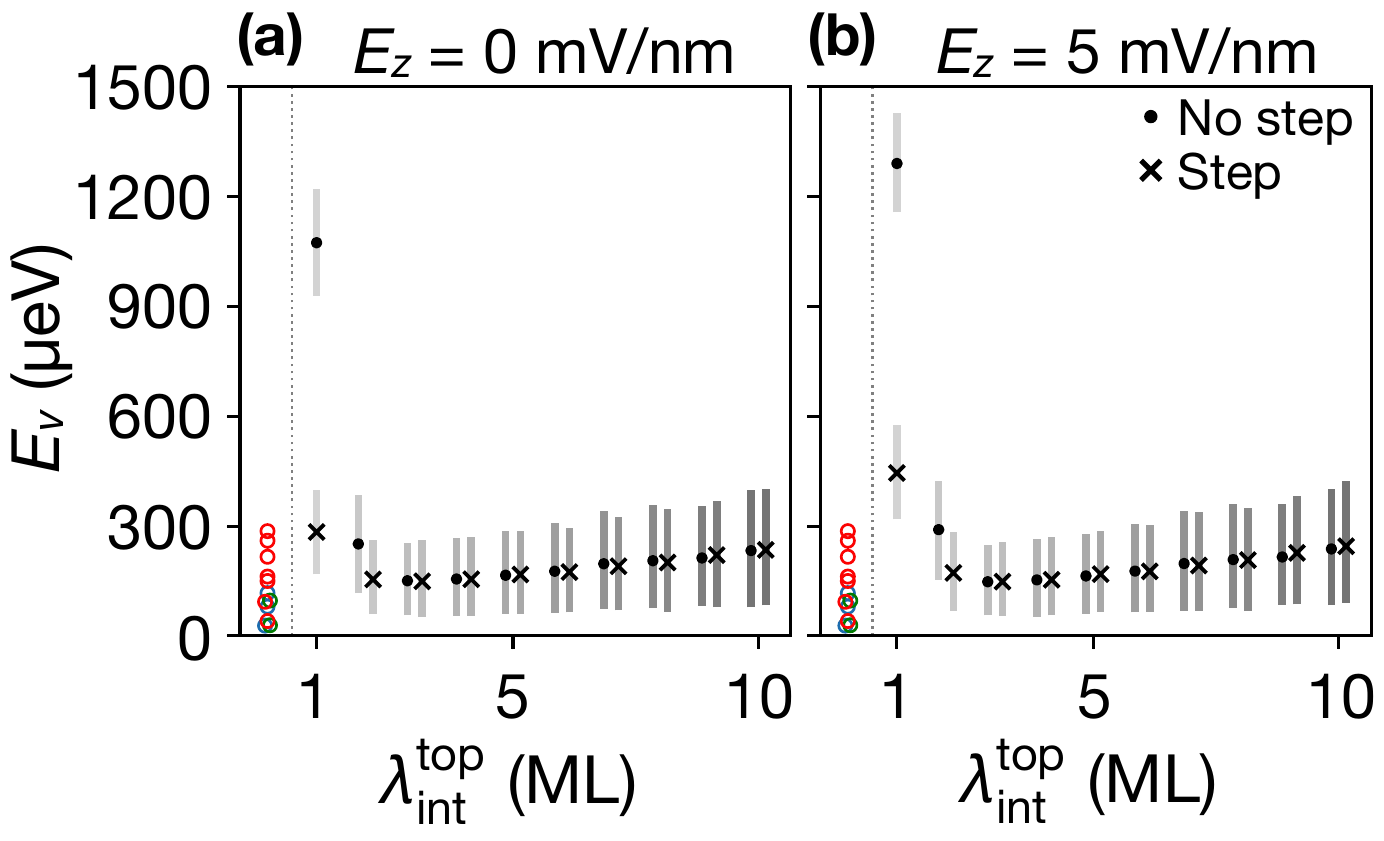}
	\centering
	\caption{
 Step disorder has the strongest effect on valley splitting in narrow wells with sharp top interfaces. 
 (a), (b) The mean values (dots and $\times$~markers) and the 10-90 percentile range (bars) for 1,000 2D two-band tight-binding valley splitting simulations of the quantum wells described in Section~\ref{sec:narrowWells}, with varying top interface widths $\lambda_\text{int}^\text{top}$, performed with no step at the interface (dots), or with a single step at the center of the dot confinement potential ($\times$~markers). 
 We use electric fields of (a) $E_z = 0$~mV/nm, or (b) $E_z = 5$~mV/nm, a well width $W = 3$~nm, and orbital energy splittings $\hbar \omega_\mathrm{orb} = 1.5$~meV. 
 $E_v$ data for the 3~nm quantum wells in Ref.~\cite{Chen:2021p044033} are included for comparison (open circles).}
	\label{fig:hrlSuppStepNostep}
\end{figure}

We can also derive a continuum approximation for the intervalley coupling by transforming the sum in Eq.~(\ref{discreteDelta}) to an integral and approximating the envelope function as linear near the interface, yielding
\begin{widetext} 
\begin{equation} \label{delta0narrow}
\begin{split}
\Delta_0 &\approx \int_{z_0}^{\lambda_\text{int}} dz\; e^{-2 i k_0 z} U_\text{qw}(z) \psi_\text{env}^2 (z) \\
& \approx {|\Delta E_c| \pi^2 \over k_0^3 L_z^3} 
\Bigg[
e^{-2 i k_0 \lambda_\text{int}} \left( {i } + {3 \over 4 k_0 \lambda_\text{int}} - { \lambda_\text{int} k_0 \over 2 } + { z_0 k_0 } - {i  z_0 \over  \lambda_\text{int} }  - { z_0^2 k_0 \over 2  \lambda_\text{int}} \right)  +
\begin{cases}
{i \over 2} e^{-2 i k_0 z_0} + \left( {i z_0 \over  \lambda_\text{int}} + { z_0^2 k_0 \over 2  \lambda_\text{int}}  -{3 \over 4 k_0  \lambda_\text{int} } \right), & z_0 < 0 \\
e^{-2ik_0z_0} \left( {i \over 2} - {3 \over 4 k_0 \lambda_\text{int}} - {i z_0 \over 2 \lambda_\text{int}} \right), & z_0 \geq 0
\end{cases}
\Bigg].
\end{split}
\end{equation}
\end{widetext}
Using $z_0$ and $L_z$ defined above and taking $E_v = 2|\Delta_0|$ gives an analytical expression for the valley splitting, plotted as a green dashed line in Fig.~\ref{fig:narrowInt}(a). 
Again, this model captures the significant decay of $\Delta_0$ for wide interfaces, although we find the continuum result lacks some of the structure captured by the discrete sum. 
This is due to the finite spacing between layers, as explained in Sec.~\ref{sec:interfaces}.

\subsection{Including alloy disorder}

Here, we study the same variational system with linear interfaces as above, using it to derive a scaling law for the average valley splitting $\bar E_v$ in the presence of alloy disorder, due to the overlap of the wavefunction with Ge in wide quantum well interfaces. 
In the wide interface limit, we can restrict our analysis to $z_0 \geq 0$. 
As discussed in Section~\ref{sec:interfaces}, the deterministic valley splitting $\Delta_0$ is suppressed for wide interfaces, so we only need to consider the contributions due to alloy disorder, $\delta \Delta$. 
Using Eq.~(\ref{varDelta}) for $\mathrm{Var} \left[\Delta\right]$ and approximating the discrete sum as an integral, we obtain
\begin{equation}
\begin{split}
    \mathrm{Var} \left[ \Delta \right] \approx {1 \over \pi} & \left[ {a_0^2 \Delta E_c \over 8 a_\mathrm{dot} (1-X_s) } \right]^2 \\ & {4 \over a_0}
    \int_{z_0}^{\lambda_\text{int}} dz \psi_\text{env}^4 (z) \bar X(z) \left[ 1 - \bar X(z) \right] .
\end{split}
\end{equation}
We can further simplify the calculation by approximating $\bar X(1-\bar X) \approx 1-\bar X$, for $\bar X\approx 1$. 
We then introduce the variational solution for the envelope function, Eq.~(\ref{varEnv}), and again apply a linear approximation near the interface, yielding
\begin{equation}
\sigma_\Delta^2=\mathrm{Var} \left[ \Delta \right] \approx {3 \over 160 \pi} {a_0^3 m_t \omega_\text{orb} e^2 E_z^2  \over (1 - X_s) \hbar} \lambda_\text{int}.
\label{eq:sigvar}
\end{equation}
We then finally obtain the result
\begin{equation} \label{wideInterfaces}
\bar E_v \approx  \sqrt{\pi \mathrm{Var}\left[\Delta\right]} = {1 \over 4} \sqrt{3 \over 10} \left[ {a_0^3 m_t \omega_\text{orb} e^2 E_z^2  \over (1 - X_s) \hbar} \lambda_\text{int} \right]^{1/2} .
\end{equation}

Equation~(\ref{wideInterfaces}) is plotted in gray in Fig.~\ref{fig:narrowInt}(b), giving good agreement with simulation data for smoothed linear interfaces. 
For perfectly linear interfaces, this formula is still valid and acts as a lower bound; however, Fourier components arising from the sharp corners raise $\bar E_v$ above this bound for the simulation results.

\section{Simulations of narrow quantum wells} \label{appendix:hrl}

\begin{figure*}[] 
\begin{minipage}{\linewidth}
\begin{algorithm}[H]
\caption{for optimizing Ge distributions, to maximize the valley splitting.}
\label{alg:maximizeEv}
\begin{algorithmic}[1]

\Require An array ($Y^\text{init}_l$) of minimum Ge concentrations for each layer $l$
\Require A maximum amount of Ge ($G_\text{max}$) that can be added to the total heterostructure, in units of atoms/$\text{nm}^2$ 
\Require $\epsilon > 0$ 

\State $Y^\text{curr} \gets Y^\text{init}$
\While{not converged}
    \State Estimate $\text{grad}_l$
    \State $Y^\text{next}_l \gets 
        \text{min} \left[1, \text{max}\left(Y^\text{curr}_l + \epsilon \cdot \text{grad}_l, Y^\text{init}_l \right) \right]$ 
    \Comment{Ensure that $Y_l^\text{next}$ is a valid concentration}
    \State $\delta Y^\text{next}_l \gets Y^\text{next}_l - Y^\text{init}_l$
    \State $Y_l^\text{next} \gets Y^\text{init}_l + \delta Y^\text{next}_l \times \text{min} \left[1, \left({  \frac{1}{2}(10^9 a_0)^2 G_\text{max} / \sum_l \delta Y^\text{next}_l} \right) \right]$ 
    \Comment{Limit the additional Ge added to $G_\text{max}$}
    \State $Y_l^\text{curr} \gets Y_l^\text{next}$
\EndWhile
\end{algorithmic}
\end{algorithm}
\end{minipage}
\end{figure*}

In this Appendix, we present additional simulations of the narrow wells considered in Sec.~\ref{sec:narrowWells}, justifying the parameter choices we made there. 
Figure~\ref{fig:hrlSuppOther}(a) illustrates how the choice of orbital energy affects the simulation results. 
Here we show the 10-90 percentile range for $E_v$, from 1,000 tight-binding simulations of a 3~nm quantum well with $\lambda_\text{int}^\text{top} = 3$~ML interfaces, for various orbital energy splittings $\hbar \omega_\mathrm{orb}$, assuming an isotropic dot. 
Larger orbital energies lead to larger average valley splittings because they give smaller dots, for which the Ge concentration fluctuations are larger. 
According to Eq.~(\ref{varDelta}), we expect the valley splitting in the disordered regime to scale as $\sqrt{\omega_\mathrm{orb}}$. 
We find that using $\hbar \omega_\text{orb} = 1.5$~meV yields results consistent with the experimental data.

\begin{figure}[b] 
	\includegraphics[width=8cm]{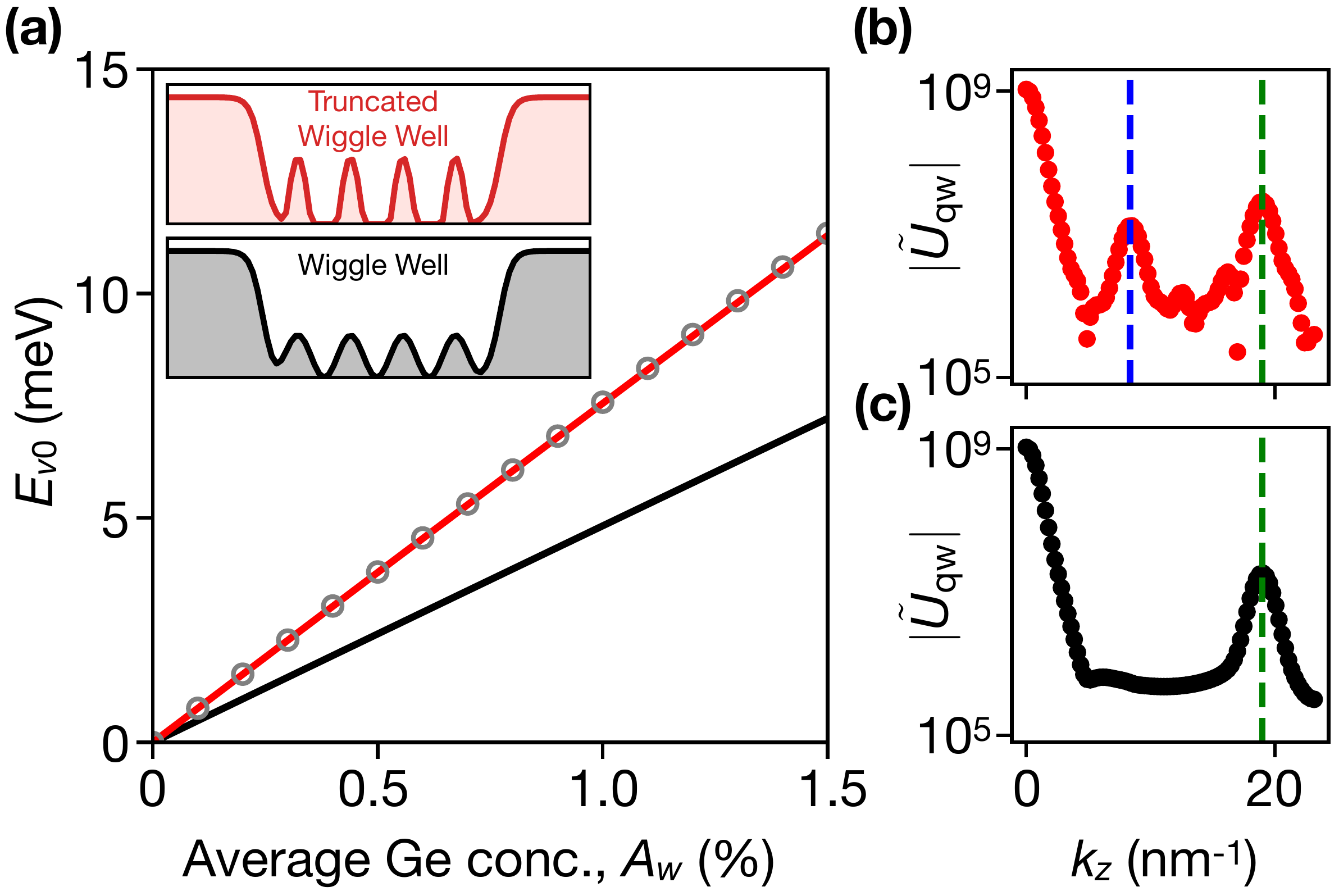}
	\centering
	\caption{
 For a fixed average Ge concentration in the quantum well $A_w$, the truncated Wiggle Well yields larger $E_v$ than the normal Wiggle Well. 
 (a) Deterministic valley splitting values $E_{v0}$, for normal ($E_{v0}^w$, black) and truncated ($E_{v0}^\text{tr}$, red) Wiggle Wells, plotted against the average Ge concentration in the quantum well, $A_w$. 
 The theoretical estimate $E_{v0}^\text{tr} = (\pi/2)E_{v0}^w$ is also shown with gray circles. 
 Inset: schematic illustrations of the truncated (red) and normal (black) Wiggle Wells.
 (b), (c) The weighted Fourier transforms of the weighted quantum well potentials $|\tilde U_\text{qw}|$ for (b) truncated and (c) normal Wiggle Wells, for the Ge concentrations $A_w = 0.01$.
 Vertical green dashed lines in (b) and (c) highlight the wavevector $k_z = 2k_0$, and the vertical blue dashed line in (b) highlights the wavevector $k_z = 4k_0$, aliased down to a lower value by the discrete lattice spacing. 
 All simulations are performed with a vertical electric field of $E_z = 5$~mV/nm, a quantum well width of $W=80$~ML, a sigmoidal interface of width $\lambda_\text{int} = 10$~ML, and Ge barrier concentration of $Y_s$ = 0.3.
 }
	\label{fig:truncatedWiggleWell}
\end{figure}

Figures~\ref{fig:hrlSuppOther}(b) and \ref{fig:hrlSuppOther}(c) show how the choice of vertical electric field, interface width, and well width interact. 
Larger vertical fields have a larger impact on wide wells, where they strongly increase the penetration of the wavefunction into the top barrier. 
On the other hand, the wavefunction in narrower wells is already strongly confined, so increasing the field has a smaller effect. 
Figures~\ref{fig:hrlSuppOther}(b) and \ref{fig:hrlSuppOther}(c) show the 10-90 percentile ranges of $E_v$, for quantum wells of width 3~nm and 8~nm, respectively, with various top interface widths $\lambda_\text{int}^\text{top}$ and vertical fields $E_z = 0$ and 5~mV/nm. 
We find that the field only modestly affects $E_v$ in the 3~nm well, but a larger field significantly increases $E_v$ in the 8~nm well. 
In particular, we find that the parameters $\lambda_\text{int}^\text{top}$ = 2-3~ML and $E_z = 5$~mV/nm yield results consistent with the data. 

Figures~\ref{fig:hrlSuppStepNostep}(a) and \ref{fig:hrlSuppStepNostep}(b) illustrate the effect of steps on a narrow quantum well. 
In this case, we consider a 3~nm well with varying $\lambda_\text{int}^\text{top}$, and the electric fields (a) $E_z=0$, or (b) $E_z=5$~mV/nm, where simulations were performed with and without a step through the center of the confinement potential.
Results show that steps strongly affect the valley splitting for very narrow interfaces, but the effect of a step becomes weak for $\lambda_\text{int}^\text{top} \geq 3$~ML. 
For the range of parameters simulated in Sec.~\ref{sec:narrowWells}, steps are found to have a modest impact on $E_v$.
Nonetheless, we find that alloy disorder is capable of explaining the full range of $E_v$ variations observed in the experimental data.

\section{Optimizing the Ge distribution} \label{appendix:algorithm}

In this Appendix, we provide details about the algorithm used in Sec.~\ref{algorithm} to optimize Ge concentrations in the quantum well. 
The algorithm pseudocode is outlined in Algorithm~\ref{alg:maximizeEv}. 
We begin with a realistic heterostructure profile $Y_l^\text{init}$ [shaded in gray in Figs.~\ref{fig:optimizePureVS}(a) and \ref{fig:optimize}(a)], and we allow the algorithm to only add Ge to this initial profile. 
We use an algorithm based on the method of projected gradient ascent. 
At each iteration, the gradient of the reward function with respect to the Ge concentration ($\text{grad}_l$) is computed for each layer $l$, and the corresponding Ge concentrations ($Y_l$) are adjusted by a small amount in the direction of the gradient. 
At each iteration, the resulting concentrations are then projected onto an acceptable parameter space as follows.
First, we ensure that the Ge concentration is never reduced below its initial value: $Y_l^\text{init} \leq  Y_l \leq 1$; this is accomplished by setting $Y_l=Y_l^\text{init}$ if $Y_l$ is too small, or $Y_l = 1$ if $Y_l$ is too large.
Second, we ensure that the total $Y_l$ never exceeds the maximum allowed density of additional Ge atoms, $G_\text{max}$, defined in units of atoms/$\text{nm}^2$.
This is accomplished by scaling the additional Ge added at each layer ($\delta Y_l$) by a common factor: 
\begin{equation} \label{eq:concScalingAlg}
    \delta Y_l \leftarrow \delta Y_l \times \min \left[1, \left( \frac{1}{2} \frac{\left(10^9 a_0\right)^2 G_\text{max}}{  \sum_l \delta Y_l }\right) \right].
\end{equation}
If less Ge is added than $G_\text{max}$, this factor is equal to 1, and nothing is changed. 
However, if the added Ge is greater than $G_\text{max}$, the added Ge at each layer is rescaled such that the density remains fixed at $G_\text{max}$.
[Note that the other factors appearing in Eq.~(\ref{eq:concScalingAlg}) convert $G_\text{max}$, in units of atoms/$\text{nm}^2$, to units consistent with $\sum_l \delta Y_l$.]
Convergence is achieved when the change in the reward function is no longer positive, for a small enough step size.

When we optimize $Y_l$ in the deterministic regime, as in Fig.~\ref{fig:optimizePureVS}, the tight-binding valley splitting is used as the reward function. 
Here, the gradient function is defined as $\text{grad}_l = \delta E_v^{\delta l} / \delta Y$, and is estimated as follows.\
First, we compute $E_v$ for the existing concentration profile. 
Then, separately for each layer $l$, we modify the concentration by a small amount, $\delta Y$, and recompute the resulting valley splitting $E_v^{\delta l}$, which includes this change. 
The ratio $(E_v^{\delta l} - E_v)/\delta Y$ provides an estimate of the gradient for each layer. 
In this work we choose $\delta Y = 10^{-8}$, and we find the algorithm step factor $\epsilon = 10^{-3}$ to be effective for this protocol. (See Algorithm~\ref{alg:maximizeEv}.)

When we optimize $Y_l$ in the disordered regime, as in Fig.~\ref{fig:optimize}, $\text{Var}\left[\Delta\right]$ is used as the reward function, as explained in the main text.
We can re-express Eq.~(\ref{varDelta}) in terms of Ge concentrations as follows:
\begin{equation}
    \text{Var}\left[\Delta\right] = {1 \over \pi} \left[{a_0^2 \Delta E_c \over 8 a_\mathrm{dot} (Y_s - Y_w) } \right]^2 \sum_l |\psi_l|^4 Y_l (1 - Y_l) ,
\end{equation}
where $Y_s$ is the Ge concentration in the barriers, $Y_w$ is the Ge concentration in the quantum well before adding extra Ge (set to 0 in this case), $Y_l$ is the Ge concentration at layer $l$, and $\psi_l$ is the value of the envelope function at layer $l$. 
In this case, we again consider small layer-by-layer variations of $\text{Var}[\Delta]$.
We then define the gradient functions as $\text{grad}_l = \delta \text{Var}[\Delta]^{\delta l} / \delta Y$ and proceed as in the previous paragraph.
In this case, we also find the algorithm step factor $\epsilon = 10^3$ yields suitable results.

\section{The truncated Wiggle Well} \label{app:truncWiggleWell}

In this Appendix, we validate the performance of the truncated Wiggle Well, which was the learned outcome of our optimization algorithm in Section~\ref{algorithm}. 
To do this, we consider both normal and truncated Wiggle Wells, where the Ge concentration oscillations are given by Eqs.~(\ref{eq:YWW}) and (\ref{eq:Ytrunc}), respectively.
The inset to Fig.~\ref{fig:truncatedWiggleWell}(a) schematically illustrates both types of wells. 
Figures~\ref{fig:truncatedWiggleWell}(b) and \ref{fig:truncatedWiggleWell}(c) show the Fourier components of the weighted quantum well potential $|\tilde U_\text{qw}|$, for the same average Ge concentration $A_w=0.01$, as described in the main text. 
We see in Fig.~\ref{fig:truncatedWiggleWell}(b) that the truncated Wiggle Well has a Fourier peak at $k_z = 2k_0$, and that the truncation also introduces higher harmonics into the spectrum, including a large peak at $4k_0$.
In the figure, the peak appears at lower $k_z$ values due to an aliasing effect caused by the finite spacing between layers. 
By zone folding, the aliased $4k_0$ peak location is given by $8\pi /a_0 - 4k_0$.
On the other hand, the normal Wiggle Well shown in Fig.~\ref{fig:truncatedWiggleWell}(c) has a peak at $k_z = 2k_0$ but no additional harmonics. 
As demonstrated in Fig.~\ref{fig:truncatedWiggleWell}(a), for a fixed average Ge concentration $A_w$, the truncated Wiggle Well produces larger $E_v$ values than the normal Wiggle Well. 
The relation between the valley splittings for truncated ($E_v^\text{tr}$) vs.\ normal ($E_v^w$) Wiggle Wells is very well represented by the theoretical estimate given in the main text, $E_v^\text{tr} = (\pi/2) E_v^w$, which is shown in Fig.~\ref{fig:truncatedWiggleWell}(a) as gray circles.

%

\end{document}